\documentstyle[preprint,aps,eqsecnum,epsfig]{revtex}
\tighten
\begin{document}
\draft
\title{\bf OUT OF EQUILIBRIUM FIELDS IN INFLATIONARY DYNAMICS. DENSITY
FLUCTUATIONS}
\author{{\bf D. Boyanovsky$^{(a)}$, D. Cormier$^{(b)}$,
H. J. de Vega$^{(c)}$, R. Holman$^{(b)}$, S. P. Kumar$^{(b)}$}}
\address
{(a)  Department of Physics and Astronomy, University of
Pittsburgh, Pittsburgh, PA. 15260, U.S.A. \\
 (b) Department of Physics, Carnegie Mellon University, Pittsburgh,
PA. 15213, U. S. A. \\
 (c)  LPTHE, 
Universit\'e Pierre et Marie Curie (Paris VI) 
et Denis Diderot  (Paris VII), Tour 16, 1er. \'etage, 4, Place Jussieu
75252 Paris, Cedex 05, France}
\date{December 1997}
\maketitle
\begin{abstract}
The energy and time scales during the inflationary stage of the
universe calls for an out of equilibrium quantum field
treatment. Moreover, the high energy densities involved ($ \sim 1/g
\sim 10^{12} $) make necessary the use of non-perturbative approaches
as the large $ N $ and Hartree methods. We start these lectures by 
introducing the such non-perturbative out of equilibrium methods in
cosmological universes. We discuss the renormalization procedure and
the choice of initial conditions. We then study with these methods 
the non-linear dynamics of quantum fields in matter and radiation
dominated FRW  and de Sitter universes. For a variety of initial
conditions, we compute the evolution of the inflaton, its quantum
fluctuations and the equation of state. We investigate
the phenomenon of explosive particle production due to spinodal instabilities
and parametric amplification in FRW and de Sitter universes with and
without symmetry breaking. We find that the particle production
is somewhat sensitive to the expansion of the universe.  
In the large $N$ limit for symmetry breaking scenarios, we determine
generic late time fields behavior for any flat FRW and de Sitter cosmology. 
We find that quantum fluctuations damp in FRW as the square of the
scale factor while the order parameter approaches a minimum of the
potential at the same rate. We  present a complete
and numerically accessible renormalization scheme for the
equation of motion and the energy momentum tensor in flat
cosmologies. In this scheme the renormalization constants are  
independent  of time and of the initial conditions. 
Furthermore, we consider an $ O(N) $ inflaton model coupled
self-consistently to gravity in the 
semiclassical approximation, where the field is subject to `new inflation'
type initial conditions. We study the dynamics  {\bf self-consistently and
non-perturbatively} with non-equilibrium field theory methods in
the large $ N $ limit. We find that spinodal instabilities drive the growth
of non-perturbatively large quantum fluctuations which shut off the
inflationary growth of the scale factor. We find that 
a very specific combination of  these large
fluctuations plus the inflaton zero mode assemble into a new effective field.
This new field behaves classically and it is the object which actually
rolls down. We show how  this reinterpretation qualitatively saves the standard
picture of how metric perturbations are 
generated during inflation and that the spinodal growth of fluctuations
dominates the time dependence of the Bardeen variable for superhorizon modes
during inflation.  We compute the amplitude and index for the spectrum of
scalar density and tensor perturbations and argue that in all models of this
type the spinodal instabilities are responsible for a `red' spectrum of
primordial scalar density perturbations.  The
decoherence aspects and the quantum to classical transition through inflation
are studied in detail by following the full evolution of the density matrix.
\end{abstract}

\section{Introduction and Motivation}

Inflationary cosmology has come of age. From its beginnings as a solution to
deep problems such as the horizon, flatness,  entropy and monopole
problems\cite{guth}, it has grown into the main contender for the source of
primordial fluctuations giving rise to large scale
structure\cite{defe}. There is evidence from the
measurements of temperature anisotropies in the cosmic microwave background
radiation (CMBR) that the scale invariant power spectrum predicted by generic
inflationary models is at least consistent with
observations\cite{kolb,turner,lyth1} and we can expect further
and more exacting tests of the inflationary power spectrum when the MAP and
PLANCK missions are flown. In particular, if the fluctuations that are
responsible for the temperature anisotropies of the CMB truly originate from
quantum fluctuations during inflation, determinations of the spectrum of scalar
and tensor perturbations will constrain inflationary models based on particle
physics scenarios and probably will validate or rule out specific
proposals\cite{turner,lyth2}. 

The tasks for inflationary universe researchers are then two-fold. First,
models of inflation must be constructed on the basis of 
a realistic particle physics model. This is in contrast
to the current situation where most, if not all acceptable inflationary models
are ad-hoc in nature, with fields and potentials put in for the sole purpose of
generating an inflationary epoch. Second, and equally important, we must be
sure that the quantum dynamics of inflation is well understood. This is
extremely important, especially in light of the fact that it is {\it
exactly} this 
quantum behavior that is supposed to give rise to the primordial metric
perturbations which presumably have imprinted themselves in the CMBR. This
latter problem is the focus of this review. 

The inflaton must be treated as a {\it non-equilibrium} quantum field . The
simplest way to see this comes from the requirement of having small enough
metric perturbation amplitudes which in turn requires that the quartic self
coupling 
$ \lambda $ of the inflaton be extremely small, typically of order $ \sim
10^{-12} $. Such a small coupling cannot establish local thermodynamic
equilibrium (LTE) for {\it all} field modes; typically the long wavelength
modes will respond too slowly to be able to enter LTE. In fact, the
superhorizon sized modes will be out of the region of causal contact and
cannot thermalize. We see then that if we
want to gain a deeper understanding of inflation, non-equilibrium tools must be
developed. Such tools exist and have now been developed to the point that they
can give quantitative answers to these questions in 
cosmology\cite{barrabajo} -\cite{grure},\cite{ctp,hu,motola,largen}. 
These methods permit us to follow the {\bf dynamics} of quantum fields
in situations where the energy  
density is non-perturbatively large ($ \sim 1/\lambda $). That is, they allow
the computation of the time evolution of non-stationary states and of
non-thermal  density matrices.

Our programme on non-equilibrium dynamics of quantum field theory, started in
 1992\cite{barrabajo}, is naturally poised to provide a framework to
 study these  problems. The larger goal of the program is to study the
 dynamics of  non-equilibrium processes from a fundamental
 field-theoretical description, by solving  the dynamical 
equations of motion of the underlying four dimensional quantum field
 theory for physically relevant problems:  the early universe
 dynamics, high energy particle collisions, phase transitions out of
 equilibrium,  symmetry breaking and  dissipative processes.  

The focus of our work is to describe the quantum field dynamics when
the  energy density is {\bf high}. That is, a large number of particles per
volume $ m^{-3} $, where $ m $ is the typical mass scale in the
theory. Usual S-matrix calculations apply in the opposite limit of low
energy density and since they only provide information on {\em in}
$\rightarrow$ {\em out} matrix elements,  are unsuitable for calculations of
expectation values. 

In high  energy density situations such as in the early universe,
the particle propagator (or Green function) depends on the particle
distribution in momenta in a nontrivial way. This makes the 
quantum dynamics intrinsically nonlinear and calls to the use of
self-consistent non-perturbative approaches as the large $N$ limit,
Hartree and self-consistent one-loop approximations. 

There are basically three different levels to study the early universe
dynamics:
\begin{enumerate}

\item To work out the nonlinear dynamics of quantum fields in
Minkowski spacetime. By non-linear dynamics we understand to solve the
quantum equations of motion {\bf  including} the quantum back-reaction
quantitatively\cite{barrabajo} -\cite{late},
\cite{kaiser,baacke,motola,largen}.
This level is in fact appropriate to describe high
energy particle collisions \cite{dcc}. 

\item To work out the nonlinear dynamics of quantum fields in fixed
cosmological backgrounds\cite{frw2,De Sitter}. New phenomena arise
then compared  with 1.  showing that a Minkowski analysis is not
quantitatively precise for expanding universes. 

\item A self-consistent treatment of the quantum fields and the
cosmological background\cite{din,ramsey}. That is, the metric is
obtained dynamically from the quantum fields (matter source)
propagating in the that metric.  

\end{enumerate}

We shall successively present the three levels of study. The first
stage was reviewed in the 1996 Chalonge School \cite{mink} (see
\cite{late} for further progress). The second level is the subject of
secs. VII and VIII. We study the parametric  and spinodal
resonances both in FRW and de Sitter backgrounds
wide range of initial conditions both in FRW and de Sitter backgrounds
\cite{frw,De Sitter}. [Parametric resonance appears in chaotic
inflationary scenarios for unbroken symmetry whereas spinodal
unstabilities show up in new inflation scenarios with broken symmetry].
Both types of  unstabilities shut-off through the
non-linear quantum evolution as described in secs.  VII and VIII
\cite{frw,De Sitter} both analytically and numerically. 
We follow the equation of state of the quantum matter during the evolution
and analyze its properties. 

The third stage of 
our approach is to apply non-equilibrium quantum field theory techniques to the
situation of a scalar field coupled to {\it semiclassical} gravity, where the
source of the gravitational field is the expectation value of the stress energy
tensor in the relevant, dynamically changing, quantum state. In this way
we can go beyond the standard analyses\cite{linde2,vilenkin,stein,guthpi} which
treat the background as fixed and do not consider the non-linear
quantum field dynamics. 

In all cases 1. - 3. , the quantum fields energy- momentum tensor is
covariantly conserved both at the regularized as well as the
renormalized levels \cite{us1} - \cite{din}.

We  mainly consider for the stage 3.   new inflation scenarios where a
scalar field 
$\phi$ evolves under the action of a typical symmetry breaking potential. The
initial conditions will be taken so that the initial value of the order 
parameter is near the top of the potential
(the disordered state) with essentially zero time derivative. 
What we find is that the existence of spinodal instabilities, i.e. the fact
that eventually (in an expanding universe) all modes will act as if they have a
{\it negative} mass squared, drives the quantum fluctuations to grow {\it
non-perturbatively} large. We have the picture of an initial wave-function or
density matrix peaked near the unstable state and then spreading until it
samples the stable vacua. Since these vacua are non-perturbatively far from
the initial state (typically $\sim m\slash \sqrt{\lambda}$, where $m$ is the
mass scale of the field and $\lambda$ the quartic self-coupling), the spinodal
instabilities will persist until the quantum fluctuations, as encoded in the
equal time two-point function $\langle \Phi(\vec{x}, t)^2 \rangle$, grow to
${\cal O}( m^2\slash \lambda$).

This growth eventually shuts off the inflationary behavior of the scale factor
as well as the growth of the quantum fluctuations (this last also happens
in Minkowski spacetime \cite{us1,mink,late}).

The scenario envisaged here is that of a quenched or super-cooled phase 
transition where the order parameter is zero or very small. Therefore one is 
led to ask: 

a) What is rolling down?. 

b) Since the quantum fluctuations are non-perturbatively large ( $ \sim  
1/\lambda $), will not they modify drastically the FRW dynamics?.

c) How can one extract (small?) metric perturbations from non-perturbatively 
large field fluctuations?

We address the questions a)-c) as well as  other issues in sec. IX. 

We choose such type of new inflationary scenario because the issue of
large quantum fluctuations is particularly dramatic there. However,
our methods do apply to any inflationary scenario as  chaotic,
extended and hybrid inflation.

\section{Quantum Field theory around an excited state with non-perturbative
energy density} 

To start we present a simple study of quantum field theory around an
excited state. The relevant situation both in cosmology and high
energy particle scattering is when the energy density is large
($ \sim {\cal O}(1/\lambda)$,  non-perturbative)  with respect to the
ground state. For simplicity we hall make the derivation in Minkowski
spacetime. The  generalization to cosmological spacetimes is
straightforward. 

Let us consider a scalar field model with quartic selfcoupling
\begin{equation}
{\cal L} = \frac12 (\partial_{\mu}\Phi)^2 - {m^2 \over 2}\; \Phi^2 -
{\lambda \over 4!}\; \Phi^4 \; .
\end{equation}
Here $ \Phi(x) $ is a real field. The equations of motion take the
form
\begin{equation}\label{ecmov}
\partial^2 \Phi + m^2\; \Phi + \frac{\lambda}{6}\;\Phi^3 = 0\; ,
\end{equation}
and the canonical momentum is
\begin{equation}\label{momca}
\Pi(x) = {{\partial {\cal L}}\over {\partial\dot \Phi}} =  {\dot \Phi}(x)\; .
\end{equation}
For  classical fields, the solutions of the equations of motion
(\ref{ecmov}) depend on the coupling constant $  \lambda $ as follows,
$$
\Phi_c(x) = {m \over \sqrt{\lambda}}\;  F(mx)
$$
where the index $ _c $ indicates that we are considering c-number
solutions of eqs.(\ref{ecmov}) and 
$ F $ is a dimensionless functions. Therefore, the classical
solutions are large for small coupling and the energy density
$$
{\cal H} = \frac12 \left[ {\dot \Phi}^2+ (\nabla \Phi)^2 \right] +
{m^2 \over 2}\;\Phi^2 +{\lambda \over 4!}\; \Phi^4
$$
scales as $ m^4/\lambda$ and is large too. Notice that here $ \lambda \sim
\hbar $. Therefore, for small $\lambda$ we should expect some
semiclassical behavior. 

Let us now consider the quantum field $ \Phi(x) $. 
$ \Phi(x) $ and $ \Pi(x) $ will be operators in a Fock space obeying
the canonical commutation rules,
\begin{equation}\label{ccr}
\left[\Phi({\vec x},t),  \Pi({\vec y},t)\right] = i \;\delta( {\vec
x}-{\vec y}) \; .
\end{equation}

Let us consider a quantum state $ |> $ which is not the vacuum. The
expectation value of $ \Phi(x) $ there
$$
\phi(x) \equiv \; <\Phi(x) >
$$
will in general be a function of space and time. $\phi(x)$ will be
typically of order $ {m \over \sqrt{\lambda}} $ for small $ \lambda $
just for correspondence with the classical theory.

We write the operator $ \Phi(x) $ as,
$$
\Phi(x) = \phi(x) + \psi(x)\; ,
$$
where $ \psi(x) $ is a new quantum operator. $ \psi(x) $ must obey the
constraint
$$
<\psi(x)> = 0 \; .
$$
The equations of motion (\ref{ecmov}) hold in the quantum theory for
the operator  $ \Phi(x) $. Taking the expectation value of
eq.(\ref{ecmov}) in the state $ |> $ yields
\begin{equation}\label{ecfi}
\left(\partial^2  + m^2\right)\phi(x) + \frac{\lambda}{6}\phi(x)^3 +
\frac{\lambda}{2} 
\phi(x)  <\psi(x)^2> +  \frac{\lambda}{6}  <\psi(x)^3> = 0\; .
\end{equation}
Subtracting eq.(\ref{ecfi}) from eq.(\ref{ecmov}) yields,
\begin{equation}\label{ecmopsi}
\left[\partial^2  + m^2  + \frac{\lambda}{2}\psi(x)^2 \right]\psi(x) +
 \frac{\lambda}{2} \phi(x) \left[ \psi(x)^2 - <\psi(x)^2> \right] + 
 \frac{\lambda}{6} \left[ \psi(x)^3 - <\psi(x)^3> \right] = 0 \; .
\end{equation}
In order to solve (\ref{ecmopsi}) perturbatively in  $ \lambda $ but
keeping $ \phi(x) $ {\bf arbitrary}, we introduce the Green function,
$$
\left[\partial^2  + m^2  + \frac{\lambda}{2}\phi(x)^2 \right]G(x,y) =
\delta^4(x-y) \; .
$$
We consider the retarded Green function which vanishes for $ x^0 < y^0
$. This Green function (propagator) depends on the field expectation value $
\phi(x) $ showing that the particle propagation depends here on the
properties of the state. As is well known, this is not the case on the
vacuum where the propagator is explicitly known.

With the help of this retarded Green function, we  can  write the
exact evolution equation (\ref{ecmopsi})  as follows
\begin{eqnarray}\label{ecint}
\psi(x)= \psi_0(x) - \frac{\lambda}{6} \int d^4y \, G(x,y) \left\{ 3 \phi(y)
\left[ \psi(y)^2 - <\psi(y)^2> \right] +\psi(y)^3 - <\psi(y)^3>
\right\}
\end{eqnarray}
Here, $ \psi_0(x) $ stands for the general operator solution of the
homogeneous equation:
$$
\left[\partial^2  + m^2  + \frac{\lambda}{2}\phi(x)^2 \right]\psi_0(x) = 0
$$
with $ <\psi_0(x)> =0 $.  $ \psi_0(x) $ will be expressed as a sum of
c-number eigenfunctions times creation and annihilation operators (see
below eq.(\ref{psiF}). It then follows that
$$
 <\psi_0(x)^2> \neq 0 \quad \mbox{but} \quad <\psi_0(x)^3> =0 \; .
$$
For small $ \lambda $ we see from eq.(\ref{ecint}) that
$$
\psi(x)= \psi_0(x) + {\cal O}(\lambda) \; .
$$
Inserting that information into eq.(\ref{ecfi}) yields,
\begin{equation}\label{ecfi0}
\left[\partial^2  + m^2 +  \frac{\lambda}{2} <\psi_0(x)^2>
\right]\phi_0(x) + \lambda\phi_0(x)^3 = 0\; ,
\end{equation}
where,
$$
\phi(x) = \phi_0(x) + {\cal O}(\lambda)\; .
$$
In addition, we can consider $ \psi_0(x) $ as the general operator
solution of the equation
\begin{equation}\label{ecpsi0}
\left[\partial^2  + m^2  + \frac{\lambda}{2}\; \phi_0(x)^2 \right]\psi_0(x) = 0
\end{equation}
Let us consider from now on spatially homogeneous states $ |> $. Then,
$$
 <\Phi(x) > = \phi(t) 
$$
only depends on time. This is the relevant situation for early
universe investigations. We can now Fourier expand $ \psi_0(x) $ as
follows,
\begin{equation}\label{psiF}
\psi_0(\vec x, t) = \int\frac{d^3k}{\sqrt{2}(2\pi)^{3/2}}
\left[a_{\vec k} \; f_k(t) \; e^{i\vec{k}\cdot \vec x} + a^{\dagger}_{\vec k}  
 \; f^*_k(t) \; e^{-i\vec{k}\cdot \vec x} \right] \; .
\end{equation}
Here,  $ a^{\dagger}_{\vec k} $ and $ a_{\vec k} $  stand for creation
and annihilation operators on the state $ |> $. That is,
\begin{equation}\label{avac}
 a_{\vec k} |> = 0 \quad \mbox{and} \quad a^{\dagger}_{\vec k}|> \neq 0
 \quad \mbox{for $\;$ all}\;  \vec{k} 
\end{equation}
and
\begin{equation}\label{aadaga}
[ a_{\vec k}, a^{\dagger}_{\vec k'} ] = \delta({\vec k}-{\vec k'}) \;.
\end{equation}

The mode functions $  f_k(t) $ have here a non-trivial time
dependence. [Recall that they have just harmonic time dependence
around the vacuum]. Actually, most of the field dynamics is encoded in
their evolution defined by eq.(\ref{ecpsi0})
$$
\left[\frac{d^2}{dt^2}+k^2 + m^2+ \frac{\lambda}{2} \; \phi_0(t)^2
\right]f_k(t)= 0\; ,  
$$
The canonical momentum (\ref{momca}) is Fourier expanded as follows,
\begin{equation}
\pi_0(\vec x, t) = \int\frac{d^3k}{\sqrt{2}(2\pi)^{3/2}}
\left[a_{\vec k} \; {\dot f}_k(t) \; e^{i\vec{k}\cdot \vec x} +
a^{\dagger}_{\vec k}   
 \; {\dot f}^*_k(t) \; e^{-i\vec{k}\cdot \vec x} \right] \; .
\end{equation}
It is easy to show  now that the canonical commutation rules (\ref{ccr})
hold using eq.(\ref{aadaga}) and the time independence of the
Wronskian that we normalize as follows:
$$
W[f_k(t),f^*_k(t)] \equiv  f_k(t) {\dot f}^*_k(t)- f^*_k(t){\dot
f}_k(t) = 2 i \; .
$$
It is then convenient to choose as initial conditions for the mode
functions:
\begin{equation}\label{conini}
 f_k(0) = {1 \over \sqrt{\omega_k}} \quad , \quad {\dot f}_k(0) = -i
 \sqrt{\omega_k} 
\end{equation}
with
$$
\omega_k \equiv \sqrt{ k^2 + m^2 + \frac{\lambda}{2}  \;\phi(0)^2}
$$
One must specify in addition the values of 
\begin{equation}\label{conini0}
\phi(0) = \phi_0 \quad \mbox{and} \quad {\dot \phi}(0)= p_0 \; .
\end{equation}

We can now compute the expectation value $  <\psi_0(x)^2> $ that
appears in eq.(\ref{ecfi0}). We get using eqs.(\ref{psiF}),
(\ref{avac}) and (\ref{aadaga},
$$
<\psi_0(x)^2> =  \int \frac{d^3k}{ (2\pi)^3}\; |f_k(t)|^2 =
\int \frac{k^2 \, dk}{ 2\pi^2}\; |f_k(t)|^2
\; .
$$
This integral diverges in the ultraviolet. We put for the moment a
cutoff $ \Lambda $ in the momentum. We shall discuss below (sec. V) how the
cutoff dependence can be absorbed into mass a coupling constant
renormalization. 

In summary, we have obtained a set of self-consistent evolution
equations for the quantum state $ |> $:
\begin{eqnarray}\label{suma}
&&\left[\frac{d^2}{dt^2}+k^2 + m^2 + \frac{\lambda}{2}\phi_0(t)^2
\right]f_k(t)= 0\; , \cr \cr 
&&\left[\frac{d^2}{dt^2}+ m^2+\frac{\lambda}{2}\int_0^{\Lambda} {{k'^2
dk'}\over {2 \pi^2}} |f_{k'}(t)|^2\right] \phi_0(t) + \lambda\phi_0(t)^3
= 0 \; ,
\end{eqnarray}
with the initial conditions defined by
eqs.(\ref{conini})-(\ref{conini0}). The choice of the initial
conditions defines the state $ |> $. We have an infinite number of
unknowns: 
$$
f_k(t), \phi_0(t) \quad \mbox{for} \quad 0 \leq k < \infty \quad
\mbox{and}  \quad t>0 \; .
$$
defined by the coupled nonlinear equations (\ref{suma}).

All physical quantities can be computed in terms of the mode functions
$ f_k(t) , \; 0 \leq k < \infty $ and the order parameter $ \phi_0(t) $. 
We have described a small coupling or self-consistent   one-loop
approximation. The large $ N $ and Hartree approximations are discussed
in sec. IV. Large $ N $ and Hartree contain definitely more
information than the self-consistent one-loop approximation
(\ref{suma}). However, the 
simplicity of its derivation makes it a very useful and pedagogical
exercise.    

\section{Non-Equilibrium Quantum Field Theory, Semiclassical Gravity and
Inflation}

We present here the framework of the non-equilibrium closed time path
formalism. For a more complete discussion, the reader is referred to
\cite{us1}, or the alternative approach given in \cite{frw}. 

The time evolution of a system is determined in the Schr\"odinger picture
by the functional Liouville equation
\begin{equation}
i\frac{\partial \rho (t)}{\partial t} = [H(t),\rho(t)],
\label{liouville}
\end{equation}
where $\rho$ is the density matrix and we allow for an explicitly time
dependent Hamiltonian as is necessary to treat quantum fields in a time
dependent background.  Formally, the solutions to this equation for the time
evolving density matrix are given by the time evolution operator, $U(t,t^{'})$,
in the form
\begin{equation}
\rho(t) = U(t,t_0)\rho(t_0)U^{-1}(t,t_0).
\label{rhoevol}
\end{equation}
The quantity $\rho(t_0)$ determines the initial condition for the evolution. We
choose this initial condition to describe a state of local equilibrium in
conformal time, which is also identified with the conformal adiabatic vacuum
for short wavelengths.  In the appendix we provide an analysis and discussion
of different initial conditions and their physical content within the context
of expanding cosmologies.

Given the evolution of the density matrix (\ref{rhoevol}), ensemble averages
of operators are given by the expression (again in the Schr\"odinger picture)
\begin{equation}
\langle{\cal O}(t)\rangle = 
\frac{Tr[U(t_0,t){\cal O}U(t,t^{'})U(t^{'},t_0)\rho(t_0)]}{Tr\rho(t_0)},
\label{expect}
\end{equation}
where we have inserted the identity, $U(t,t^{'})U(t^{'},t)$ with $t^{'}$ an
arbitrary time which will be taken to infinity. The state is first evolved
forward from the initial time $t_0$ to $t$ when the operator is inserted.  We
then evolve this state forward to time $t^{'}$ and back again to the initial
time.

The actual evolution of various quantities in the theory can now be evaluated
by either constructing the appropriate Green functions as in \cite{us1}, or by
choosing an explicit Ansatz for the functional form of the time dependent
density matrix so that the trace in (\ref{expect}) may be explicitly evaluated
as a functional integral (see \cite{frw}).  The methods are equivalent, and
provide the results which will be presented below for the cases of interest.

We shall study the inflationary dynamics in a spatially flat
Friedmann-Robertson-Walker background with scale factor $a(t)$ and
line element: 
\begin{equation}
ds^2 = dt^2 - a^2(t)\; d\vec{x}^2.
\label{metric}
\end{equation}
Our Lagrangian density has the form
\begin{equation}
{\cal L} =\sqrt{-g}\left[ \frac12 \nabla_\mu\Phi\nabla^\mu\Phi -
V(\Phi)\right]\; . \label{lagrangian}
\end{equation}
Our approach can be generalized to open as well as closed cosmologies.

Our program incorporates the non-equilibrium
behavior of the quantum fields involved in inflation into a framework
where the geometry (gravity) is dynamical and is  
treated  self consistently. We do this
via the use of semiclassical gravity\cite{birrell} where we say that
the metric is classical and determined through  the Einstein equations 
using the expectation value of the stress energy tensor $\langle
T_{\mu \nu} \rangle$. Such
 expectation value is taken in the dynamically determined state
described by the density matrix $\rho(t)$. This dynamical problem can be
described schematically as follows: 

\begin{enumerate}

\item{The dynamics of the scale factor $a(t)$ is driven by the 
semiclassical Einstein equations 

\begin{equation}
\frac{1}{8\pi G_R} G_{\mu \nu} + \frac{\Lambda_R}{8\pi G_R} \; g_{\mu \nu} +
\left(\rm{higher\  curvature}\right)= -\langle T_{\mu \nu}\rangle_R .
\end{equation}
Here $G_R, \Lambda_R$ are the renormalized values of Newton's constant and the
cosmological constant, respectively and $G_{\mu \nu}$ is the Einstein
tensor. The higher curvature terms must be included to absorb
ultraviolet divergences.}

\item{On the other hand, the density matrix $\rho(t)$ of the matter
(that determines $\langle T_{\mu \nu}\rangle_R$) obeys the Liouville equation

\begin{equation}
i\frac{\partial \rho(t)}{\partial t} = \left[H, \rho(t)\right],
\end{equation}
where $H$ is the evolution Hamiltonian, which is dependent on the scale factor,
 $a(t)$.}
\end{enumerate}

It is this set of equations we must try to solve; it is clear that initial
conditions must be appended to these equations for us to be able to arrive at
unique solutions to them. Let us discuss some aspects of the initial state of
the field theory first.

\subsection{On the initial state: dynamics of phase transitions}

The situations we consider are 
\begin{enumerate}
\item{the theory
admits a symmetry breaking potential and in which the field expectation value
starts its evolution near the unstable point.}
\item{The symmetry is not broken and  the field expectation value
starts its evolution at a finite distance from the absolute minimum.}
\end{enumerate}

 There is an
issue as to how the field got to have an expectation value near the
unstable point (typically at $\Phi=0$) as well as an issue concerning
the initial state of the non-zero momentum modes.

Since our background is an FRW spacetime, it is spatially homogeneous and we
can choose our state $\rho(t)$ to respect this symmetry. Starting from the full
quantum field $\Phi(\vec{x},t)$ we can extract a part that has a natural
interpretation as the zero momentum, c-number part of the field by writing:

\begin{eqnarray}
\Phi(\vec{x}, t)& = &\phi(t) + \Psi(\vec{x}, t) \nonumber \\ \phi(t) & = & {\rm
Tr}[\rho(t)\Phi(\vec{x}, t)]\equiv \langle \Phi(\vec{x}, t)) \rangle.
\end{eqnarray}
The quantity $\Psi(\vec{x}, t)$ represents the quantum fluctuations about the
zero mode $\phi(t)$ and clearly satisfies $\langle \Psi(\vec{x}, t) \rangle=0$.

We need to choose a basis to represent the density matrix. A natural choice
consistent with the translational invariance of our quantum state is that given
by the Fourier modes, in {\it comoving} momentum space, of the quantum
fluctuations $\Psi(\vec{x}, t)$:

\begin{equation}
\Psi(\vec{x}, t) = \int \frac{d^3 k}{(2 \pi)^3} \exp(-i\ \vec{k} \cdot \vec{x})\
\psi_k (t) .
\end{equation}

In this language we can state our ansatz for the initial condition of the
quantum state as follows. We take the zero mode $\phi(t=0)=\phi_0,\
\dot{\phi}(t=0)=0$, where $\phi_0$ will typically be very near the
origin for broken symmetry and at a finite distance from it in the
unbroken symmetry case. The initial conditions on the the nonzero
modes $\psi_k (t=0)$ will be chosen such that the initial density
matrix $\rho(t=0)$ describes a vacuum state 
(i.e. an initial state in local thermal equilibrium at a temperature $T_i=0$).
There are some subtleties involved in this choice. First, as
explained in \cite{frw2}, in order for the density matrix to commute with the
initial Hamiltonian, we must choose the modes to be initially in the conformal
adiabatic vacuum (these statements will be made more precise below). This
choice has the added benefit of allowing for time independent renormalization
counterterms to be used in renormalizing the theory.

We are making the assumption of an initial vacuum state
in order to be able to proceed with the calculation. It would be
interesting to understand what forms of the density matrix can be used for
other more general initial conditions. 

The assumptions of an initial equilibrium vacuum state are essentially 
the same used in refs. \cite{linde2}, \cite{vilenkin} and
\cite{guthpi} in the analysis of the quantum mechanics of inflation in
a fixed de Sitter background. 

As discussed in the introduction, if we start from such an initial state,
spinodal or parametric instabilities will drive the growth of
non-perturbatively large 
quantum fluctuations. In order to deal with these, we need to be able to
perform calculations that take these large fluctuations into account. 
Although the quantitative features of the dynamics will depend on the
initial state, the qualitative features associated with spinodal or
parametric unstabilities are fairly robust for a wide choice of
initial states that describe a phase transition.

\section{The Inflaton Model and the Equations of Motion}

Having recognized the appearance of large quantum fluctuations driven
by parametric or spinodal unstabilities, we need to study the dynamics
within a non-perturbative framework. That is, a framework allowing calculations
for non-perturbatively large energy densities. 
We require that such a framework be: i) renormalizable, ii)
covariant energy conserving, iii) numerically implementable.  There are very
few schemes that fulfill all of these criteria: the Hartree and the large $ N $
approximation\cite{us1}-\cite{din}. Whereas the Hartree approximation
is basically a Gaussian variational approximation\cite{jackiwetal} that
in general cannot be consistently improved upon, the large $ N $ approximation
can be consistently implemented beyond leading order\cite{motola,largen}.
In addition, the presence of a large number of fields in most of the
GUT's models suggest that the large $ N $ limit will be actually a
realistic one. Moreover, for the case of broken symmetry
it has the added bonus of providing many light fields (associated with
Goldstone modes) that will permit the study of the effects of other fields
which are lighter than the inflaton on the dynamics. Thus we will
study the inflationary dynamics  within the framework of
the large $ N $ limit of a scalar theory in the vector representation of $ O(N)
$ both for unbroken and broken symmetry. In the second case we will
have a quenched phase transition. 

We assume that the universe is spatially flat with a metric given by
eq.(\ref{metric}). The matter action and Lagrangian density are given
by eq.(\ref{lagrangian}), 
\begin{equation}
S_m  =  \int d^4x\; {\cal L}_m = \int d^4x \;
a^3(t)\left[\frac{1}{2}\dot{\vec{\Phi}}^2(x)-\frac{1}{2} 
\frac{(\vec{\nabla}\vec{\Phi}(x))^2}{a^2(t)}-V(\vec{\Phi}(x))\right]
\label{action}
\end{equation}
\begin{equation}
V(\vec{\Phi})  =  \frac{m^2}2\; \vec{\Phi}^2 +
\frac{\lambda}{8N}\left(\vec{\Phi}^2\right)^2 +\frac12 \, \xi\; {\cal
R} \;\vec{\Phi}^2  \;, \label{potential}
\end{equation}
where $ m^2 > 0 $ for unbroken symmetry and  $ m^2 < 0 $ for broken
symmetry. Here $ {\cal R}(t) $ stands for the scalar curvature
\begin{equation}
{\cal R}(t)  =  6\left(\frac{\ddot{a}(t)}{a(t)}+
\frac{\dot{a}^2(t)}{a^2(t)}\right), \label{ricciscalar}
\end{equation}
The coupling of $\Phi(x)$ to the scalar curvature ${\cal
R}(t)$ has been included since  arises anyhow as a consequence of
renormalization\cite{frw}.

The gravitational sector includes the usual Einstein term in addition
to a higher order curvature term and a cosmological constant term 
which are necessary to renormalize the theory. The action for
the gravitational sector is therefore:
\begin{equation}
S_g  =  \int d^4x\; {\cal L}_g = \int d^4x \, a^3(t) \left[\frac{{\cal
R}(t)}{16\pi G}  
+ \frac{\alpha}{2}\; {\cal R}^2(t) - K\right].
\end{equation}
with $K$ being the cosmological constant (we use $K$ rather than the
conventional 
$\Lambda/8\pi G$ to distinguish the cosmological constant from the 
ultraviolet cutoff $\Lambda$ we introduce to regularize the theory;
see section V).
In principle, we also need to include the terms $R^{\mu\nu}R_{\mu\nu}$
and $R^{\alpha\beta\mu\nu}R_{\alpha\beta\mu\nu}$ as they are also terms
of fourth order in derivatives of the metric (fourth adiabatic order),
but the variations resulting from these terms turn out not to be 
independent of that of ${\cal R}^2$ in the flat
FRW cosmology we are considering.

The variation of the action $S = S_g + S_m$ with respect to the 
metric $g_{\mu\nu}$ gives us Einstein's equation
\begin{equation}
\frac{G_{\mu\nu}}{8\pi G} + \alpha H_{\mu\nu} + K g_{\mu\nu}
= - <T_{\mu\nu}>\; ,
\label{extendEinstein}
\end{equation}
where $G_{\mu\nu}$ is the Einstein tensor given by the variation of
$\sqrt{-g}{\cal R}$, $H_{\mu\nu}$ is the higher order curvature term given
by the variation of $\sqrt{-g}{\cal R}^2$, and $T_{\mu\nu}$ is the
contribution  from the matter Lagrangian. 
 With the metric (\ref{metric}), the various components
of the curvature tensors in terms of the scale factor are:
\begin{eqnarray}
G^{0}_{0} & = & -3(\dot{a}/a)^2\; , \;
G^{\mu}_{\mu}  =  -{\cal R} = -6\left(\frac{\ddot{a}}{a}
+\frac{\dot{a}^2}{a^2}\right)\; ,\nonumber \\
H^{0}_{0} & = & -6\left(\frac{\dot{a}}{a}\dot{{\cal R}} + 
\frac{\dot{a}^2}{a^2}{\cal R} - \frac{1}{12}{\cal R}^2\right)\; , \;
H^{\mu}_{\mu}  =  -6\left(\ddot{{\cal R}} + 
3\frac{\dot{a}}{a}\dot{{\cal R}}\right)\; .\nonumber
\end{eqnarray}
Eventually, when we have fully renormalized the theory,
we will set $\alpha_R=0$ and keep as our only contribution to
$K_R$ a piece related to the matter fields which we shall 
incorporate into $T_{\mu\nu}$.  

\subsection{\bf The Large $N$ Approximation} 
To obtain the proper large $N$ limit, the inflaton  field is written as
$$
\vec{\Phi}(\vec x, t) = (\sigma(\vec x,t), \vec{\pi}(\vec x,t)),
$$ 
with $\vec{\pi}$ an $(N-1)$-plet, and we write
\begin{equation}
\sigma(\vec x,t) = \sqrt{N}\phi(t) + \chi(\vec x,t) \; \; ; \; \; \langle
\sigma(\vec x, t) \rangle= \sqrt{N}\phi(t) \; \; ; \; \; 
\langle \chi(\vec x,t) \rangle = 0.
\label{largenzeromode} 
\end{equation}
To implement the large $N$ limit in a consistent manner, one may introduce an
auxiliary field as in\cite{largen}.  However, the leading order
contribution can be obtained equivalently by invoking the 
factorization\cite{De Sitter,frw2}:

\begin{eqnarray}
\chi^4 & \rightarrow & 6 \langle \chi^2 \rangle \chi^2 +\mbox{
constant }\; , \;
\chi^3  \rightarrow  3 \langle \chi^2 \rangle \chi\;,  \nonumber \\ 
\left( \vec{\pi} \cdot \vec{\pi} \right)^2
& \rightarrow & 2 \langle \vec{\pi}^2 \rangle \vec{\pi}^2 - \langle
\vec{\pi}^2 \rangle^2+ {\cal{O}}(1/N)\; ,  \nonumber \\ 
\vec{\pi}^2 \chi^2 & \rightarrow & \langle \vec{\pi}^2 \rangle \chi^2
+\vec{\pi}^2 \langle \chi^2 \rangle\; , \;
\vec{\pi}^2 \chi  \rightarrow  \langle \vec{\pi}^2 
\rangle \chi\; . 
\end{eqnarray}

To obtain a large $N$ limit, we define\cite{De Sitter,frw2}
\begin{equation} 
\vec{\pi}(\vec x, t) = \psi(\vec x, t)
\overbrace{\left(1,1,\cdots,1\right)}^{N-1}, \label{filargeN}
\end{equation} 
where the large $N$ limit is implemented by the requirement that
\begin{equation}
\langle \psi^2 \rangle \approx {\cal{O}} (1) \; , \; \langle \chi^2 \rangle
\approx {\cal{O}} (1) \; , \; \phi \approx {\cal{O}} (1).
\label{order1}
\end{equation}
The leading contribution is obtained by neglecting the $ {\cal{O}} ({1}\slash
{N})$ terms in the $ N \to \infty $  limit. The resulting Lagrangian density is
quadratic, with  linear terms in $\chi$ and $\vec{\pi}$.  
The equations of motion are obtained by imposing the tadpole conditions
$<\chi(\vec x,t)>=0$ and $<\vec{\pi}(\vec x,t)> =0$ which in this case are
tantamount to requiring that the linear terms in $\chi$ and $\vec{\pi}$ in
the Lagrangian density vanish. 
Since the action is quadratic, the quantum fields can be expanded in terms
of creation and annihilation operators and mode functions that obey the
Heisenberg equations of motion
\begin{equation}
\vec{\pi}(\vec x, t) = \int\frac{d^3k}{(2\pi)^3}
\left[{\vec a}_k \; f_k(t) \; e^{i\vec{k}\cdot \vec x} + {\vec a}^{\dagger}_k
 \; f^*_k(t) \; e^{-i\vec{k}\cdot \vec x} \right] .
\end{equation}
We see that since there are $N-1$ `pion' fields, contributions from the field
$\chi$ can be neglected in the $ N \to \infty $ limit as they are of
order $1/N$ with respect those of $\psi$ and $\phi$.

The tadpole condition leads to the following equations of
motion\cite{De Sitter,frw2}: 

\begin{equation}
\ddot{\phi}(t)+3H(t) \; \dot{\phi}(t)+{\cal M}^2(t) \; \phi(t)=0,
\label{modcer}
\end{equation}
with the mode functions
\begin{equation}
\left[\frac{d^2}{dt^2}+3H(t)\frac{d}{dt}+\frac{k^2}{a^2(t)}+{\cal
M}^2(t) \right]f_k(t)= 0, 
\label{modkN}
\end{equation}
where
\begin{equation}
{\cal M}^2(t) =  m^2+\xi{\cal R}(t)+ \frac{\lambda}{2}\phi^2(t)+
\frac{\lambda}{2}\langle \psi^2(t) \rangle \; .
\label{Ngranmass}
\end{equation}
An important point to note in the large $N$ equations of motion is that the
form of the equation for the zero mode (\ref{modcer}) is the same as for the
$k=0$ mode function (\ref{modkN}). It will be this identity that allows
solutions of these equations in a symmetry broken scenario to satisfy
Goldstone's theorem.

In this leading order in $ 1/N $ the theory becomes Gaussian, but with the
self-consistency condition
\begin{equation}
\langle \psi^2(t) \rangle = \int
\frac{d^3k}{2(2\pi)^3}\;|f_k(t)|^2 \; .
\label{largenfluc}
\end{equation}

The initial conditions on the modes $f_k(t)$ must now be determined. 
At this stage it proves illuminating to pass to conformal time variables
in terms of the conformally rescaled fields (see \cite{frw2} and
section VI for a discussion)
in which the mode functions obey an equation which is very similar to that
of harmonic oscillators with time dependent frequencies in Minkowski
space-time.  
It has been realized that different initial conditions on the mode functions
lead to different renormalization counterterms\cite{frw2};
 in particular imposing initial conditions in comoving time leads to
counterterms that depend on these initial conditions. Thus we chose to
impose initial conditions in conformal time in terms of the conformally
rescaled mode functions leading to the following initial conditions
in comoving time:
\begin{equation}
f_k(t_0)=\frac{1}{\sqrt{W_k}}, \;\;\; 
\dot{f}_k(t_0)=\left[-\frac{\dot{a}(t_0)}{a(t_0)}-iW_k\right]f_k(t_0),
\label{condini}
\end{equation}
with
\begin{equation}
W_k^2 \equiv k^2 + {\cal M}^2(t_0) - \frac{{\cal R}(t_0)}{6}.
\label{frec}
\end{equation}
For convenience, we have set $a(t_0)=1$ in eq.(\ref{frec}).
At this point we recognize that when ${\cal M}^2(t_0) - {\cal R}(t_0)/6 <0$ 
the above initial condition must be modified to avoid imaginary frequencies,
which are the signal of instabilities for long wavelength modes in the
broken symmetry case. Thus
we {\em define} the initial frequencies that determine the initial conditions
(\ref{condini}) as
\begin{eqnarray}
W_k^2 & \equiv &  k^2 + \left|{\cal M}^2(t_0) - \frac{{\cal
R}(t_0)}{6}\right| \; \mbox{ for } 
k^2 < \left |{\cal M}^2(t_0) - \frac{{\cal R}(t_0)}{6} \right|\; ,
\label{unstcond1} \\ 
W_k^2 & \equiv &  k^2 + {\cal M}^2(t_0) - \frac{{\cal R}(t_0)}{6} \;
\mbox{ for } 
k^2 \geq \left |{\cal M}^2(t_0) - \frac{{\cal R}(t_0)}{6} \right|\;
. \label{unstcond2}  
\end{eqnarray}
In the unbroken symmetry case ($ m^2 > 0 $ ) we use
eq.(\ref{unstcond2}) for all $ k $. 

As an alternative we have also used initial conditions which smoothly
interpolate from positive frequencies for the unstable modes to the adiabatic
vacuum initial conditions defined by (\ref{condini})-(\ref{frec}) for
the high $k$ modes. While the alternative choices of initial
conditions result in small quantitative differences in the results (a
few percent in quantities which depend strongly on these low-$k$
modes), all of the qualitative features we will examine are
independent of this choice. 

In the large $N$ limit we find the energy density and pressure 
density to be given by\cite{De Sitter,frw2}
\begin{eqnarray}
\frac{\varepsilon}{N} & = & \frac12\dot{\phi}^2 + \frac12m^2\phi^2 +
\frac{\lambda}{8}\phi^4 + \frac{m^4}{2\lambda} - \xi\;
G^{0}_{0}\;\phi^2 +  6\xi\;\frac{\dot{a}}{a}\;\phi\;\dot{\phi}
\nonumber \\
& + & \frac12\langle\dot{\psi}^2\rangle + 
\frac{1}{2a^2}\langle(\nabla\psi)^2\rangle + \frac12 m^2 
\langle\psi^2\rangle 
+ \frac{\lambda}{8}[2\phi^2\;\langle\psi^2\rangle + \langle\psi^2\rangle^2] 
\nonumber \\
& - & \xi G^{0}_{0}\langle\psi^2\rangle + 
6\xi\;\frac{\dot{a}}{a}\;\langle\psi\dot{\psi}\rangle, \label{energy} \\
\frac{\varepsilon-3p}{N} & = & -\dot{\phi}^2 + 2m^2\;\phi^2 + 
\frac{\lambda}{2}\;\phi^4 + \frac{2m^4}{\lambda} - \xi
G^{\mu}_{\mu}\phi^2 + 
6\xi\left(\phi\ddot{\phi} + \dot{\phi}^2 +
3\frac{\dot{a}}{a}\phi\dot{\phi}\right) 
\nonumber \\
& - & \langle\dot{\psi}^2\rangle +
\frac{1}{a^2}\langle(\nabla\psi)^2\rangle + 
2m^2\;\langle\psi^2\rangle - \xi\;G^{\mu}_{\mu}\;\langle\psi^2\rangle  
\nonumber \\
& + & \frac{\lambda}{2}[2\phi^2\;\langle\psi^2\rangle +
\langle\psi^2\rangle^2] + 
6\xi\left(\langle \psi\;\ddot{\psi}\rangle + \langle \dot{\psi}^2 \rangle+
3\,\frac{\dot{a}}{a}\;\langle\psi\dot{\psi}\rangle \right),
\label{trace}
\end{eqnarray}
where $\langle\psi^2\rangle$ is given by equation (\ref{largenfluc}) 
and we have defined the following integrals:
\begin{eqnarray}
\langle(\nabla\psi)^2\rangle & = & \int \frac{d^3k}{2 (2\pi)^3}\; k^2\;
|f_k(t)|^2 \; , \label{delpsi} \\
\langle\dot{\psi}^2\rangle & = & \int \frac{d^3k}{2(2\pi)^3}\;
|\dot{f}_k(t)|^2 \;. \label{dotpsi}
\end{eqnarray}
The composite operators $\langle \psi \dot{\psi} \rangle$ and 
$\langle \psi \ddot{\psi} \rangle$ are symmetrized by
removing a normal ordering constant to yield
\begin{eqnarray}
\frac{1}{2}(\langle\psi\dot{\psi}\rangle + \langle\dot{\psi}\psi\rangle)
 &=&  \frac{1}{4} \int \frac{d^3k}{(2\pi)^3}
\frac{d |f_k(t)|^2}{dt}\; , \label{psidotpsi} \\
\frac{1}{2}(\langle\psi\ddot{\psi}\rangle + \langle\ddot{\psi}\psi\rangle)
 &=&  \frac{1}{4} \int \frac{d^3k}{(2\pi)^3}
\left[f_k(t)\ddot{f}_k^*(t)+\ddot{f}_k(t)f_k^*(t)\right] \; . 
\label{psiddotpsi} 
\end{eqnarray}
The last of these integrals, (\ref{psiddotpsi}), may be rewritten using the
equation of motion (\ref{modkN}):
\begin{equation}
\langle \psi \ddot{\psi} \rangle = -3\frac{\dot{a}}{a}\langle \;
\psi \dot{\psi} \rangle
-\frac{\langle(\nabla\psi)^2\rangle}{a^2}-{\cal M}^2 \langle\psi^2\rangle \; .
\end{equation}
It is straightforward to show that the bare energy is
covariantly conserved by using the equations of motion for the zero mode and
the mode functions. 

\subsection{Hartree Approximation}
In the Hartree approximation, our theory is that of a single component scalar
field, $\Phi(\vec{x},t)$, with the $Z_2$ symmetry $\Phi \to -\Phi$.  The
potential can be written as:
\begin{equation}
V(\Phi) = \frac12 (m^2+\xi{\cal R}) \Phi^2 + \frac{\lambda}{4!}\Phi^4,
\label{hartpot}
\end{equation}
where ${\cal R}$ is the Ricci scalar. We have rescaled the  coupling
constant here by a factor 3 compared with the one used in the large
$N$ limit [eq.(\ref{potential})].

We decompose the field into its zero mode, $\phi(t) =
\langle\Phi(\vec{x},t)\rangle$, fluctuations $\psi(\vec{x}, t)$ about it:
$$
\Phi(\vec{x},t) = \phi(t) + \psi(\vec{x},t).
$$
The potential (\ref{hartpot}) may then be expanded in terms of these fields.

The Hartree approximation is achieved by making the potential quadratic in
the fluctuation field $\psi$ by invoking the factorization
\begin{eqnarray}
\psi^3(\vec{x},t) & \to &
3\langle\psi^2(\vec{x},t)\rangle\; \psi(\vec{x},t)\; , \nonumber \\
\psi^4(\vec{x},t) & \to & 6\langle\psi^2(\vec{x},t)\rangle\;\psi^2(\vec{x},t)
	-3\langle\psi^2(\vec{x},t)\rangle^2\; . \nonumber
\end{eqnarray}
This factorization yields a quadratic theory in which the effects of
interactions are encoded in the time dependent mass which is determined
self-consistently.

The equations of motion for the zero mode and the fluctuations are
given by the  tadpole equation
$$
\langle\psi(\vec{x},t)\rangle = 0.
$$
Introducing the Fourier mode functions, $U_k(t)$, they can be written as:
\begin{eqnarray}
\ddot{\phi}(t)+3\frac{\dot{a}(t)}{a(t)}\; \dot{\phi}(t)+(m^2+\xi{\cal
R}(t))\;\phi(t) 
+\frac{\lambda}{6}\;\phi^3(t)+\frac{\lambda}{2}\;\phi(t)
\;\langle\psi^2(t)\rangle 
& = & 0\; ,
\label{hartphieq} \\  \nonumber \\
\left[\frac{d^2}{dt^2}+3\frac{\dot{a}(t)}{a(t)}\frac{d}{dt}+\frac{k^2}{a^2(t)}
+m^2+\xi{\cal R}(t)+\frac{\lambda}{2}\; \phi^2(t)+
\frac{\lambda}{2}\langle\psi^2(t)\rangle\right]
U_k(t) & = & 0\; .
\label{hartukeq}
\end{eqnarray}
\begin{eqnarray}
\langle\psi^2(t)\rangle &=& \int \frac{d^3k}{2(2\pi)^3}\, |U_k(t)|^2
\; ,
\label{fluct} \\ \nonumber
\end{eqnarray}
The initial conditions on the mode functions are the same as
eq.(\ref{condini}) in the large $ N $ limit,
\begin{equation}
U_k(t_0) = \frac{1}{\sqrt{W_k}}, \quad \dot{U}_k(t_0) =\left[
-\frac{\dot{a}(t_0)}{a(t_0)}-iW_k\right]U_k(t_0),
\label{initcond}
\end{equation}
with the frequencies $ W_k $ given by eq.(\ref{frec})

A detailed analysis and discussion of the choice of initial conditions and the
frequencies (\ref{frec}) is provided in the sec. VI. As discussed there, this
choice corresponds to the large-$k$ modes being in the conformal adiabatic
vacuum state. In what follows we will subtract the composite operator
$\psi^2(t)$ at the initial time and absorb the term
$\frac{\lambda}{2}\langle\psi^2(t_0)\rangle$ in a renormalization of the mass.

Notice that we have used identical notations in the large $ N $ and
Hartree cases to avoid cluttering and also to stress the similarity
between the two approximations. In particular, we note that the only
difference in the expressions for the two cases [eqs.(\ref{hartukeq})
and (\ref{modcer}), respectively] is a factor of three appearing in
the self interaction term in the equations for the zero mode.

It is instructive to compare the Hartree approximation with the
self-consistently one-loop approximation presented in sec. II. We see
comparing eq.(\ref{suma}) with eq.(\ref{hartphieq})-(\ref{hartukeq})
that they are rather similar. The only difference being that the
quantum fluctuations are present in both Hartree equations whereas
they are only present in the zero mode equation in the 
self-consistently one-loop approximation. In fact, in the weak
coupling limit, the Hartree approximation becomes the
self-consistently one-loop approximation.

\section{Renormalization}

Renormalization is a very subtle but important issue in gravitational
backgrounds\cite{birrell}. The fluctuation contribution
$\langle \psi^2(\vec x,t) \rangle$, the energy, and the pressure all need to be
renormalized. The renormalization aspects in curved space times have been
discussed at length in the literature\cite{birrell} and have
been extended to the large $N$ self-consistent approximations for
the non-equilibrium backreaction problem 
in\cite{largen,frw2,De Sitter,din,ramsey}. More recently, a consistent
and covariant  regularization scheme that can be implemented
numerically has been proposed\cite{baacke}.  

In terms of the effective mass term for the large $ N $ limit given by
(\ref{Ngranmass}) and defining the quantity
\begin{eqnarray}
B(t) &\equiv& a^2(t)\left({\cal M}^2(t)-{\cal{R}}/6 \right),\label{boft}\\
{\cal M}^2(t) &=& m^2_B+\xi_B {\cal R}(t)+\frac{\lambda_B}{2}\phi^2(t)
+\frac{\lambda_B}{2}\langle\psi^2(t)\rangle_B \; , \label{masso}
\end{eqnarray}
where the subscript $B$ stands for bare quantities,  
we find the following large $k$ behavior for the case of an {\em arbitrary}
scale factor $a(t)$ (with $a(0)=1$):
\begin{eqnarray}
|f_k(t)|^2 &=& \frac{1}{ka^2(t)}- \frac{1}{2k^3
a^2(t)}\;B(t)  \cr \cr &+&
{1 \over {8 k^5 \; a^2(t) }}\left\{  3 B(t)^2 + a(t)
\frac{d}{dt} \left[ a(t) {\dot B}(t) \right]  \right\} +
{\cal{O}}(1/k^7)  \cr \cr
& = & {\cal S}^{(2)}+ {\cal{O}}(1/k^5) \; ,
\label{sub1}\\ 
|\dot{f}_k(t)|^2 &=&
\frac{k}{a^4(t)}+\frac{1}{2ka^4(t)}\left[B(t)+2\dot{a}^2 \right] \cr \cr 
& + & {1 \over {8 k^3 \; a^4(t) }}\left\{ - B(t)^2 - a(t)^2 {\ddot
B}(t) + 3 a(t) 
{\dot a}(t)
{\dot B}(t) - 4 {\dot a}^2(t) B(t) \right\} +  {\cal{O}}(1/k^5) \cr \cr
& = & {\cal S}^{(1)}+ {\cal{O}}(1/k^5) \; ,
\label{sub2}  
\end{eqnarray}
\begin{equation}
\frac12 \left[f_k(t)\dot{f}_k^*(t)+\dot{f}_k(t)f_k^*(t)\right] =
-\frac{1}{k \; a^2(t)}\frac{\dot{a}(t)}{a(t)} - \frac{1}{4 k^3 a^2(t)}
\left[\dot{B}(t) - 2\frac{\dot{a}(t)}{a(t)}\;B(t)\right] +
{\cal{O}}(1/k^5) \; .
\label{sub3}
\end{equation}

Although the divergences can be dealt with by dimensional
regularization, this procedure is not well suited to numerical
analysis (see however ref.\cite{baacke}).  We will make our 
subtractions using an ultraviolet cutoff, $\Lambda a(t)$, constant in {\em physical
coordinates}. This guarantees that the counterterms will be time
independent. The renormalization then proceeds much in the same manner as in
reference\cite{frw}; the quadratic divergences renormalize the mass 
and the logarithmic terms renormalize the quartic coupling and the coupling
to the Ricci scalar. In addition, there is a quartic divergence which
renormalizes the cosmological constant while the leading renormalizations
of Newton's constant and the higher order curvature coupling are quadratic and
logarithmic respectively.
The renormalization conditions on the mass, coupling to the Ricci
scalar and coupling constant are obtained from the requirement that the
frequencies that appear in the mode equations are finite\cite{frw}, i.e:
\begin{equation}
m^2_B+\xi_B {\cal R}(t)+\frac{ \lambda_B}{2}\phi^2(t)
+\frac{\lambda_B}{2}\langle\psi^2(t)\rangle_B=
m^2_R+\xi_R {\cal R}(t)+\frac{ \lambda_R}{2}\phi^2(t)
+\frac{\lambda_R}{2}\langle\psi^2(t)\rangle_R, \label{rencond}
\end{equation}
while the renormalizations of Newton's constant, the higher order curvature
coupling, and the cosmological constant are given by the condition of
finiteness of the semi-classical Einstein-Friedmann equation:
\begin{equation}
\frac{G^0_0}{8\pi G_B} + \alpha_B H^0_0 + K_B g^0_0 + 
\langle T^0_0 \rangle_B = \frac{G^0_0}{8\pi G_R} + 
\alpha_R H^0_0 + K_R g^0_0 + \langle T^0_0 \rangle_R \; .
\end{equation}

Finally we arrive at the following set of renormalizations\cite{frw2}:
\begin{eqnarray}
\frac{1}{8\pi N G_R} &=& \frac{1}{8\pi N G_B} 
- 2\left(\xi_R-\frac16\right)\frac{\Lambda^2}{16\pi^2}
+ 2\left(\xi_R-\frac16\right)m_R^2\; \frac{\ln(\Lambda/\kappa)}{16\pi^2}, \\
\frac{\alpha_R}{N} &=& \frac{\alpha_B}{N} 
- \left(\xi_R-\frac16\right)^2\frac{\ln(\Lambda/\kappa)}{16\pi^2}, \\
\frac{K_R}{N} &=& \frac{K_B}{N} - \frac{\Lambda^4}{16\pi^2}
- m_R^2\;\frac{\Lambda^2}{16\pi^2}
+ \frac{m_R^4}{2}\;\frac{\ln(\Lambda/\kappa)}{16\pi^2}, \\
m_R^2 &=& m_B^2 + \lambda_R\;\frac{\Lambda^2}{16\pi^2}
-\lambda_R\; m_R^2\;\frac{\ln(\Lambda/\kappa)}{16\pi^2}, \\
\xi_R &=& \xi_B 
- \lambda_R\left(\xi_R-\frac16\right)\frac{\ln(\Lambda/\kappa)}{16\pi^2}, \\
\lambda_R &=& \lambda_B - \lambda_R\;\frac{\ln(\Lambda/\kappa)}{16\pi^2},\\
\langle \psi^2(t) \rangle_R &=& \int
\frac{d^3k}{2(2\pi)^3}\left\{|f_k(t)|^2-
\frac{1}{ka^2(t)} + \frac{\Theta(k-\kappa)}{2k^3}
\left[{\cal M}^2(t)-{{{\cal R}(t)}\over 6} \right] \right\} \; .
\end{eqnarray}
Here, $\kappa$ is the renormalization point.  As expected, the logarithmic 
terms are consistent with the renormalizations found using dimensional
regularization\cite{baacke,ramsey}.  Again, we set $\alpha_R=0$ and
choose the renormalized cosmological constant such that the vacuum
energy is zero in the true vacuum.  We emphasize that while 
the regulator we have chosen does not respect the covariance of 
the theory, the renormalized energy momentum tensor defined in this 
way nevertheless retains the property of covariant conservation in the
limit when the cutoff is taken to infinity.

The logarithmic subtractions can be neglected because of the coupling
$\lambda \leq 10^{-12}$.  Using the Planck scale as the cutoff and the
inflaton mass $m_R$ as a renormalization point, these terms are of order
$\lambda \ln[M_{pl}/m_R] \leq 10^{-10}$, for $m \geq 10^9 \mbox{ GeV }$. An
equivalent statement is that for these values of the coupling and inflaton
masses, the Landau pole is well beyond the physical cutoff $M_{pl}$.
Our relative 
error in the numerical analysis is of order $10^{-8}$, therefore our numerical
study is insensitive to the logarithmic corrections. Though these corrections
are fundamentally important, numerically they can be neglected. Therefore, in
the numerical computations that follow, we will neglect logarithmic 
renormalization and subtract only
quartic and quadratic divergences in the energy and pressure, and quadratic
divergences in the fluctuation contribution. 

\subsection{Renormalized Equations of Motion for Dynamical Evolution in the 
Large N limit}

It is convenient to introduce the following dimensionless quantities
and definitions,
\begin{equation}
\tau = m_R t \quad ; \quad h= \frac{H}{m_R} \quad ; 
\quad q=\frac{k}{m_R} \quad \; \quad
\omega_q = \frac{W_k}{m_R} \quad ; \quad g= \frac{\lambda_R}{8\pi^2} \; ,
\label{dimvars1}
\end{equation}
\begin{equation}
\eta^2(\tau) = \frac{\lambda_R}{2m^2_R} \; \phi^2(t)
\quad ; \quad  g\Sigma(\tau) = \frac{\lambda}{2m^2_R}\; \langle \psi^2(t)
\rangle_R  \quad ; \quad f_q(\tau) \equiv \sqrt{m_R} \; f_k(t) \; .
\label{dimvars3}
\end{equation}

Choosing $\xi_R=0$ (minimal coupling)  and the renormalization
 point $\kappa = |m_R|$ and setting $a(0)=1$, 
the equations of motion become for unbroken symmetry:

\begin{eqnarray}
&&\left[\frac{d^2}{d \tau^2}+ 3h \frac{d}{d\tau}+1+\eta^2(\tau)+
g\Sigma(\tau)\right]\eta(\tau) = 0\; , \label{modcnr} \cr \cr
& &\left[\frac{d^2}{d \tau^2}+3h
\frac{d}{d\tau}+\frac{q^2}{a^2(\tau)}+1+\eta^2+g\Sigma(\tau)
\right]f_q(\tau)=0\;, \nonumber \\ 
& &  f_q(0)  =  \frac{1}{\sqrt{\omega_q}} \quad ; \quad 
\dot{f}_q(0)  = \left[-h(0)-i\omega_q\right]f_q(0)\; , \nonumber \\
& & \omega_q  =  
\left[q^2+1+\eta^2(0)-\frac{{\cal
R}(0)}{6m^2_R}+g\Sigma(0)\right]^{\frac{1}{2}} \; .
\label{modknr}  
\end{eqnarray}

We find for broken symmetry,
\begin{eqnarray}
&&\left[\frac{d^2}{d \tau^2}+ 3h \frac{d}{d\tau}-1+\eta^2(\tau)+
g\Sigma(\tau)\right]\eta(\tau) = 0\; , \label{modcr}\cr \cr
& &\left[\frac{d^2}{d \tau^2}+3h
\frac{d}{d\tau}+\frac{q^2}{a^2(\tau)}-1+\eta^2+g\Sigma(\tau)
\right]f_q(\tau)=0\; ,  \\ 
& &  f_q(0)  =  \frac{1}{\sqrt{\omega_q}} \quad ; \quad 
\dot{f}_q(0)  = \left[-h(0)-i\omega_q\right]f_q(0)\;, \nonumber \\
& & \omega_q  =  
\left[q^2-1+\eta^2(0)-\frac{{\cal
R}(0)}{6m^2_R}+g\Sigma(0)\right]^{\frac{1}{2}} \; \mbox{ for } \; q^2
> 1-\eta^2(0)+\frac{{\cal R}(0)}{6m^2_R}-g\Sigma(0)\;, \nonumber \\ 
& & \omega_q  =  
\left[q^2+1-\eta^2(0)+\frac{{\cal
R}(0)}{6m^2_R}-g\Sigma(0)\right]^{\frac{1}{2}} \; \mbox{ for } \; q^2
< 1-\eta^2(0)+\frac{{\cal R}(0)}{6m^2_R}-g\Sigma(0) \; . 
\label{modkr}  
\end{eqnarray}

Here,
$$
\Sigma(\tau)= \int_0^{\infty} q^2 dq \left[ | f_q(\tau)|^2 - {1 \over
{q\; a(\tau)^2}} + {{\Theta(q - 1)}\over {2 q^3}} \left(\frac{{\cal
M}^2(\tau)}{m^2_R}-{{{\cal{R}(\tau)}}\over{6 m^2_R}}\right)\right] \; .
$$
both for unbroken and broken symmetry.

The initial conditions for $\eta(\tau)$ will be specified later. 
An important point to notice is that the equation of
motion for the $q=0$ mode coincides with that of the zero mode
(\ref{modcr}). Furthermore, for $\eta(\tau \rightarrow \infty) \neq
0$, a stationary (equilibrium) solution of the eq.(\ref{modcr})  
is obtained for broken symmetry
when the sum rule\cite{us1,mink,De Sitter,frw2}
\begin{equation}
-1+\eta^2(\infty)+g\Sigma(\infty) = 0 \label{sumrule}
\end{equation}
is fulfilled. This sum rule is nothing but a consequence of Goldstone's
theorem and is a result of the fact that the large $ N $ approximation 
satisfies the Ward identities associated with the $ O(N) $ symmetry, since
the term  $-1+\eta^2+g\Sigma$ is seen to be the effective mass of the
modes transverse to the symmetry breaking direction, i.e. the Goldstone
modes in the broken symmetry phase.

The renormalized dimensionless evolution equations in the Hartree approximation
are very similar to eqs.(\ref{modknr})-(\ref{modkr}). They can be obtained
 just dividing by three the $ \eta^2 $ term in the zero mode equation.
[Compare with eqs.(\ref{modkN})-(\ref{Ngranmass}) and 
(\ref{hartphieq})-(\ref{hartukeq})].

\bigskip

In terms
of the zero mode $\eta(\tau)$ and the quantum mode function given
by eq.(\ref{modcr}) we find that the Friedmann equation for the dynamics
of the scale factor in dimensionless variables is given by

\begin{equation}
 h^2(\tau)    =   4h^2_0 \; \epsilon_R(\tau) \quad  ;   \quad h^2_0 =
 \frac{4\pi N m^2_R}{3M^2_{Pl}\lambda_R} \; .\label{eif} 
 \end{equation}
and the renormalized energy and pressure are given by: 

\begin{eqnarray}
 \epsilon_R(\tau) &  =  &  
\frac{1}{2}\dot{\eta}^2+\frac{1}{4}\left(-1+\eta^2+g\Sigma \right)^2
 \nonumber \\ 
&+ &  \frac{g}{2}\int q^2 dq \left[|\dot{f_q}|^2-
{\cal S}^{(1)}(q,\tau)
+\frac{q^2}{a^2}\left(|f_q|^2-\Theta(q - 1)\; {\cal
S}^{(2)}(q,\tau)\right) \right]\; , 
\label{hubble} \\
 (p+\varepsilon)_R(\tau)  & = & \frac{2Nm^4_R}{\lambda_R}\left\{
\dot{\eta}^2 \right. \nonumber \\
& + & \left. g \int q^2 dq \left[|\dot{f_q}|^2-
{\cal S}^{(1)}(q,\tau)
+\frac{q^2}{3a^2}\left(|f_q|^2-\Theta(q - 1)\;{\cal S}^{(2)}(q,\tau)\right)
\right]\right\} \;, \label{ppluse} 
\end{eqnarray}
where the subtractions ${\cal S}^{(1)}$ and ${\cal S}^{(2)}$ are given
by the right hand sides of eqs.(\ref{sub2}) and (\ref{sub1}) respectively.
 
The renormalized energy and pressure are covariantly conserved:
\begin{equation}\label{concov}
{\dot  \epsilon}_R(\tau) + 3 \, h(\tau)\, (p+\varepsilon)_R(\tau) = 0 \; .
\end{equation}

From the evolution of the mode functions that determine the quantum
fluctuations, we can study the growth of correlated domains with the equal time
correlation function,

\begin{eqnarray}\label{correlator}
S(\vec{x},t) & = & \langle\psi(\vec{x},t)\psi(\vec{0},t)\rangle
=  \int\frac{d^3k}{2(2\pi)^3}e^{i\vec{k}\cdot\vec{x}}\; |f_k(t)|^2\; ,
\end{eqnarray}
which can be written in terms of the  power spectrum of quantum fluctuations,
$ |f_q(\tau)|^2 $. It is convenient to define the dimensionless
correlation function, 
\begin{equation}
{\cal S}(\rho,\tau) = \frac{S(|\vec x|, t)}{m^2_R}= \frac{1}{{4\pi^2 }\rho}
\int_0^{\infty} q\; dq \; \sin[q\rho] \; |f_q(\tau)|^2  \; ; \; \rho= m_R|\vec
x| \; .  \label{rhocorr} 
\end{equation}

We now have all the ingredients to study the particular cases of interest.

\section{Conformal time analysis and initial Conditions}

The issue of renormalization and initial conditions is best understood in
conformal time which is a natural framework for adiabatic renormalization and
regularization.

Quantization in conformal time proceeds by writing the metric element as
\begin{equation}
ds^2= C^2({\cal T})(d{\cal T}^2 - d\vec{x}^2). 
\end{equation}
Under a conformal rescaling of the field
\begin{equation}\label{camco}
\Phi(\vec x, t) = \chi(\vec x, {\cal T})/ C({\cal T}),
\end{equation}
the action for a scalar field (with the obvious generalization to $ N $
components) becomes, after an integration by parts and dropping a
 surface term
\begin{equation}
S= \int d^3x d{\cal T} \left\{\frac12 (\chi')^2-\frac12 (\vec{\nabla}\chi)^2-
{\cal{V}}(\chi)\right\},
\end{equation}
with
\begin{equation}
{\cal{V}}(\chi) = C^4({\cal T})V(\Phi/C({\cal T}))-C^2({\cal T})\frac{{\cal{R}}}{12}\chi^2,
\end{equation}
where ${\cal{R}}= 6C''({\cal T})/C^3({\cal T})$ is the Ricci scalar,
and primes stand for derivatives with respect to conformal time ${\cal T}$.

The conformal time Hamiltonian operator, which is the generator of translations
in ${\cal T}$, is given by
\begin{equation}
H_{{\cal T}}= \int d^3x \left\{ \frac{1}{2}\Pi^2_{\chi}+\frac{1}{2}
(\vec{\nabla}\chi)^2+{\cal{V}}(\chi) \right \}, \label{confham}
\end{equation}
with $\Pi_{\chi}$ being the canonical momentum conjugate to $\chi$,
$\Pi_{\chi} = \chi'$. 
Separating the zero mode of the field $\chi$ 
\begin{equation}
\chi(\vec x, {\cal T}) = \chi_0({\cal T}) + \bar{\chi}(\vec x,{\cal T}),
\end{equation}
and performing the large $N$ or Hartree factorization on the
fluctuations we find 
that the Hamiltonian becomes linear plus quadratic in the fluctuations, and
similar to a Minkowski space-time Hamiltonian with 
a ${\cal T}$ dependent mass term
given by
\begin{equation}
{\cal{M}}^2({\cal T}) = C^2({\cal T}) \left[m^2+(\xi-\frac{1}{6}){\cal{R}}
+ \frac{\lambda}{2}\chi_0^2({\cal T}) + \frac{\lambda}{2}
\langle \bar{\chi}^2 \rangle\right]. \label{masseff}
\end{equation}

We can now follow the steps and use the results of reference\cite{frw} for the
conformal time evolution of the density matrix by setting $a(t)=1$ in the
proper equations of that reference and replacing the frequencies by
\begin{equation}
\omega^2_k({\cal T}) = \vec{k}^2 + {\cal{M}}^2({\cal T}), \label{freqs}
\end{equation}
and the expectation value in (\ref{masseff}) is obtained in this
${\cal T}$ evolved density matrix.
The time evolution of the kernels in the density matrix (see\cite{frw})
is determined by the mode functions that obey
\begin{equation}
\left[ \frac{d^2}{d{\cal T}^2}+k^2+{\cal{M}}^2({\cal T})\right] 
F_k({\cal T})=0.
\label{fmodeqn}
\end{equation}
The Wronskian of these mode functions
\begin{equation}\label{wff}
{\cal{W}}(F,F^*)= F'_k F^*_k-F_k F^{*'}_k
\end{equation} 
is a constant. It is natural to impose initial conditions such that at the
initial ${\cal T}$ the density matrix describes a situation of local thermodynamic
equilibrium and therefore commutes with the conformal time Hamiltonian at the
initial time. This implies that the initial conditions of the mode functions
$F_k({\cal T})$ be chosen to be (see\cite{frw})
\begin{equation}
F_k({\cal T}_o)= \frac{1}{\sqrt{\omega_k({\cal T}_o)}} \; \; ; \; 
F'_k({\cal T}_o)= -i\sqrt{\omega_k({\cal T}_o)} F_k({\cal
T}_o). \label{inicond} 
\end{equation}
With such initial conditions, the Wronskian (\ref{wff}) takes the value
\begin{equation}\label{wro}
{\cal{W}}(F,F^*)= -2i \; .
\end{equation}
These initial conditions correspond to the choice of mode functions which
coincide with the first order adiabatic modes and those of the Bunch-Davies
vacuum for large momentum\cite{birrell}.  To see this clearly, we
write the solution of eq.(\ref{fmodeqn}) in the form,
\begin{equation}
D_k({\cal T}) = e^{\int^{{\cal T}}_{{\cal T}_o} R_k({\cal T}')d{\cal T}'},
\end{equation} 
with the function $R_k({\cal T})$ obeying the Riccati equation
\begin{equation}
R'_k + R^2_k+ k^2+{\cal{M}}^2({\cal T})=0.
\end{equation} 
This equation posses the solution
\begin{equation}
R_k({\cal T}) = -ik+R_{0,k}({\cal T})-i\frac{R_{1,k}({\cal T})}{k}+
\frac{R_{2,k}({\cal T})}{k^2}-i\frac{R_{3,k}({\cal T})}{k^3}+
\frac{R_{4,k}({\cal T})}{k^4}+ {\cal O}\frac{1}{k^5} 
\end{equation}
and its complex conjugate. We find for the  coefficients:
\begin{eqnarray}
&& R_{0,k} = 0 \; \; ; \; \; R_{1,k} = \frac{1}{2} {\cal{M}}^2({\cal T}) \; \; ; \;
\; R_{2,k} = -\frac{1}{2} R'_{1,k} \nonumber \\ &&R_{3,k} = \frac{1}{2}\left(
R'_{2,k}-R^2_{1,k} \right) \; \; ; \; \; R_{4,k} = -\frac{1}{2}\left(
R'_{3,k}+2R_{1,k}R_{2,k} \right).
\end{eqnarray}
The solutions $F_k({\cal T})$ obeying the boundary conditions (\ref{inicond}) are
obtained as linear combinations of this WKB solution and its complex conjugate
\begin{equation}
F_k({\cal T}) = \frac{1}{2\sqrt{\omega_k({\cal T}_o)}}\left[
(1+\gamma)D_k({\cal T})+(1-\gamma)D^*_k({\cal T}) \right],
\end{equation}
where the coefficient $\gamma$ is obtained from the initial conditions. It is
straightforward to find that the real and imaginary parts are given by
\begin{equation}
\gamma_R= 1+ {\cal{O}}(1/k^4) \; \; ; \; \; \gamma_I= {\cal{O}}(1/k^3).
\end{equation}
Therefore the large-$k$ mode functions satisfy the adiabatic vacuum initial
conditions\cite{birrell}. This, in fact, is the rationale for the choice of the
initial conditions (\ref{inicond}).

Following the analysis presented in \cite{frw} we find, in conformal time that
\begin{equation}
\langle \bar{\chi}^2(\vec x,{\cal T}) \rangle = \int \frac{d^3k}{2(2\pi)^3}
\; |F_k({\cal T})|^2.
\end{equation}
The Heisenberg field operators $\bar{\chi}(\vec x, {\cal T})$  and 
their canonical momenta $\Pi_{\chi}(\vec x, {\cal T})$ can be expanded as:
\begin{eqnarray}
&& \bar{\chi}(\vec x, {\cal T}) = \int {{d^3k}\over {\sqrt2 \, (2\pi)^{3/2}}} 
\left[ a_k F_k({\cal T})+ a^{\dagger}_{-k}F^*_k({\cal T}) \right]
 e^{i \vec k \cdot \vec x}, \label{heisop}\\
&& \Pi_{\chi}(\vec x, {\cal T}) = 
\int {{d^3k}\over {\sqrt2 \, (2\pi)^{3/2}}} 
\left[ a_k F'_k({\cal T})+ a^{\dagger}_{-k}F^{*'}_k({\cal T}) \right]
 e^{i \vec k \cdot \vec x}, \label{canheisop}
\end{eqnarray}
with the time independent creation and
annihilation operators $ a_k$ and $ a^{\dagger}_k $ 
obeying canonical commutation relations. Since the fluctuation fields
in comoving and conformal time are related by the conformal rescaling
(\ref{camco}), 
it is straightforward to see that the mode functions in comoving time are
related to those in conformal time simply as
\begin{equation}
f_k(t) = \frac{F_k({\cal T})}{C({\cal T})}.
\end{equation}
Therefore the initial conditions (\ref{inicond}) on the conformal time mode
functions imply the initial conditions for the mode functions in comoving time
are given by eq.(\ref{condini}).
 
For renormalization purposes we need the large-$k$ behavior of $|f_k(t)|^2 \; ,
|\dot{f}_k(t)|^2$, which are determined by the large-$k$ behavior of the
conformal time mode functions and its derivative. These are given by
\begin{eqnarray}
|F_k({\cal T})|^2 &=& \frac{1}{k}\left[ 1-\frac{R_{1,k}({\cal T})}{k^2}+
\frac{1}{k^4}\left(\frac{R''_{1,k}({\cal T})}{4}+\frac{3}{2}R^2_{1,k}({\cal T})
\right) + {\cal O}\left(\frac{1}{k^6}\right) \right]\; , 
\cr \cr
|F'_k({\cal T})|^2 &=& {k}\left[ 1+\frac{R_{1,k}({\cal T})}{k^2}+
\frac{1}{k^4}\left(-\frac{R''_{1,k}({\cal
T})}{4}+\frac{3}{2}R^2_{1,k}({\cal T}) \right) + {\cal
O}\left(\frac{1}{k^6}\right) \right] \; . \label{largekf}
\end{eqnarray}

We note that the large $k$ behavior of the mode functions to the order
needed to 
renormalize the quadratic and logarithmic divergences is insensitive to the
initial conditions. This is not the case when the initial conditions
are imposed as described 
in\cite{frw,De Sitter}. Thus the merit in considering the initial conditions in
conformal time \cite{frw2}.

The correspondence with the comoving time mode functions is given by:
\begin{eqnarray}
|f_k(t)|^2 & = & \frac{|F_k({\cal T})|^2}{C^2({\cal T})} \nonumber \\
|\dot{f}_k(t)|^2 & = & \frac{1}{C^2({\cal T})}
\left[ \frac{|F'_k({\cal T})|^2}{C^2({\cal T})}+
\left(H^2-\frac{H}{C({\cal T})} 
\frac{d}{d{\cal T}}\right)|F_k({\cal T})|^2 \right] \; .
\end{eqnarray}
There is an important physical consequence of this choice of initial
conditions, which is revealed by analyzing the evolution of the density matrix.

In the large $N$ or Hartree (also to one-loop) approximation, the
density matrix 
is Gaussian, and defined by a normalization factor, a complex covariance that
determines the diagonal matrix elements and a real covariance that determines
the mixing in the Schr\"odinger representation as discussed in
reference\cite{frw} (and references therein).

In conformal time quantization and in the Schr\"odinger representation in which
the field $\chi$ is diagonal the conformal time evolution of the density matrix
is via the conformal time Hamiltonian (\ref{confham}). The evolution equations
for the covariances is obtained from those given in reference\cite{frw} by
setting $a(t) = 1$ and using the frequencies $\omega^2_k({\cal T})=
k^2+{\cal{M}}^2({\cal T})$. In particular, by setting the covariance of the
diagonal elements (given by equation (2.20) in\cite{frw}; see also equation
(2.44) of\cite{frw}),
\begin{equation}
{\cal{A}}_k({\cal T}) = -i \frac{F^{'*}_k({\cal T})}{F^*_k({\cal T})},
\end{equation}
we find that with the initial conditions (\ref{inicond}), the
conformal time density matrix is that of local equilibrium at ${\cal T}_0$
in the sense that it commutes with the conformal time Hamiltonian. 
However, it is straightforward to see, that the comoving time density
matrix {\em does not} commute with the {\it comoving time} Hamiltonian at
the initial time $t_0$.  

An important corollary of this analysis and comparison with other initial
conditions used in comoving time is that assuming initial conditions of local
equilibrium in comoving time leads to divergences that depend on the initial
condition as discussed at length in\cite{frw}.  This dependence of the
renormalization counterterms on the initial condition was also
realized in ref.\cite{leutwyler} within the context of the CTP formulation.
Imposing the initial conditions corresponding to local thermal equilibrium in
{\em conformal} time, we see that: i) the renormalization counterterms do not
depend on the initial conditions and ii) the mode functions are identified with
those corresponding to the adiabatic vacuum for large momenta.
This is why we prefer the  initial conditions (\ref{inicond}).


For our main analysis we choose this initial temperature to be zero so that the
resulting density matrix describes a pure state, which for the large
momentum modes coincides with the conformal adiabatic vacuum. Such
zero temperature choice seems appropriate after the exponential
inflation of the universe.

{\bf Particle Number:}

We write the Fourier components of the field $\chi$ and its canonical
momentum $\Pi_{\chi}$ given by (\ref{heisop}) -(\ref{canheisop}) as:
\begin{eqnarray}
&& \bar{\chi}_k({\cal T}) = \frac{1}{\sqrt{2}}\left[
a_k \; F_k({\cal T})+ a^{\dagger}_{-k} \;F^*_k({\cal T}) \right],
  \label{heisopfu}\\
&& \Pi_{\chi,k}({\cal T}) = \frac{1}{2}\left[
a_k \; F'_k({\cal T})+ a^{\dagger}_{-k} \;F^{*'}_k({\cal T}) \right].
  \label{canheisopfu}
\end{eqnarray}
These (conformal time) Heisenberg operators can be written equivalently
in terms of the ${\cal T}$ dependent creation and annihilation operators

\begin{eqnarray}
&& \bar{\chi}_k({\cal T}) = \frac{1}{\sqrt{2 \omega_k({\cal T}_0)}}\left[
\tilde{a}_k({\cal T}) \; e^{-i\omega_k({\cal T}_0){\cal T}}+ 
\tilde{a}^{\dagger}_k({\cal T}) \; e^{i\omega_k({\cal T}_0){\cal T}}
 \right],
  \label{newheis}\\
&& \Pi_{\chi,k}({\cal T}) = -i\sqrt{\frac{\omega_k({\cal T}_0)}{2}}\left[
\tilde{a}_k({\cal T}) \; e^{-i\omega_k({\cal T}_0){\cal T}}-
\tilde{a}^{\dagger}_k({\cal T}) \; e^{i\omega_k({\cal T}_0){\cal T}}
 \right].
  \label{newcanheisopfu}
\end{eqnarray}

The operators $\tilde{a}_k({\cal T}) \; ; a_k({\cal T}) $ are related by a 
Bogoliubov transformation. The number of particles referred to the
initial Fock vacuum of the modes $F_k$, is given by
\begin{equation}
N_k({\cal T}) = \langle \tilde{a}^{\dagger}_k({\cal T}) 
\tilde{a}_k({\cal T}) \rangle 
= \frac{1}{4} \left|\frac{F_k({\cal T})}{F_k(0)}\right|^2 \left [
1+ \frac{1}{\omega^2_k({\cal T}_0)} \left| 
\frac{F'_k({\cal T})}{F_k({\cal T})} 
\right|^2 \right] - \frac{1}{2} \; , \label{partnumb}
\end{equation}
 or alternatively, in terms of the comoving mode functions $f_k(t) =
F_k({\cal T})/C({\cal T})$ we find

\begin{equation}
N_k(t) = \frac{a^2(t)}{4} \left| \frac{f_k(t)}{f_k(0)}\right|^2
\left[ 1+ \frac{1}{\omega^2_k(0)} 
\left| \frac{ \dot{f}_k(t) + H f_k(t)}{f_k(t)}
\right|^2 \right] - \frac{1}{2} \; . \label{partnumbcomo}
\end{equation}

Using the large $k$-expansion of the conformal mode functions given by
eqs. (\ref{largekf})  we find the large-$k$ behavior
of the particle number to be $N_k \buildrel {k\to \infty} \over =
{\cal{O}}(1/k^4) $, and the 
total number of particles (with reference to the initial state at
${\cal T}_0$) is therefore finite.  

\section{Fields Evolution on a fixed FRW background}

We consider in this section 
the matter evolution on radiation or matter dominated FRW
cosmologies\cite{frw2}. The case for de Sitter expansion will be
discussed in sec. VIII \cite{De Sitter}. 

We write the scale factor as $ a(t) = (t/t_0)^n $ with $ n = 1/2 $ and
$ n = 2/3 $ corresponding to radiation and matter dominated
backgrounds, respectively. Note that the value of $ t_0 $ determines
the initial Hubble constant since
$$
H(t_0) = {{{\dot a}(t_0)} \over {a(t_0)}} = {n \over t_0} \; .
$$
We now solve the system of equations (\ref{hartphieq}) - (\ref{fluct}) in the
Hartree approximation, with (\ref{modcer}) replacing (\ref{hartphieq}) in the
large $N$ limit.  We begin by presenting an early time analysis of the
slow roll scenario.  We then undertake a thorough numerical investigation of
various cases of interest.  For the symmetry broken case, we also provide an
investigation of the late time behavior of the zero mode and the quantum
fluctuations. We use the dimensionless variables 
(\ref{dimvars1})-(\ref{dimvars3}).

We will assume minimal coupling to
the curvature, $\xi_r = 0$.  In the cases of interest, ${\cal R} \ll \mu^2$, so
that finite $\xi_r$ has little effect.

\subsection{Early Time Solutions for Slow Roll}

For early times in a slow roll scenario [$m^2=-\mu^2$, $\eta(t_0) \ll 1$], we
can neglect in eqs.(\ref{hartphieq}) or (\ref{modcer}) and in
eq.(\ref{hartukeq}) both the quadratic and cubic terms in $\eta(t)$ as well as
the quantum fluctuations $\langle\psi^2(t)\rangle_r $ [recall that
$\langle\psi^2(t_0)\rangle_r = 0 $]. Thus, the differential equations for the
zero mode (\ref{hartphieq}) or (\ref{modcer}) and the mode functions
(\ref{hartukeq}) become linear equations. In terms of the scaled variables
introduced above, with $ a(t)= t^n $ ($ n=2/3 $ for a matter dominated
cosmology while $ n=1/2 $ for a radiation dominated cosmology) we have:
\begin{eqnarray}
\ddot{\eta}(t)+\frac{3n}{t} \;\dot{\eta}(t)-\eta(t) & = & 0 \; ,
\label{earlyeta} \\ \nonumber \\
\left[\frac{d^2}{dt^2}+\frac{3n}{t}\frac{d}{dt}
+\frac{k^2}{t^{2n}}-1\right]U_k(t) & = & 0 \;. \label{earlyuk}
\end{eqnarray}

The solutions to the zero mode equation (\ref{earlyeta}) are
\begin{equation}
\eta(t)=c\; t^{-\nu}I_{\nu}(t)+d\; t^{-\nu}K_{\nu}(t) \; ,
\label{earlyetasoln}
\end{equation}
where $\nu \equiv (3n-1)/2$, and $I_{\nu}(t)$ and $K_{\nu}(t)$ are modified
Bessel functions.  The coefficients, $c$ and $d$, are determined by the initial
conditions on $\eta$.  For $\eta(t_0)=\eta_0$ and $\dot{\eta}(t_0)=0$, we have:
\begin{eqnarray}
c & = & \eta_0  \; t_0^{\nu+1}
\left[ \dot{K}_{\nu}(t_0)-\frac{\nu}{t_0}K_{\nu}(t_0)\right] \; ,
\label{coeffc} \\
d & = & -\eta_0  \; t_0^{\nu+1}
\left[\dot{I}_{\nu}(t_0)-\frac{\nu}{t_0}I_{\nu}(t_0)\right] \; .
\label{coeffd}
\end{eqnarray}
Taking the asymptotic forms of the modified Bessel functions, we find that for
intermediate times $\eta(t)$ grows as
\begin{equation}
\eta(t) \stackrel{t \gg 1}{=} {c \over {\sqrt{2 \pi}}}
\; t^{-3n/2}\; e^t\left[1-\frac{9n^2-6n}{8t}+
{\cal O}({1\over{t^2}})\right].
\label{asymeta}
\end{equation}
We see that $\eta(t)$ grows very quickly in time, and the approximations
(\ref{earlyeta}) and (\ref{earlyuk}) will quickly break down.  For the case
shown in fig.1 (with $n=2/3$, $\eta(t_0)=10^{-7}$, and $\dot{\eta}(t_0)=0$),
we find that this approximation is valid up to $t-t_0 \simeq 10$.

The equations for the mode functions ({\ref{earlyuk}) can be solved in closed
form for the modes in the case of a radiation dominated cosmology with $n=1/2$.
The solutions are

\begin{equation}
U_k(t) = c_k \; e^{-t}\, U\left(\frac34-\frac{k^2}{2},\frac32,2t\right) +d_k \;
e^{-t}\, M\left(\frac34-\frac{k^2}{2},\frac32,2t\right).
\label{earlyuksoln}
\end{equation}
Here, $U(\cdot)$ and $M(\cdot)$ are confluent hypergeometric functions
\cite{aands} (in another common notation, $M(\cdot) \equiv\; _1F_1(\cdot)$),
and the $c_k$ and $d_k$ are coefficients determined by the initial conditions
(\ref{initcond}) on the modes.  The solutions can also be written in terms of
parabolic cylinder functions.

For large $ t $ we have the asymptotic form
\begin{equation}
U_k(t) \stackrel{t \gg 1}{=} d_k \; e^t \; (2t)^{-(3/4+k^2 t_0 /2)} \;
\frac{\sqrt{\pi}}{2\;\Gamma\left(\frac34-\frac{k^2 t_0 }{2}\right)}
\left[1+{\cal O}({1\over{t}})\right]
+ c_k \; e^{-t} \;  (2t)^{(-3/4+k^2 t_0 /2)}\left[1+{\cal
O}({1\over{t}})\right]\; 
. \label{asymuk}
\end{equation}
Again, these expressions only apply for intermediate times before the
nonlinearities have grown significantly.

\subsection{Numerical Analysis}

We now present the numerical analysis of the dynamical evolution of
scalar fields in time dependent, matter and radiation dominated cosmological
backgrounds.  We use initial values of the Hubble constant such that 
$H(t_0) \geq 0.1$.  For expansion
rates much less than this value the evolution will look similar to
Minkowski space, which has been studied in great detail elsewhere
\cite{us1,mink}.  As will be seen, the equation of state found
numerically is, in the majority of cases, that of cold matter.  We therefore
use matter dominated expansion for the evolution in much of the analysis that
follows.  While it presents some inconsistency at late times, 
the evolution in radiation
dominated universes remains largely unchanged, although there is greater
initial growth of quantum fluctuations due to the scale factor growing more
slowly in time.  Using the large $N$ and Hartree approximations to study 
theories with continuous and discrete symmetries respectively, we treat three
important cases.  They are 1) $m^2<0$, $\eta(t_0)\ll 1$; 2) $m^2<0$,
$\eta(t_0)\gg 1$; 3) $m^2>0$, $\eta(t_0)\gg 1$.

In presenting the figures, we have shifted the origin of time such that
$t \to t'=t-t_0$.  This places the initial time, $t_0$, at the origin.
In these shifted coordinates, the scale factor is given by 
$$
a(t)=\left(\frac{t+\tau}{\tau}\right)^n,
$$ 
where, once again, $n=2/3$ and $n=1/2$ in matter and radiation dominated
backgrounds respectively, and the value of $\tau$ is determined by the 
Hubble constant at the initial time:
$$
H(t_0=0)=\frac{n}{\tau}.
$$

{\bf Case 1: $m^2<0$, $\eta(t_0)\ll 1$}.  This is the case of an early universe
phase transition in which there is little or no biasing in the initial
configuration (by biasing we mean that the initial conditions break the $ \eta
\to -\eta $ symmetry).  The transition occurs from an initial temperature above
the critical temperature, $T>T_c$, which is quenched at $t_0$ to the
temperature $T_f \ll T_c$.  This change in temperature due to the rapid
expansion of the universe is modeled here by an instantaneous change in the
mass from an initial value $m_i^2=T^2/T_c^2-1$ to a final value
$m_f^2=-1$.  We will use the value $m_i^2=1$ in what follows.
This quench approximation is necessary since the low momentum
frequencies (\ref{frec}) appearing in our initial conditions (\ref{initcond}) 
are complex for negative mass squared and small $\eta(t_0)$.  An alternative
choice is to use initial frequencies given by
$$
\omega_k(t_0)=\left[k^2+{\cal{M}}^2(t_0)\tanh\left(
\frac{k^2+{\cal{M}}^2(t_0)}{|{\cal{M}}^2(t_0)|}\right)\right]^{1/2}.
$$
These frequencies have the attractive feature that they match the conformal
adiabatic frequencies given by (\ref{frec}) for large values of $k$ while 
remaining positive for small $k$.  We find that such a choice of initial
conditions changes the quantitative value of the particle number by a few 
percent, but leaves the qualitative results unchanged.

We plot the the zero mode $\eta(t)$, the equal time correlator $g\Sigma(t)$, 
the total number of produced particles $gN(t)$  (see sec. VI for
a discussion of our definition of particles), the number of particles
$gN_k(t)$ as a function of wavenumber for both intermediate and
late times, and the ratio of the pressure
and energy densities $p(t)/\varepsilon(t)$ (giving the equation of state).

Figs. 1a-e shows these quantities in the large $N$ approximation for a matter
dominated cosmology with an initial condition on the zero mode given by
$\eta(t_0\! =\! 0)=10^{-7}$, $\dot{\eta}(t_0\! =\! 0)=0$ and for an initial
expansion rate of $H(t_0)=0.1$.  This choice for
the initial value of $\eta$ stems from the fact that the quantum fluctuations
only have time to grow significantly for initial values satisfying
$\eta(t_0) \ll  \sqrt{g}$; for values $\eta(t_0) \gg \sqrt{g}$ the evolution
is essentially classical.  This result is clear from the intermediate time
dependence of the zero mode and the low momentum mode functions given by 
the expressions (\ref{asymeta}) and (\ref{asymuk}) respectively.

\begin{figure}
\epsfig{file=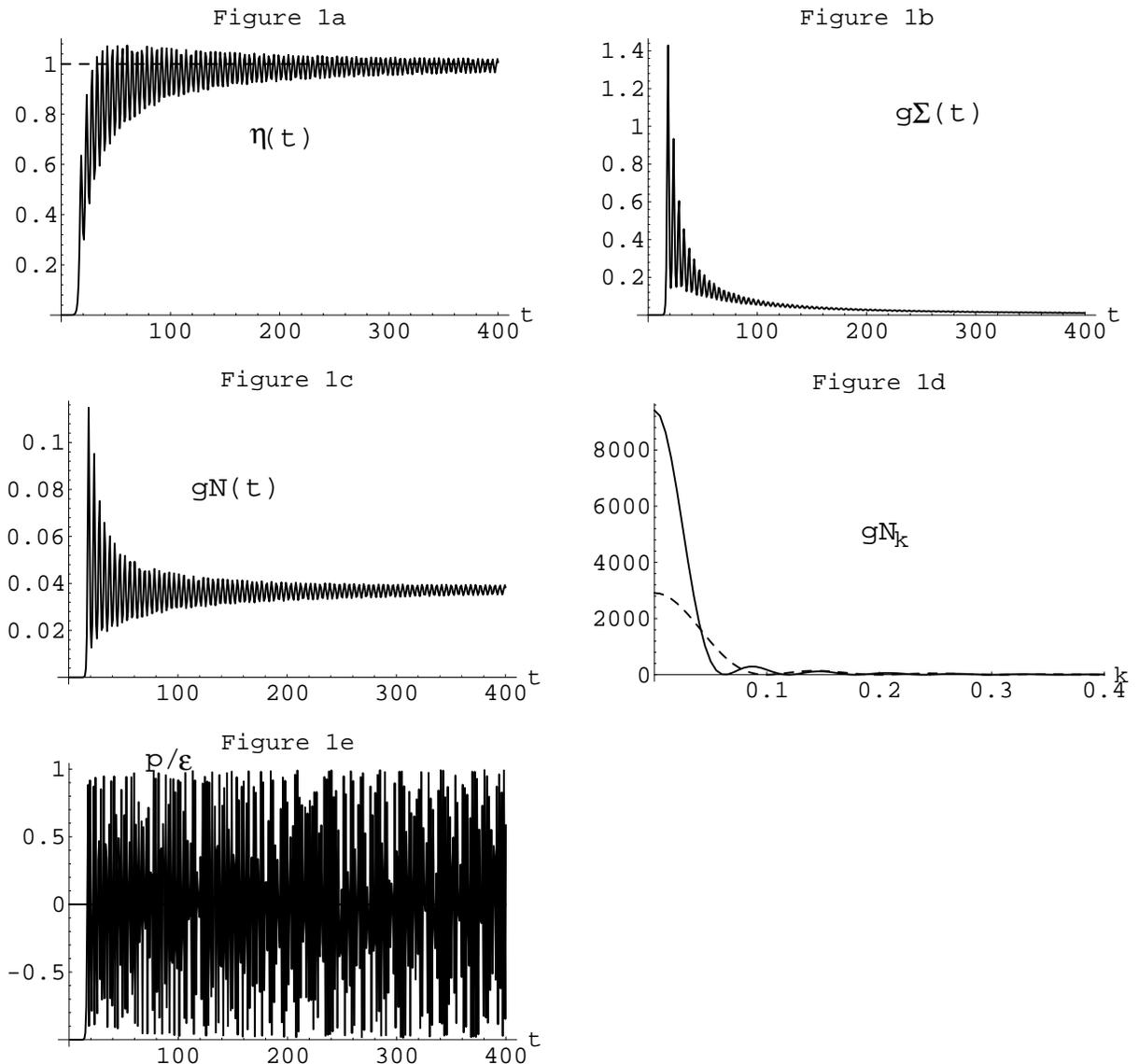}
\caption{Symmetry broken, slow roll, large $N$, matter dominated
evolution of (a) the zero mode $\eta(t)$ vs. $t$, (b) the quantum fluctuation
operator $g\Sigma(t)$ vs. $t$, (c) the number of particles $gN(t)$ vs. $t$,
(d) the particle distribution $gN_k(t)$ vs. $k$ at $t=149.1$ (dashed line)
and $t=398.2$ (solid line),  and (e) the ratio of the pressure and
energy density 
$p(t)/\varepsilon(t)$ vs. $t$ for the parameter values $m^2=-1$, $\eta(t_0) =
10^{-7}$, $\dot{\eta}(t_0)=0$, $g = 10^{-12}$, $H(t_0) = 0.1$. \label{fig1}}
\end{figure}
After the initial
growth of the fluctuation $g\Sigma(t)$ (fig.1b) we see that the zero mode
(fig.1a) approaches the value given by the minimum of the tree level
potential, $\eta=1$, while $g\Sigma(t)$ decays for late times as
$$
g\Sigma(t) \simeq {{\cal C}\over{a^2(t)}} ={{\cal C}\over{t^{4/3}}} \; .
$$
For these late times, the Ward identity corresponding to the $O(N)$ symmetry of
the field theory is satisfied, enforcing the condition
\begin{equation}
-1 + \eta^2(t) + g\Sigma(t) = 0.
\label{ward}
\end{equation} 
Hence, the zero mode approaches the classical minimum as
$$
\eta^2(t) \simeq 1 -{{\cal C}\over{a^2(t)}} \; .
$$ 

Figure 1c depicts the number of particles produced.  After an initial burst of 
particle production, the number of particles settles down to a relatively
constant value.  Notice that the number of particles produced is approximately
of order $1/g$.
In fig.1d, we show the number of particles as a function of the
wavenumber, $k$.  For intermediate times we see the simple structure depicted
by the dashed line in the figure, while for late times this quantity becomes
concentrated more at low values of the momentum $k$.  

Finally, fig.1e shows that the field begins with a de Sitter equation of 
state $p=-\varepsilon$ but evolves quickly to a state dominated by ordinary matter, 
with an equation of state (averaged over the oscillation timescale) $p=0$.
This last result is a bit surprising as one expects from the condition
(\ref{ward}) that the particles produced in the final state are massless
Goldstone bosons  which should have the equation of state of radiation.
However, as shown in fig.1d, the produced particles are of low momentum,
$ q \ll 1 $, and while the effective mass of the particles is zero to
very high  
accuracy when averaged over the oscillation timescale, the effective mass 
makes small oscillations about zero so that the dispersion relation for these
particles differs from that of radiation.  In addition, since the produced
particles have little energy, the contribution to the energy density from
the zero mode, which contributes to a cold matter equation of state, remains
significant.

In figs. 2a-e we show the same situation depicted in fig.1 using the
Hartree approximation.  The initial condition on the zero mode is $\eta(t_0\!
=\! 0)=\sqrt{3}\cdot 10^{-7}$; the factor of $\sqrt{3}$ appears due to the
different scaling in the zero mode equations, (\ref{hartphieq}) and
(\ref{modcr}), which causes the minimum of the tree level effective potential
in the Hartree approximation to have a value of $\eta=\sqrt{3}$.  Again, the
Hubble constant has the value $H(t_0)=0.1$.  Here, we see
again that there is an initial burst of particle production as $g\Sigma(t)$
(fig.2b) grows large.  However, the zero mode (fig.2a) quickly reaches the
minimum of the potential and the condition
\begin{equation}
-1 + \eta^2(t)/3 + g\Sigma(t) = 0
\label{ward2}
\end{equation}
is approximately satisfied by forcing the value of $g\Sigma(t)$
quickly to zero. 

There are somewhat fewer particles produced here compared to the large
$N$ case,  and the distribution of particles is more extended.  
Since the effective mass of the particles is nonzero, we expect a
matter dominated equation of state (fig 2e) for later times.  
The fact that the Hartree approximation does not
satisfy Goldstone's theorem means that the resulting particles must be
massive,  explaining why somewhat fewer particles are produced.

Finally, we show the special case in which
there is no initial biasing in the field, $\eta(t_0\! =\! 0)=0$, 
$\dot{\eta}(t_0\! =\! 0)=0$, and $H(t_0)=0.1$ in figs.
3a-d.  With such an initial condition, the Hartree approximation and the 
large $N$ limit are equivalent.
The zero mode remains zero for all time, so that the quantity
$ g\Sigma(t) $ (fig.3a) satisfies the sum rule (\ref{ward}) by reaching 
the value one without decaying for late times. 
Notice that many more particles are produced in this case (fig 3b); the growth
of the particle number for late times is due to the expansion of 
the universe.  The particle distribution (fig.3c) is similar to that of the 
slow roll case in fig.1.  The equation of state (fig.3d) is likewise
similar.

\begin{figure}
\epsfig{file=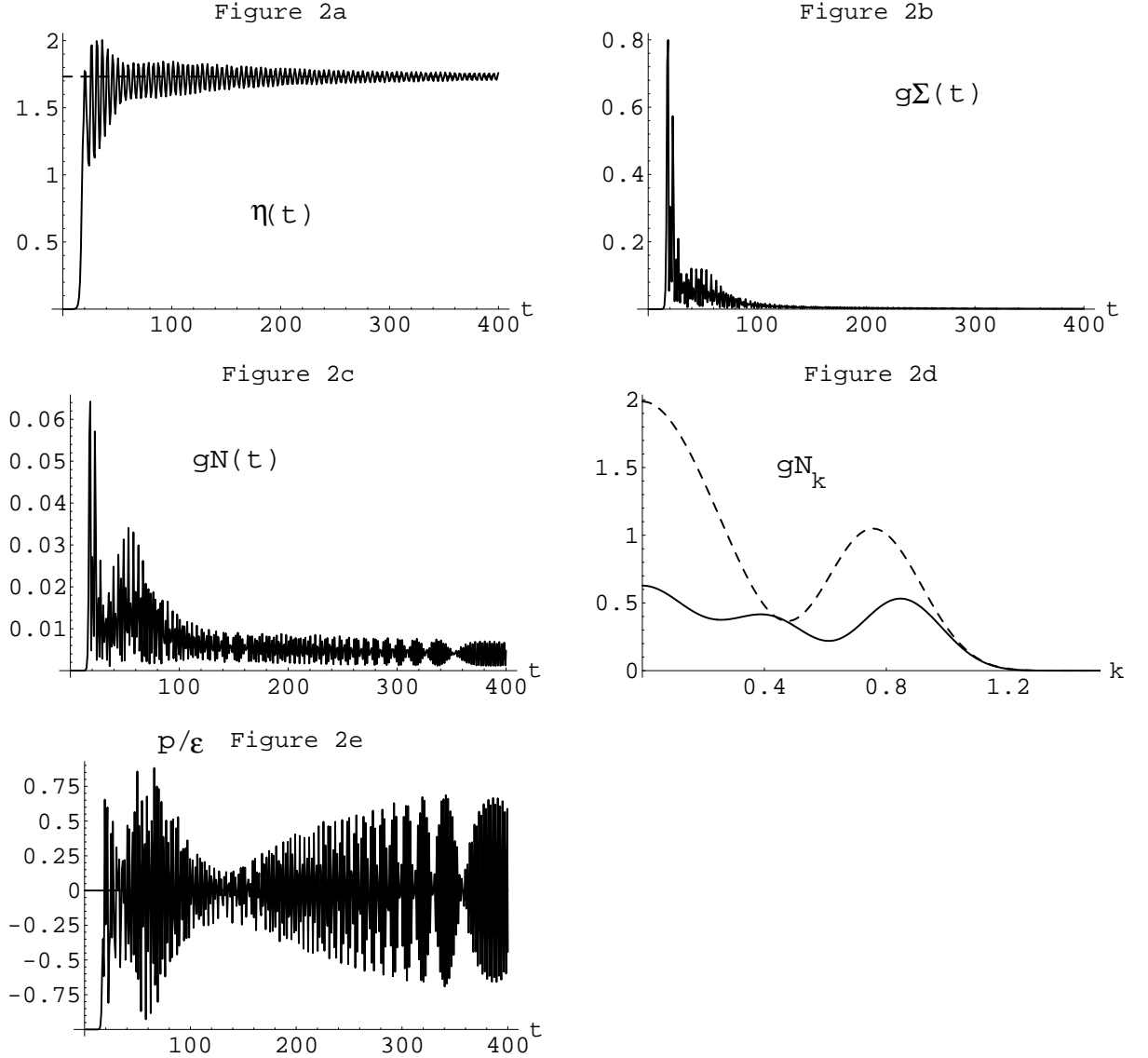}
\caption{Symmetry broken, slow roll, Hartree, matter dominated
evolution of (a) the zero mode $\eta(t)$ vs. $t$, (b) the quantum fluctuation
operator $g\Sigma(t)$ vs. $t$, (c) the number of particles $gN(t)$ vs. $t$,
(d) the particle distribution $gN_k(t)$ vs. $k$ at $t=150.7$ (dashed line)
and $t=396.1$ (solid line),  and (e) the ratio of the pressure and 
energy density
$p(t)/\varepsilon(t)$ vs. $t$ for the parameter values $m^2=-1$, $\eta(t_0) =
3^{1/2}\cdot 10^{-7}$, $\dot{\eta}(t_0)=0$, $g = 10^{-12}$, $H(t_0) =
0.1$. \label{fig2}}
\end{figure}

In each of these cases of slow roll dynamics, increasing the Hubble constant
has the effect of slowing the growth of both $\eta$ and $g\Sigma(t)$.  The equation
of state will be that of a de Sitter universe for a longer period before
moving to a matter dominated equation of state.  Otherwise, the dynamics is
much the same as in figs. 1-3.

{\bf Case 2: $m^2<0$, $\eta(t_0)\gg 1$}.  We now examine the case of 
 a chaotic inflationary scenario with a symmetry broken potential.
In chaotic inflation, the zero mode begins with a value $\eta(t)\gg 1$.  During
the de Sitter phase, $H \gg 1$, and the field initially evolves classically,
dominated by the first order derivative term appearing in the zero mode
equation (see (\ref{hartphieq}) and (\ref{hartukeq})).  
Eventually, the zero mode
rolls down the potential, ending the de Sitter phase and beginning the
FRW phase.  We consider the field dynamics in the FRW universe after
the end of inflation. We thus take the initial temperature to be zero,
 $T=0$.

\begin{figure}
\epsfig{file=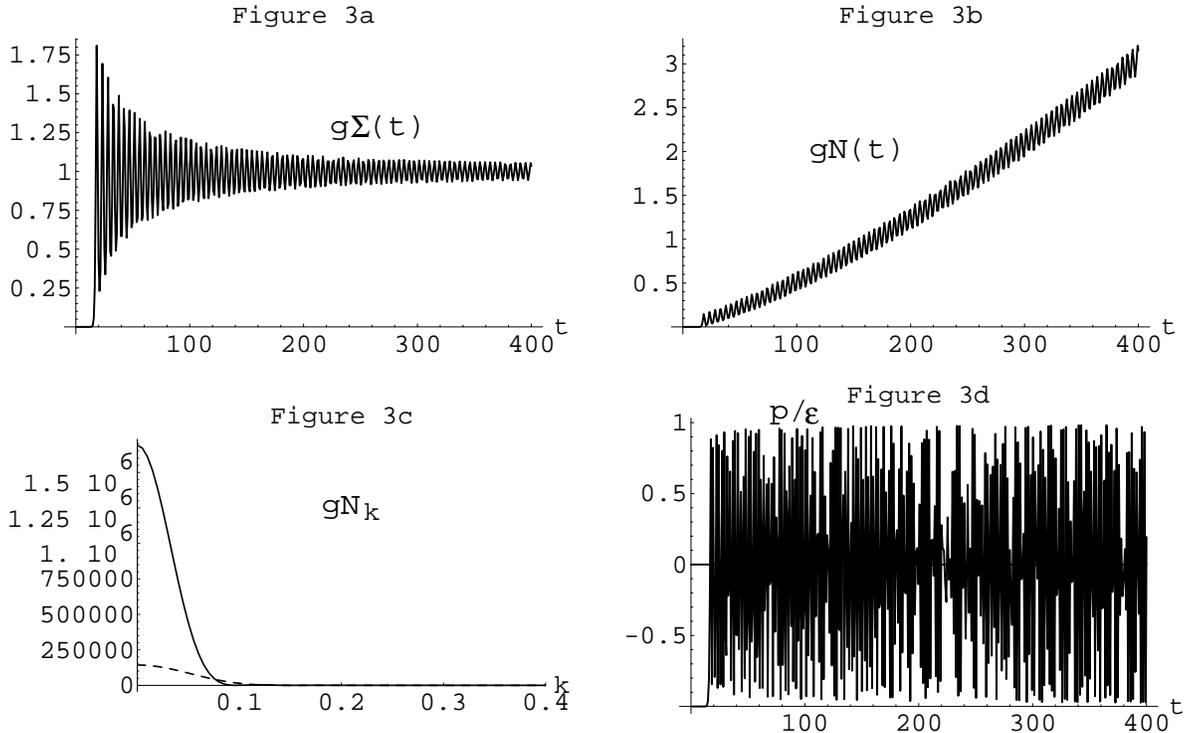}
\caption{Symmetry broken, no roll, matter dominated
evolution of (a) the quantum fluctuation operator $g\Sigma(t)$ vs. $t$, 
(b) the number of particles $gN(t)$ vs. $t$,
(c) the particle distribution $gN_k(t)$ vs. $k$ at $t=150.1$ (dashed line)
and $t=397.1$ (solid line),  and (d)
the ratio of the pressure and energy density $p(t)/\varepsilon(t)$ vs. $t$ for
the parameter values $m^2=-1$, $\eta(t_0) = 0$, $\dot{\eta}(t_0)=0$, $g =
10^{-12}$, $H(t_0) = 0.1$. \label{fig3}}
\end{figure}

Figure 4 shows our results for the quantities, $\eta(t)$, $g\Sigma(t)$,
$gN(t)$, $gN_k(t)$, and $p(t)/\varepsilon(t)$ for the evolution in
the large $N$ approximation within a {\em radiation} dominated gravitational
background with $H(t_0)=0.1$.  The initial condition on the zero mode
is chosen to have the representative value $\eta(t_0\!=\! 0)=4$ with 
$\dot{\eta}(t_0\! =\! 0)=0$.  
Initial values of the zero mode much smaller than this will not produce
significant growth of quantum fluctuations; initial values larger than this
produces qualitatively similar results, although the resulting number of 
particles will be greater and the time it takes for the zero mode to settle
into its asymptotic state will be longer.

\begin{figure}
\epsfig{file=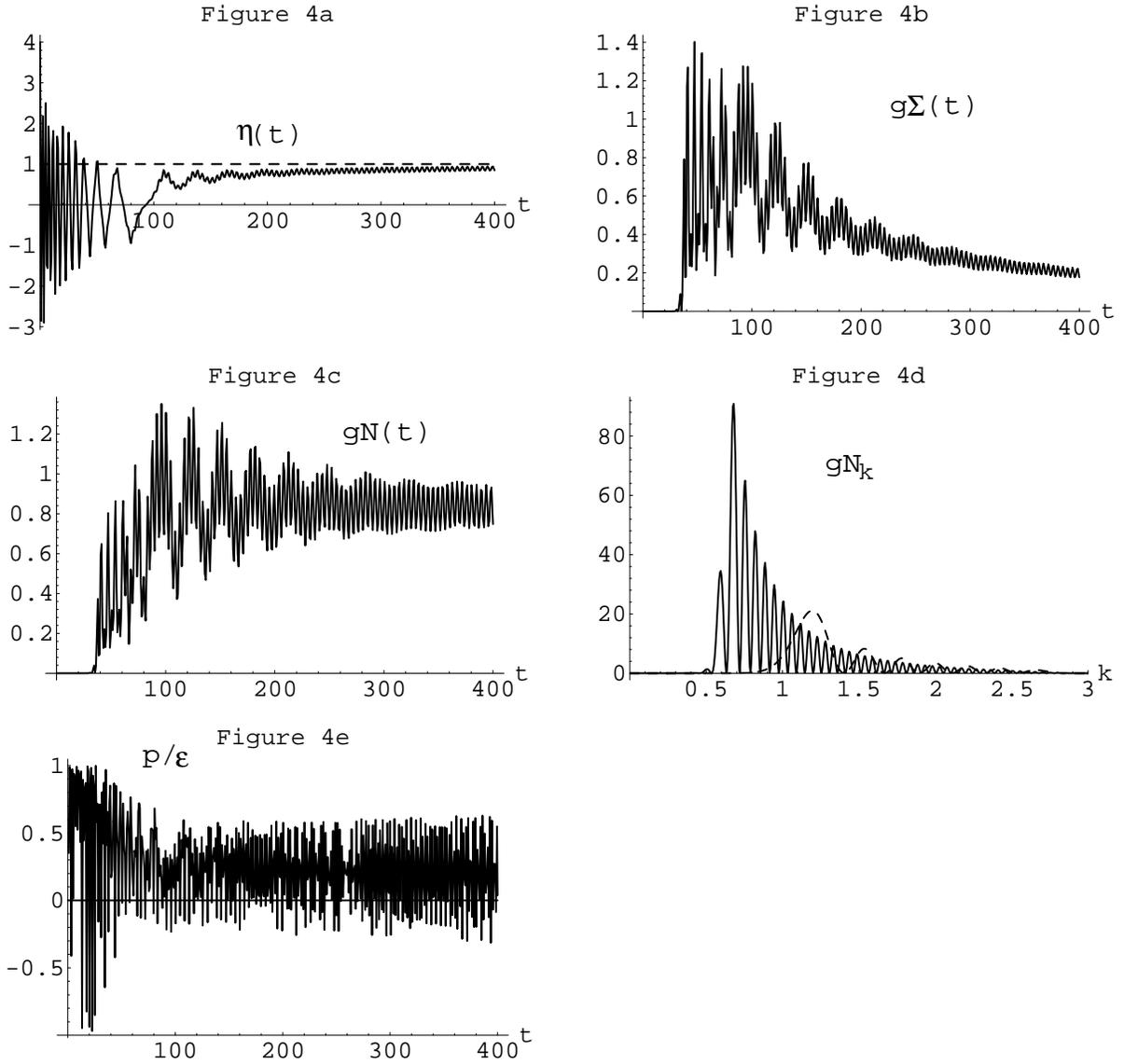}
\caption{Symmetry broken, chaotic, large $N$, radiation dominated
evolution of (a) the zero mode $\eta(t)$ vs. $t$, (b) the quantum fluctuation
operator $g\Sigma(t)$ vs. $t$, (c) the number of particles $gN(t)$ vs. $t$,
(d) the particle distribution $gN_k(t)$ vs. $k$ at $t=76.4$ (dashed line)
and $t=392.8$ (solid line),  and (e) the ratio of the pressure and energy
density $p(t)/\varepsilon(t)$ vs. $t$ for the parameter values $m^2=-1$,
$\eta(t_0) = 4$, $\dot{\eta}(t_0)=0$, $g = 10^{-12}$, $H(t_0) =
0.1$. \label{fig4}}
\end{figure}

We see from fig.4a that the zero
mode oscillates rapidly, while the amplitude of the oscillation decreases due
to the expansion of the universe.  This oscillation induces particle production
through the process of parametric amplification (fig.4c) and causes the
fluctuation $g\Sigma(t)$ to grow (fig.4b).  Eventually, the zero mode loses
enough energy that it is restricted to one of the two minima of the tree level
effective potential.  The subsequent evolution closely follows that of Case 1
above with $g\Sigma(t)$ decaying in time as $1/a^2(t) \sim 1/t$ 
with $\eta$ given by
the sum rule (\ref{ward}).  The spectrum (fig.4d) indicates a single unstable
band of particle production dominated by the modes $k=1/2$ to about $k=3$ 
for late times.  The structure within this band becomes more complex with time 
and shifts somewhat toward lower momentum modes.  
Such a shift is also  observed in Minkowski spacetimes \cite{us1,mink,late}. 
Figure 4e shows the equation of
state which we see to be somewhere between the relations for matter and
radiation for times out as far as $t=400$, but slowly moving to a matter
equation of state.  Since matter redshifts as $1/a^3(t)$ while radiation
redshifts as $1/a^4(t)$, the equation of state should eventually become matter
dominated.  Given the equation of state indicated by fig.4e, we estimate that
this occurs for times of order $t=10^4$.  The reason the equation of state
in this case differs from that of cold matter as was seen in figs. 1-3 is 
that the particle distribution produced by parametric amplification is 
concentrated at higher momenta, $k \simeq 1$. 

Figure 5 shows the corresponding case with a matter dominated background.  The
results are qualitatively very similar to those described for fig.4 above.
Due to the faster expansion, the zero mode (fig.5a) finds one of the two wells
more quickly and slightly less particles are produced.  For late times, the
fluctuation $g\Sigma(t)$ (fig.5b) decays as $1/a^2(t) \propto 1/t^{4/3}$.  
Again we see an equation of state (figs. 5e) which evolves from a
state between that of pure radiation or matter toward one of cold matter.

The Hartree case is depicted in fig.6 for a matter dominated universe, with
the initial condition on the zero mode $\eta(t_0\! =\! 0)=4\sqrt{3}$.  Again,
the evolution begins in much the same manner as in the large $N$ approximation
with oscillation of the zero mode (fig.6a), which eventually settles into one
of the two minima of the effective potential.  Whereas in the large $N$
approximation, the zero mode approaches the minimum asymptotically [as given by
(\ref{ward}) and our late time analysis below], in the Hartree approximation we
see that the zero mode finds the minimum quickly and proceeds to oscillate
about that value.  The two point correlator (fig.6b) quickly evolves toward
zero without growing large.  Particle production in the Hartree approximation
(figs. 6c-d) is again seen to be inefficient compared to that of the large $N$ 
case above.  Fig. 6e again shows that the equation of state is 
matter dominated for all but the earliest times.

A larger Hubble constant prevents significant particle production unless the 
initial amplitude of the zero mode is likewise increased such that the relation
$\eta(t_0) \gg H(t_0)$ is satisfied. For very large amplitude
$\eta(t_0) \gg 1$,  to the extent that the mass term can be neglected
and while the quantum fluctuation term has not grown to be large, the
equations of motion (\ref{hartphieq}), (\ref{hartukeq}), and
(\ref{modcer}) are scale invariant with the scaling $\eta \to \mu
\eta$, $H \to \mu H$, $t \to t/\mu$, and $k \to \mu k$, where $\mu$ is
an arbitrary scale. 

{\bf Case 3: $m^2>0$, $\eta(t_0)\gg 1$}.  The final case we examine is that of
a simple chaotic scenario with a positive mass term in the Lagrangian.  Again,
the FRW stage occurs after the inflationary expansion; this allows us
to take  zero  initial temperature. 

Figure 7 shows this situation in the large $N$ approximation for a matter
dominated cosmology.  The zero mode, $\eta(t)$, oscillates in time while
decaying in amplitude from its initial value of $\eta(t_0\! =\! 0)=5$,
$\dot{\eta}(t_0\! =\! 0)=0$ (fig.7a), while the quantum fluctuation,
$g\Sigma(t)$, grows rapidly for early times due to parametric resonance
(figs. 7b).  We choose here an initial condition on the zero mode which
differs from that of figs 4-5 above since there is no significant growth
of quantum fluctuations for smaller initial values.  From fig.7d, we see
\begin{figure}
\epsfig{file=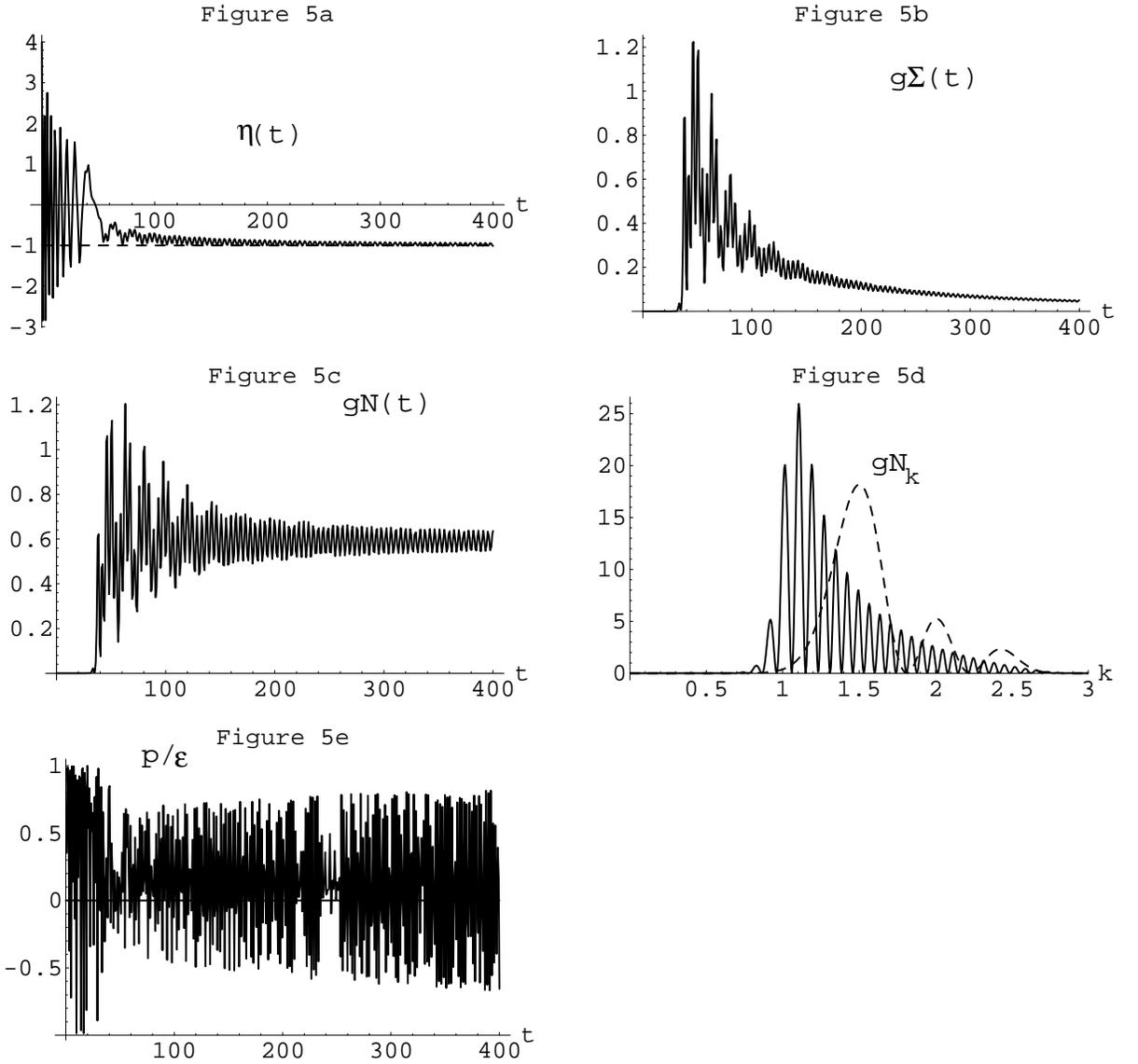}
\caption{Symmetry broken, chaotic, large $N$, matter dominated
evolution of (a) the zero mode $\eta(t)$ vs. $t$, (b) the quantum fluctuation
operator $g\Sigma(t)$ vs. $t$, (c) the number of particles $gN(t)$ vs. $t$,
(d) the particle distribution $gN_k(t)$ vs. $k$ at $t=50.8$ (dashed line)
and $t=399.4$ (solid line),  and (e) the ratio of the pressure and
energy density $ p(t)/\varepsilon(t) $ vs. $t$ for the parameter
values $ m^2=-1$, $\eta(t_0) = 4 $, $ \dot{\eta}(t_0)=0 $, $ g =
10^{-12} $, $ H(t_0) = 0.1 $. \label{fig5}}
\end{figure}
that there exists a single unstable band at values of roughly $k=1$ to
$k=3$, although careful examination reveals that the unstable band 
extends all the way to $k=0$.  The equation of state is depicted by the 
quantity $p(t)/\varepsilon(t)$ in fig.7e. As expected in this massive 
theory, the equation of state is matter dominated.

\begin{figure}
\epsfig{file=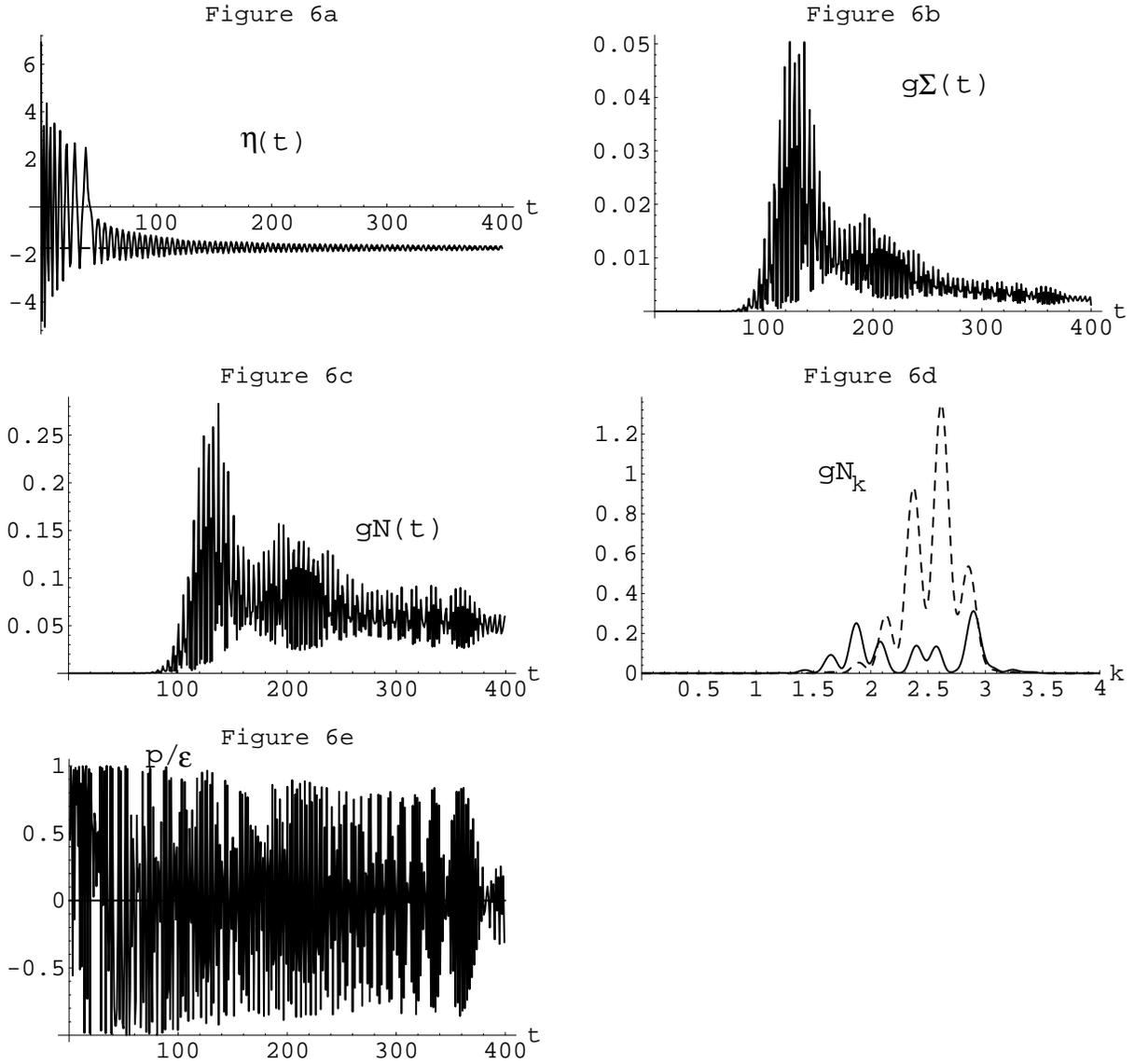}
\caption{Symmetry broken, chaotic, Hartree, matter dominated
evolution of (a) the zero mode $\eta(t)$ vs. $t$, (b) the quantum fluctuation
operator $g\Sigma(t)$ vs. $t$, (c) the number of particles $gN(t)$ vs. $t$,
(d) the particle distribution $gN_k(t)$ vs. $k$ at $t=151.3$ (dashed line)
and $t=397.0$ (solid line),  and (e) the ratio of the pressure and
energy density 
$p(t)/\varepsilon(t)$ vs. $t$ for the parameter values $m^2=-1$, $\eta(t_0) =
4\cdot 3^{1/2}$, $\dot{\eta}(t_0)=0$, $g = 10^{-12}$, $H(t_0) =
0.1$. \label{fig6}}
\end{figure}

The final case is the Hartree approximation, shown in fig.8.  Here,
parametric amplification is entirely inefficient when expansion of the universe
is included and we require an initial condition on the zero mode of
$\eta(t_0\! =\! 0)=12\sqrt{3}$ to provide even meager growth of quantum 
fluctuations.  We have used a matter dominated gravitational background with 
$H(t_0)=0.1$.  We see that while the zero mode oscillates (fig.8a), 
there is little growth in quantum fluctuations (fig.8b) and few particles
produced (fig.8c).  Examining the particle distribution (fig.8d), 
it is found that the bulk of these particles is produced within a single

\begin{figure}
\epsfig{file=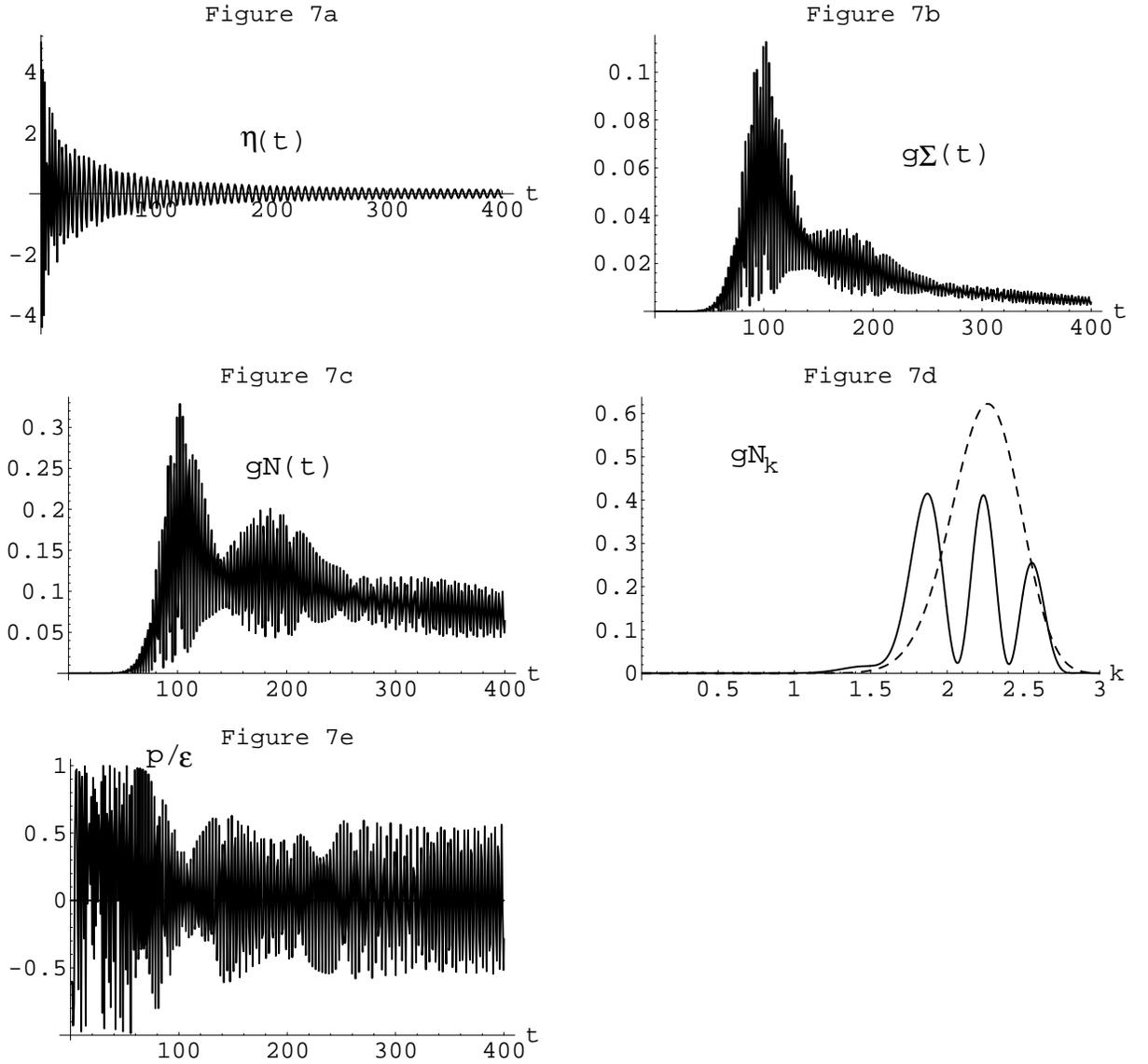}
\caption{Symmetry unbroken, chaotic, large $N$, matter dominated
evolution of (a) the zero mode $\eta(t)$ vs. $t$, (b) the quantum fluctuation
operator $g\Sigma(t)$ vs. $t$, (c) the number of particles $gN(t)$ vs. $t$,
(d) the particle distribution $gN_k(t)$ vs. $k$ at $t=77.4$ (dashed line)
and $t=399.7$ (solid line),  and (e) the ratio of the pressure and
energy density 
$p(t)/\varepsilon(t)$ vs. $t$ for the parameter values $m^2=+1$, $\eta(t_0) =
5$, $\dot{\eta}(t_0)=0$, $g = 10^{-12}$, $H(t_0) = 0.1$. \label{fig7}}
\end{figure}

resonance band extending from $k \simeq 15$ to $k \simeq 16$.  
This resonance develops at early time during the large amplitude oscillation 
of the zero mode.  These results are explained by a simple resonance band
analysis described below.

At first glance, it is not entirely clear why there are so many more particles
produced in the large $N$ case of fig.7 than in the Hartree case of fig.8.
Since in the present case the Hubble time is long compared to the oscillation
timescale of the zero mode, $H \ll 1$, we would expect a forbidden band for
early times at the location given approximately by the

\begin{figure}
\epsfig{file=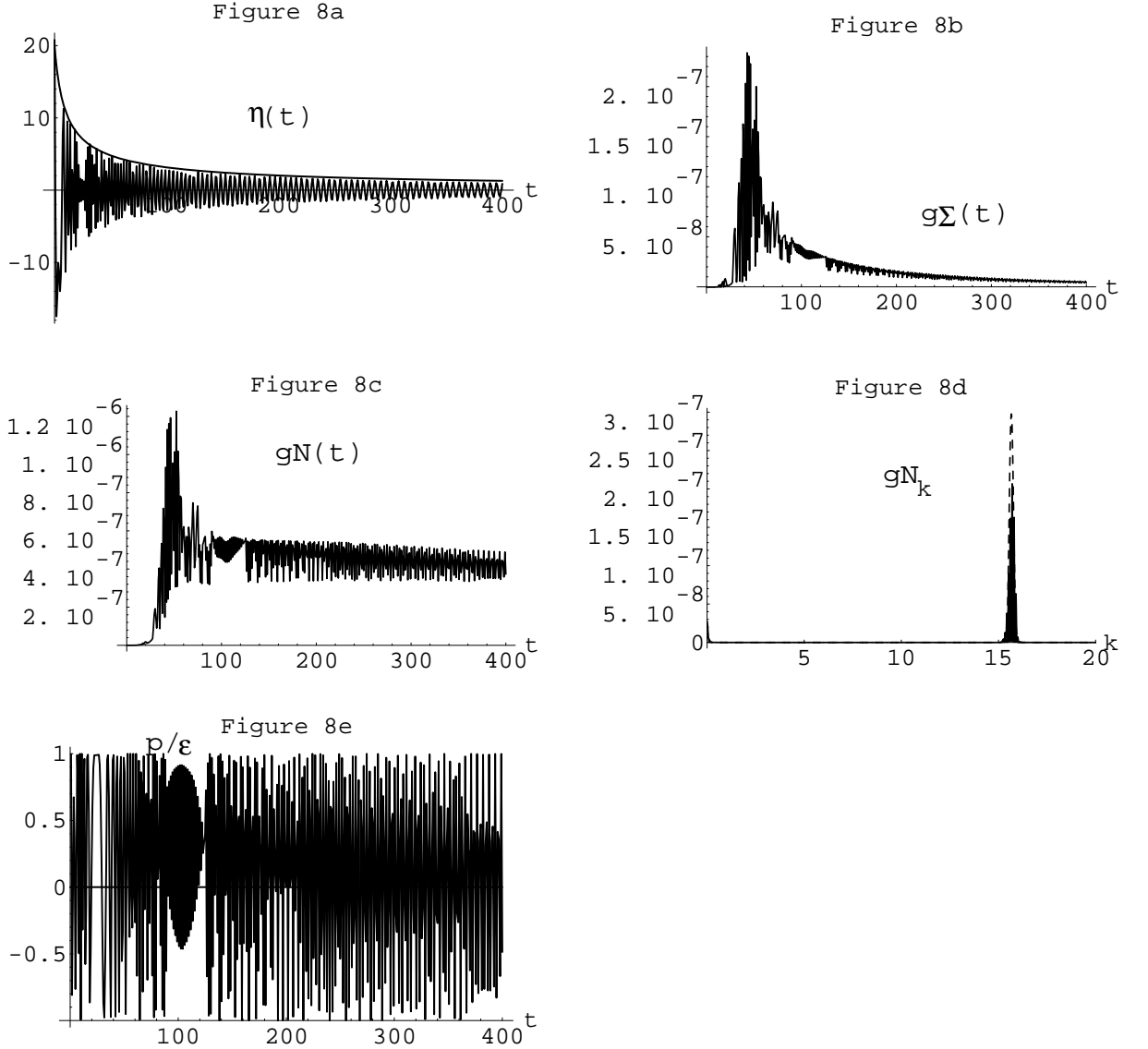}
\caption{Symmetry unbroken, chaotic, Hartree, matter dominated
evolution of (a) the zero mode $\eta(t)$ vs. $t$, (b) the quantum fluctuation
operator $g\Sigma(t)$ vs. $t$, (c) the number of particles $gN(t)$ vs. $t$,
(d) the particle distribution $gN_k(t)$ vs. $k$ at $t=50.5$ (dashed line)
and $t=391.2$ (solid line),  and (e) the ratio of the pressure and
energy density 
$p(t)/\varepsilon(t)$ vs. $t$ for the parameter values $m^2=+1$, $\eta(t_0) =
12\cdot 3^{1/2}$, $\dot{\eta}(t_0)=0$, $g = 10^{-12}$, $H(t_0) =
0.1$. \label{fig8}}
\end{figure}
Minkowski results provided in Ref. \cite{mink}.  In fact, we find this
to be the case.   

The solution to this problem is inherent in the band structure of the two cases
when combined with an understanding of the dynamics in an expanding spacetime.
First, we note that, for early times when $g\Sigma(t) \ll 1$, the zero mode 
is well fit by the function $\eta(t)=\eta_0 f(t)/a(t)$ where $ f(t) $ is an 
oscillatory function taking on values from $-1$ to $1$.  This is clearly seen
from the envelope function $\eta_0/a(t)$ shown in fig.8a (recall that 
$g\Sigma(t) \ll 1$ during the entire evolution in this case).  Second, the 
momentum that appears in the equations for the modes (\ref{hartukeq}) is the
{\em physical} momentum $k/a(t)$.  We therefore write the approximate 
expressions for the locations of the forbidden bands in FRW by using the
Minkowski results of \cite{mink} with the substitutions 
$\eta_0^2 \to \gamma\eta_0^2/a^2(t)$ (where the factor of $\gamma$ accounts
for the difference in the definition of the non-linear coupling between
this study and \cite{mink}) and $q^2 \to k^2/a^2(t)$.  

Making these substitutions, we find for the location in comoving momentum $k$ 
of the forbidden band in the large $N$ (fig.7) and Hartree (fig.8) cases:
\begin{eqnarray}
0 \leq & k^2 & \leq \frac{\eta_0^2}{2}, \; \; (\mbox{large}\; N) \\
\frac{\eta_0^2}{2}+3a^2(t) \leq & k^2 & \leq
a^2(t)\left(\sqrt{\frac{\eta_0^4}{3a^4(t)}+
\frac{2\eta_0^2}{a^2(t)}+4}+1\right) \; ,
\; \; (\mbox{Hartree}) \; .\label{hartband}
\end{eqnarray}
The important feature to notice is that while the location of the unstable
band (to a first approximation) in the case of the continuous $O(N)$ theory
is the same as in Minkowski and does not change in time, the location of 
the band is time dependent in the discrete theory described by the 
non-perturbative Hartree approximation.

While $\eta_0/a(t) \gg 1$, the Hartree relation reduces to
\begin{equation}
\frac{\eta_0^2}{2} \leq k^2 \leq \frac{\eta_0^2}{\sqrt{3}}.
\end{equation}
This is the same as the Minkowski result for large amplitude, and one finds
that this expression accurately predicts the location of the resonance band
of fig.8d.  However, with time the band shifts its location toward higher
values of comoving momentum as given by (\ref{hartband}), cutting off particle
production in that initial band.  There is continuing particle production 
for higher modes, but since the Floquet index is decreased due to the reduced
amplitude of the zero mode, since there is no enhancement of production
of particles in these modes (as these modes begin with at most of order $1$
particles), and because the band continues to shift to higher momenta while 
becoming smaller in width, this particle production never becomes significant.

As in the symmetry broken case of figs. 4-6, the equations of motion for
large amplitude and relatively early times are approximately scale invariant.
In fig.9 we show the case of the large $N$ evolution in a radiation
dominated universe with initial Hubble constant of $H(t_0)=2$ 
 with appropriately scaled initial value of the zero mode of
 $\eta(t_0)=16$.  Again, the qualitative dynamics remains 
largely unchanged from the case of a smaller Hubble constant.

\subsection{Late Time Behavior}
We see clearly from the numerical evolution that in the case of a symmetry
broken potential, the late time large $N$ solutions obey the sum rule
(\ref{sumrule}). This sum rule is a consequence of the late time Ward
identities which enforce 
Goldstone's Theorem.  Because of this sum rule, we can write down the
analytical expressions for the late time behavior of the fluctuations and the
zero mode.  Using (\ref{sumrule}), the mode equation (\ref{hartukeq}) becomes
\begin{equation}
\left[\frac{d^2}{dt^2}+3\frac{\dot{a}(t)}{a(t)}\frac{d}{dt}+\frac{k^2}{a^2(t)}
\right]U_k(t) = 0.
\label{lateuk}
\end{equation}
This equation can be solved exactly if we assume a power law dependence for the
scale factor 

\begin{figure}
\epsfig{file=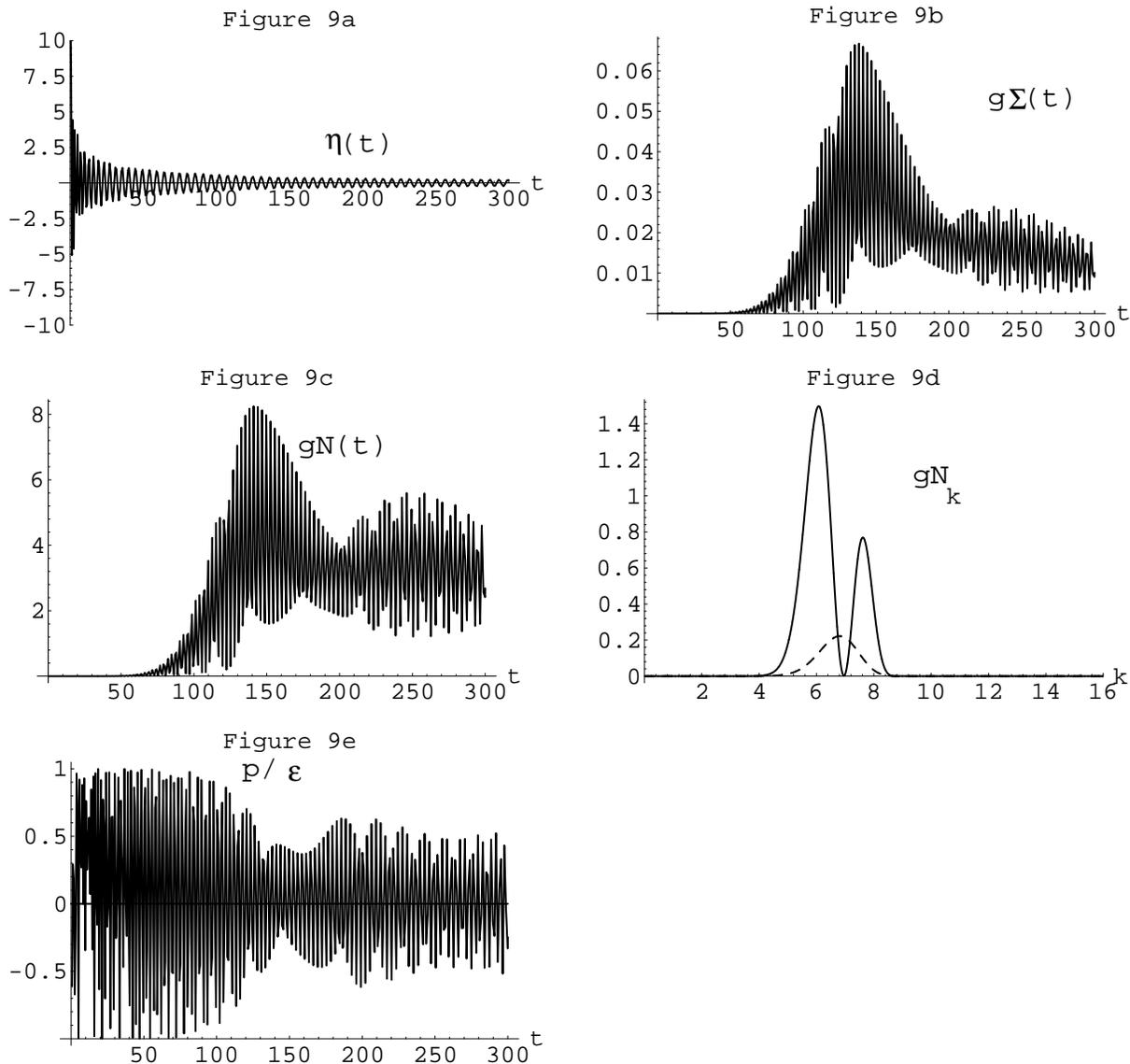}
\caption{Symmetry unbroken, chaotic, large $N$, radiation dominated
evolution of (a) the zero mode $\eta(t)$ vs. $t$, (b) the quantum fluctuation
operator $g\Sigma(t)$ vs. $t$, (c) the number of particles $gN(t)$ vs. $t$,
(d) the particle distribution $gN_k(t)$ vs. $k$ at $t=102.1$ (dashed line)
and $t=251.6$ (solid line),  and (e) the ratio of the pressure and energy density
$p(t)/\varepsilon(t)$ vs. $t$ for the parameter values $m^2=+1$, $\eta(t_0) =
16$, $\dot{\eta}(t_0)=0$, $g = 10^{-12}$, $H(t_0) = 2.0$. \label{fig12}}
\end{figure}

$a(t) = (t/t_0)^n$.  The solution is
\begin{equation}
U_k(t) = c_k \; t^{(1-3n)/2} \; J_{\frac{1-3n}{2-2n}}\left(\frac{k t_0^n
	t^{1-n}}{n-1}\right) + 
	d_k \; t^{(1-3n)/2} \; Y_{\frac{1-3n}{2-2n}}\left(\frac{k t_0^n
	t^{1-n}}{n-1}\right), 
\label{latesoln}
\end{equation}
where $J_{\nu}$ and $Y_{\nu}$ are Bessel and Neumann functions respectively, 
and the constants $c_k$ and $d_k$ 
carry dependence on the initial conditions and on the dynamics up to the point
at which the sum rule is satisfied.  

These functions have several important properties.  In particular, in radiation
or matter dominated universes, $n<1$, and for values of wavenumber satisfying
$k \gg t^{-(1-n)}/t_0^n$, the mode functions decay in time as 
$1/a(t) \sim t^{-n}$.  Since
the sum rule applies for late times, $t-t_0 \gg 1$ in dimensionless units, we see
that all values of $k$ except a very small band about $k=0$ redshift as
$1/a(t)$.  The $k=0$ mode, however, remains constant in time, explaining the
support evidenced in the numerical results for values of small $k$ (see
figs. 1,3).  These results mean that the quantum fluctuation has a late time
dependence of $\langle \psi^2(t)\rangle_r \sim 1/a^2(t)$.  The late time
dependence of the zero mode is given by this expression combined with the sum
rule (\ref{sumrule}).  These results are accurately reproduced by our numerical
analysis.  Note that qualitatively this late time dependence is independent of
the choice of initial conditions for the zero mode, except that there is no
growth of modes near $k=0$ in the case in which particles are produced via
parametric amplification (figs. 4,5).

For the radiation $ n= \frac12 $ and matter dominated $ n = \frac23 $
universes, eq.(\ref{latesoln}) reduces to elementary functions:
\begin{eqnarray}
a(t) \; U_k(t) &=&  c_k \; e^{2ik t_0^{1/2} t^{1/2} } + 
d_k\; e^{-2ik t_0^{1/2} t^{1/2} }  
\;  \; \mbox{(RD) } \; ,\cr \cr
a(t) \; U_k(t) &=&  c_k \; e^{3ik t_0^{2/3} t^{1/3} } \; 
\left[ 1 + {i \over {3k t_0^{2/3}
t^{1/3}}}\right] + d_k\; e^{-3ik t_0^{2/3} t^{1/3}}\left[ 1 - {i \over {3k
t_0^{2/3} t^{1/3}}}\right] \; \; \mbox{ (MD) }.
\end{eqnarray}

It is also of interest to examine the $n>1$ case.  Here, the modes of interest
satisfy the condition $k \ll t^{n-1}/t_0^n$ for late times.  
These modes are constant
in time and one sees that the modes are {\it frozen}.  In the case of a de
Sitter universe, we can formally take the limit $n \to \infty$ and we see that
{\it all} modes become frozen at late times.   This case is detailed in
sec. VIII \cite{De Sitter}.

\subsection{Discussion and Conclusions for the FRW background}

We have shown that there can be significant particle production through quantum
fluctuations after inflation\cite{frw2}. However, this production is somewhat 
sensitive
to the expansion of the universe. From our analysis of the equation of state,
we see that the late time dynamics is given by a matter dominated cosmology.
We have also shown that the quantum fluctuations of the inflaton decay for late
times as $1/a^2(t)$, while in the case of a symmetry broken inflationary model,
the inflaton field moves to the minimum of its tree level potential.  The
exception to this behavior is the case when the inflaton begins exactly at the
unstable extremum of its potential for which the fluctuations grow out to the
minimum of the potential and do not decay.  Initial production of particles due
to parametric amplification is significantly greater in chaotic scenarios with
symmetry broken potentials than in the corresponding theories with positive
mass terms in the Lagrangian, given similar initial conditions on the zero mode
of the inflaton.

\section{Fields Evolution on a fixed Inflationary Background 
(the de Sitter Universe)}

We describe in this section the matter evolution on a fixed de Sitter
background using large $ N $ and Hartree approaches. The evolution
with a dynamical background is treated in sec. IX.

\subsection{Evolution for $\phi(0)= \dot{\phi}(0)=0$. Analytical Results}

We begin by considering the broken symmetry 
situation in which the expectation value of the
inflaton field sits atop the potential hill with zero initial velocity. This
situation is expected to arise if the system is initially in local
thermodynamic equilibrium  an initial
temperature larger than the critical temperature and cools down
through the critical temperature in the absence of an external field or bias.

The order parameter and its time derivative vanish in the
local equilibrium high
temperature phase, and this condition is a fixed point of the evolution
equation for the zero mode of the inflaton.  There is no rolling of the
inflaton zero mode in this case, although the fluctuations will grow
and will be responsible for the dynamics. 

We can understand the early stages of the dynamics analytically as follows. For
very weak coupling and early time we can neglect the backreation in the mode
equations, which in both the large $N$ and Hartree cases become,

\begin{equation} 
\left[\frac{d^2}{d\tau^2}+3h
\frac{d}{d\tau}+\frac{q^2}{a^2(\tau)}-1\right]f_q(\tau)=0, \label{earlytime}
\end{equation}
\begin{equation}
f_q(0)=\frac{1}{\sqrt{\omega_q}} \; ; \quad \dot{f}_q(\tau) = -i
\sqrt{\omega_q} \; ; \quad \omega_q= \sqrt{q^2+r^2-1} \; .
\end{equation}
The solutions are of the form,
\begin{equation}
f_q(\tau) = \exp[-\frac{3}{2}h\tau] \left\{a(q)\; J_{\nu}(z)+b(q)\;
J_{-\nu}(z) \right\} \; ; \;
z=\frac{q}{h}\exp[-h\tau] 
\; ; \; \nu = \sqrt{\frac{1}{h^2}+\frac{9}{4}}, \label{bessel}
\end{equation}
where the coefficients $a(q)$ and $b(q)$ are determined by the initial
conditions:
\begin{equation}
b(q) = - {{\pi \, q}\over { 2 h \, \sin{\nu \pi} }} \; \left[ {{i \omega_q -
\frac32 \, h }\over q} \; J_{\nu}\left(\frac{q}{h}\right) - 
J'_{\nu}\left(\frac{q}{h}\right) \right],
\label{coefb}
\end{equation}
\begin{equation}
a(q) =   {{\pi \, q}\over { 2 h \, \sin{\nu \pi} }} \; \left[ {{i \omega_q -
\frac32 \, h }\over q} \;  J_{-\nu}\left(\frac{q}{h}\right) -  
J'_{-\nu}\left(\frac{q}{h}\right)\right] \;. \label{coefa}
\end{equation}

For long times, $e^{h\tau}\geq q/h$, these mode functions grow exponentially,
\begin{equation}\label{Uasi}
f_q(\tau) \simeq b(q) \; J_{-\nu}(z) \simeq {{b(q)} \over {\Gamma(1-\nu)}}
\; \left( {{2h\, }\over q}\right)^{\nu}e^{(\nu-3/2)h \tau}   \;.
\end{equation}

The Bessel functions appearing in the expression for the modes $ f_q(\tau) $
can be approximated by their series expansion,
\begin{equation}
f_q(\tau) = \frac12 \left[ 1 + \frac{1}{\nu} \; \left( \frac32 -{{q^2}\over{4
h^2}} - i {{\omega_q}\over h} \right) + {\cal O}\left(\frac1{\nu^2}\right) \right] \; 
e^{(\nu - 3/2) h\tau} \; .
\end{equation}
This is an expansion in powers of $ q^2/(\nu h^2) $ and we conclude that $
g\Sigma(\tau) $ is dominated by the modes with $q \leq \sqrt{h}$.

The integral for $g\Sigma(\tau)$ can be approximated by keeping only the modes
$ q \leq f \sqrt{h} $, where $ f $ is a number of order one, and by neglecting
the subtraction term which will cancel the contributions from high
momenta. Numerically, even with the backreaction taken into account, the
integral is dominated by modes $q \leq f \approx 10-20$ in all of the cases
that we studied (see fig.12 below).

The contribution to the fluctuations from these unstable modes is:
\begin{equation}\label{estim}
 g\Sigma(\tau) \simeq \sqrt{{g}\over 3}\, {{f^3\,h^{3/2} \, r m^2_R}
\over { 2 \pi \, \, M_0^2 }}\; \left(1 + {{ M_0^2}\over { m^2_R}}\right) \;
e^{(2\nu-3)h\tau} \; ,
\end{equation}
where again, we have taken the high temperature limit, $ T_i \sim T_c \gg m_R
$.

From this equation, we can estimate the value of $\tau_s$, the `spinodal
time', at which the contribution of the quantum fluctuations becomes
comparable to the contribution from the tree level terms in the equations of
motion. This time scale is obtained
 from the condition $g\Sigma(\tau_s) = {\cal O}(1)$:
\begin{equation}
\tau_s \simeq -\frac{1}{(2\nu-3)h}\ln\left[\sqrt{g\over 3}
{{f^3\,h^{3/2}}\over { 2 \pi \, m_R M_0^2 }}\;\frac{T_i}{T_c} 
\left(1 + {{ M_0^2}\over { m_R^2}}\right)\right]\label{spinoest} ,
\end{equation}
which is in good agreement with our numerical results, as will become clear
below (see figs. 10, 15 and 18). For values of $h \geq 1$,
which, as argued
below, lead to the most interesting case, an estimate for the spinodal time is,
\begin{equation} \label{tslargeh}
\tau_s \simeq  \frac{3h}{2} \ln[1 / \sqrt{g}] + {\cal{O}}(1)
\label{spinotime}
\end{equation}
which is consistent with our numerical results (see fig.10).

For $\tau > \tau_s$, the effects of backreaction become very important, and the
contribution from the quantum fluctuations competes with the tree level terms
in the equations of motion, shutting-off the instabilities. Beyond $\tau_s$,
only a full numerical analysis will capture the correct dynamics.

It is worth mentioning that had we chosen zero temperature initial conditions,
then the coupling $\bar{g} \rightarrow g $ (see (\ref{gsigma})) 
and the estimate for the spinodal
time would have been,
\begin{equation}
\tau_s \simeq \frac{3h}{2} \ln[1/{g}] + {\cal{O}}(1), \label{spinotimeT0}
\end{equation}
that is, roughly a factor 2 larger than the estimate for which the de Sitter
stage began at a temperature above the critical value. Therefore
(\ref{spinotime}) represents an {\em underestimate} of the spinodal time scale
at which fluctuations become comparable to tree level contributions.

The number of e-folds occurring during the stage of growth of spinodal
fluctuations is therefore,
\begin{equation}
{\cal{N}}_e \approx \frac{3h^2}{2} \ln[1/ \sqrt{g}], 
\end{equation}
or in the zero temperature case,
\begin{equation}
{\cal{N}}_e \approx \frac{3h^2}{2} \ln[1/ g], \label{efoldsT0}
\end{equation}
which is a factor 2 larger.  Thus, it becomes clear that with $ g \approx
10^{-12}$ and $h \geq 2$, a required number of e-folds, ${\cal{N}}_e \approx
100$ can easily be accommodated before the fluctuations become large,
modifying the dynamics and the equation of state.

The implications of these estimates are important.  The first conclusion drawn
from these estimates is that a `quench' approximation is well justified (see
fig.10). While the temperature drops from an initial value of a few
times the critical temperature to below critical in just a few e-folds, the
contribution of the quantum fluctuations needs a large number of e-folds
to grow
to compensate for the tree-level terms and overcome the instabilities. Only for
a strongly coupled theory is the time scale for the quantum fluctuations to
grow short enough to restore local thermodynamic
equilibrium during the transition.

The second conclusion is that most of the growth of spinodal fluctuations
occurs during the inflationary stage, and with $ g \approx 10^{-12}$ and
$H \geq m_R$, the quantum fluctuations become of the order of the tree-level
contributions to the equations of motion within the number of e-folds necessary
to solve the horizon and flatness problems.  
Since the fluctuations grow to become
of the order of the tree level contributions at times of the
order of this time scale, for larger times they will
modify the equation of state substantially and will be shown to provide a
graceful exit from the inflationary phase within an acceptable number of
e-folds.

For $\tau < \tau_s$, when the contribution from the renormalized quantum
fluctuations can be ignored, the Hubble constant is given by the classical
contribution to the energy density. In terms of the dimensionless quantities
introduced above (\ref{dimvars3}), we have,
\begin{equation}
H= \frac{2 m^4_R}{3\pi \, g \,
M^2_{pl}}\left[\frac{\dot{\eta}^2}{2}+\frac{1}{4}(\eta^2-1)^2\right].
\label{hubbleconst}
\end{equation}
In the situation we consider here, with $\dot{\eta}=\eta=0$, the condition
that $h \geq 2$ for $ g \simeq 10^{-12}$ translates into $m_R \simeq
10^{13}\mbox{ GeV }$, which is an acceptable bound on the inflaton mass.

To understand more clearly whether or not the effect of quantum fluctuations
and growth of unstable modes during the inflationary phase transition can
provide a graceful exit scenario, we must study in detail the contribution to
the energy and the equation of state of these quantum fluctuations.

Although we are working in a fixed de Sitter background, the energy and
pressure will evolve dynamically. A measure of the backreaction effects of
quantum fluctuations on the dynamics of the scale factor is obtained from
defining the `effective Hubble constant',
\begin{equation}
{\cal{H}}^2(\tau) = \frac{8\pi}{3M^2_{pl}}\; \varepsilon(\tau). \label{hub}
\end{equation} 
Therefore, the quantities,
$$
\frac{{\cal{H}}(\tau)}{{\cal{H}}(0)}= \sqrt{\frac{
\varepsilon(\tau)}{\varepsilon(0)}}\quad 
\mbox{and}\quad
\frac{\dot{\cal{H}}(\tau)}{{\cal{H}}^2(\tau)} = -\frac{3}{2}
\left[1+\frac{p(\tau)}{\varepsilon(\tau)}\right]\; , 
$$
give dynamical information of the effects of the backreaction of the quantum
fluctuations on the dynamics of the scale factor.  Whenever
$p(\tau)+\varepsilon(\tau) \neq 0$, ${\cal{H}}(\tau)/{\cal{H}}(0) \neq 1$, or
$\dot{\cal{H}}(\tau)/ {\cal{H}}^2(\tau) \neq 0$, the backreaction from the
quantum fluctuations will dramatically change the dynamics of the scale factor,
and it will no longer be consistent to treat the scale factor as fixed. When
${\cal H}(\tau)/ {\cal H}(0) \ll 1$ the de Sitter era will end.

From this point onwards only a full treatment of the backreaction, {\em
including} the correct dynamics of the scale factor, will describe the physics.
This will be the subject of sec. IX \cite{din}.

\subsection{\bf Numerical Analysis}

We now solve numerically the large $N$ set of equations  (\ref{modcr}) 
with the initial conditions (\ref{modkr}), taking $\dot{\eta}(0) = \eta(0) =0$.

The numerical code is based on a fourth order Runge-Kutta algorithm for the
differential equation and an 11-points Newton-Cotes algorithm for the integral,
with a typical relative errors $10^{-9}$ in the differential equation and
in the integrals. We have tested for cutoff insensitivity with
cutoffs $q_{max} \approx 50 \; ; 100 \; ; 150$ with no appreciable variation in
the numerical results.  The reason for this cutoff insensitivity is due to the
fact that only long-wavelength modes grow in amplitude to become
non-perturbatively large, whereas the short-wavelength modes always have
perturbatively small amplitudes.  We have chosen $r=T_i/T_c =2$ as a
representative value and $ g = 10^{-12} $. 

As argued previously, for $ g  \approx 10^{-12}$, the cosmologically
interesting time scales for the spinodal instabilities to grow during say the
first 60-100 e-folds of inflation occur for $h \geq 1$, leading to $H \geq m_R
\geq 10^{13} \mbox{ GeV }$ which is a phenomenologically acceptable 
range for the Hubble constant during the inflationary stage.

Fig. 10 shows the contribution from the quantum fluctuations,
$g\Sigma(\tau)$ vs. $\tau$ for $ g = 10^{-12} \; ; \; T_i/T_c=2 \; ; \;
h=2 \; ; \; \eta(0)=0\; ; \;\dot{\eta}(0)=0$. The quantum fluctuations, as
measured by $g\Sigma(\tau)$, grow to be of order 1 in a time scale $\tau
\approx 40$ which is the time scale predicted by the early time estimate
(\ref{spinotime}). 
We have checked numerically that the energy is covariantly conserved,
obeying the relation $\dot{\varepsilon}+3H(p+\varepsilon)=0$ to our
numerical accuracy of one part in $10^7$.
Fig. 11 shows ${\cal{H}}(\tau)/{\cal{H}}(0)$ 
vs. $\tau$. This
figure shows clearly that when the spinodal quantum fluctuations become
comparable to the tree level contribution to the equations of motion, the
backreaction on the scale factor becomes fairly large. At this point, the
approximation of keeping a fixed background breaks down and the full
self-consistent dynamics will have to be studied. At this time, the
inflationary stage basically ends since ${\cal{H}}$ is no longer constant. This
occurs for $\tau \approx 40$ giving about 80 e-folds of inflation during the
time in which ${\cal{H}}$ is approximately constant and equal to $H$. 
Therefore,
this new mechanism of spinodal fluctuations, with the zero mode sitting atop
the potential hill provides a graceful exit of the inflationary era without any
further assumptions on the evolution of the scalar field.

\begin{figure}
\epsfig{file=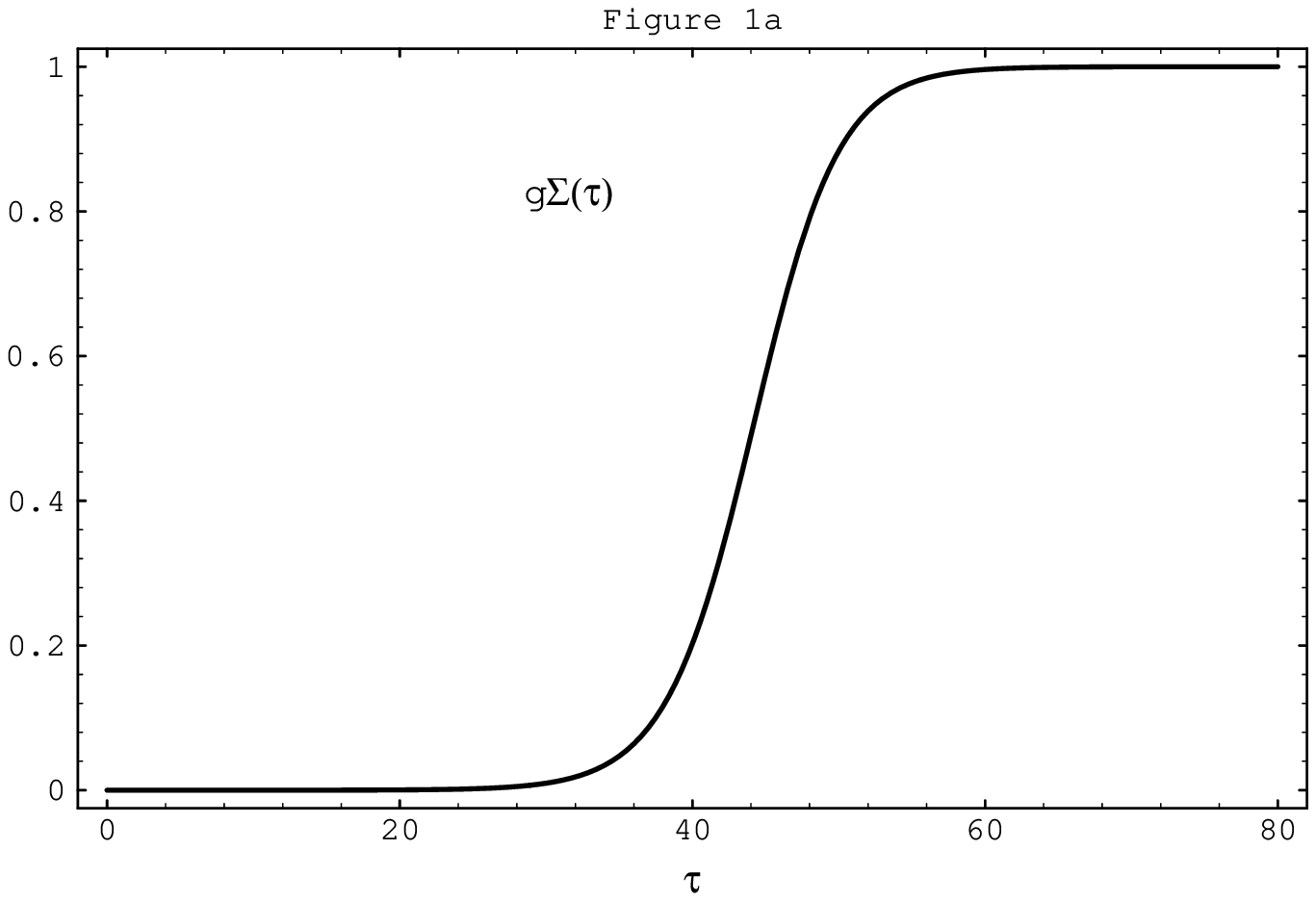,width=10.5cm,height=5.5cm}
\caption{$g\Sigma(\tau)$ vs. $\tau$ for
$\eta(0)=\dot{\eta}(0)=0; \quad \lambda=10^{-12}; \quad r=2; \quad h=2$.
\label{fig1a}}
\epsfig{file=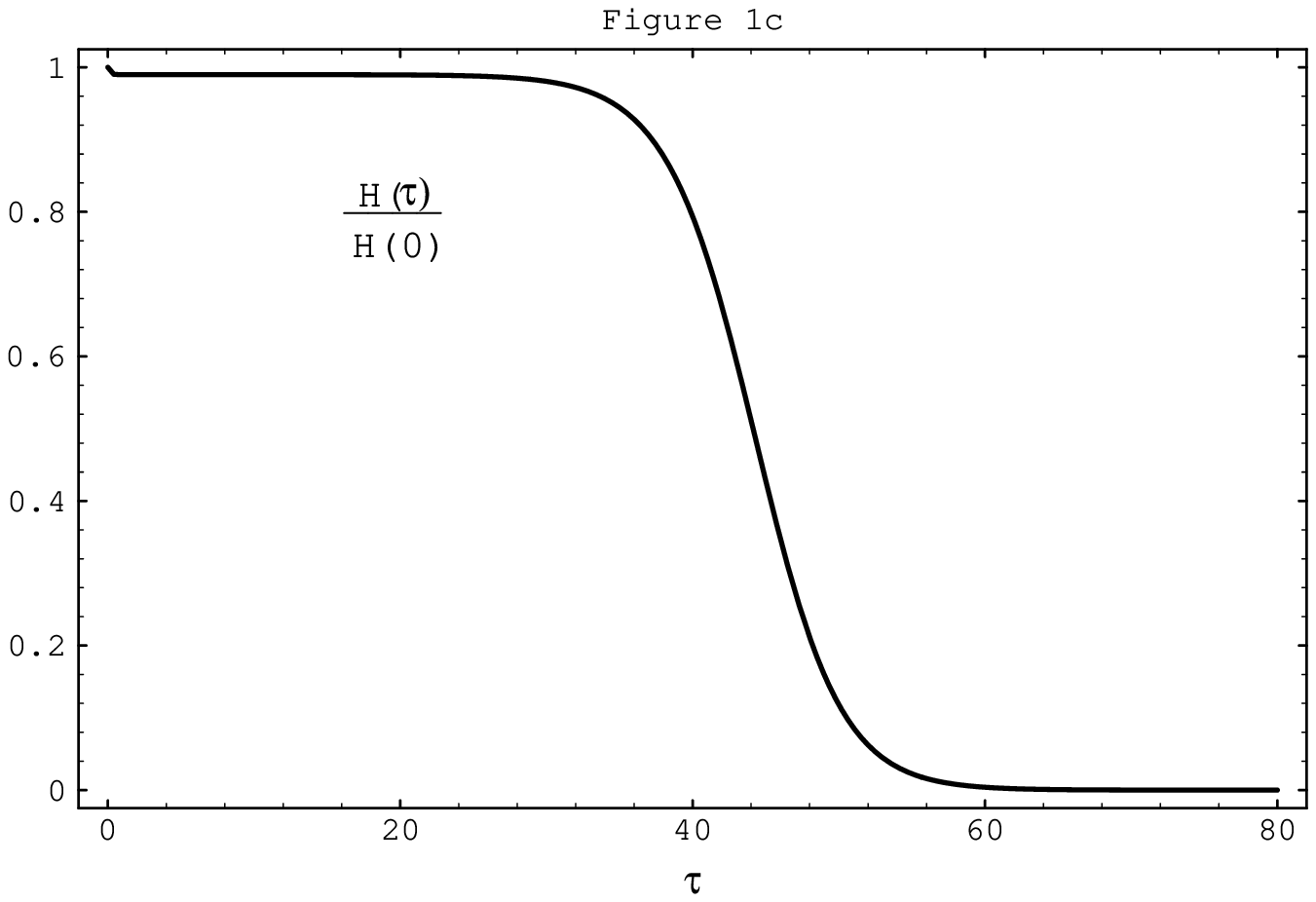,width=10.5cm,height=5.5cm}
\caption{${\cal H}(\tau) / {\cal H}(0)$ vs $\tau$ for
the same parameters as in fig.10.
\label{fig1c}}
\epsfig{file=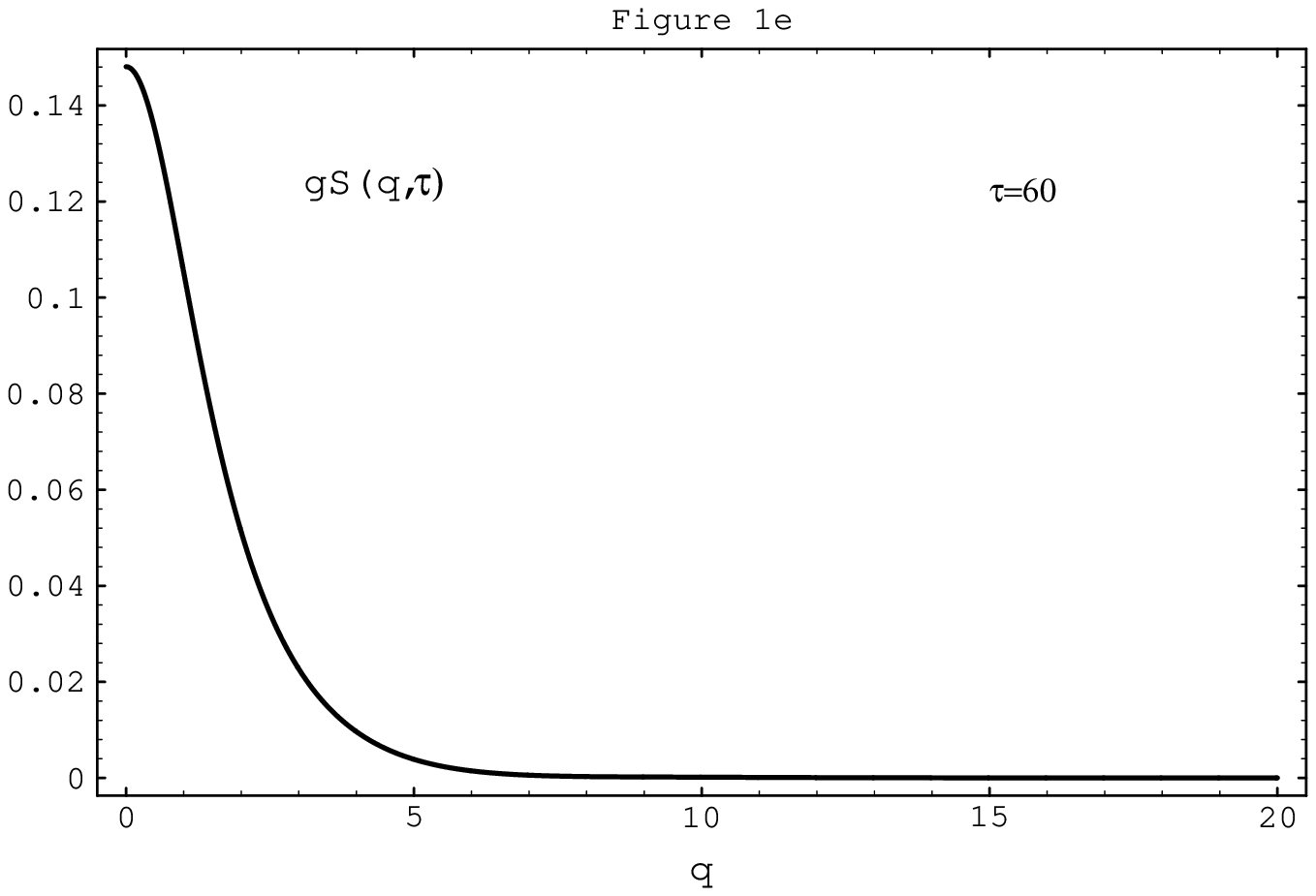,width=10.5cm,height=5.5cm}
\caption{$g{\cal S}(q,\tau)$ vs. $q$ for $\tau=
60$. Same parameters as in fig.10.
\label{fig1e}}
\end{figure}

These fluctuations translate into an amplification of the power spectrum
at long
wavelengths for $q \approx h$. To see this clearly we plot $g
{\cal{S}}(q,\tau) = g |f_q(\tau) |^2 $ 
vs. $q$ for $\tau= 60$ in fig.12. This quantity is very small, because of
the coupling constant in front, but for $\tau \approx \tau_s$ it grows to be of
order one for long wavelengths (see also fig.14) 
and vanishes very fast for $q > 10$. The
integral in $g\Sigma(\tau)$ is dominated by these long wavelengths that become
non-perturbatively large, whereas the contribution from the short wavelengths
remains always perturbatively small. This is the justification for the
approximations performed early that involved only the long-wavelength modes and
cutoffs of order $\sqrt{h}$. The equal time spatial
correlation function given by
eq.(\ref{rhocorr}) can now be computed explicitly. Figure 13 shows $S(\rho
; \tau)/S(0;\tau)$ as a function of $\rho$ for $\tau \ge 2$. We {\em define}
the correlation length $\xi(\tau)$ as the value of $\rho$ for which the ratio
is $1/e$. Figure (1.g) shows $\xi(\tau)$; notice that the correlation length
saturates to a value $\xi(\infty) \approx 1/h$, and that the correlated
regions are of horizon size.

We have performed numerical analysis varying $h$ with the same values of
$ g $ and for the same initial conditions, and found that the only
quantitative change is in the time scale for $g\Sigma(\tau)$ to be of order
one.  We find that the spinodal time scale grows almost linearly with $h$ and
its numerical value is accurately described by the estimate (\ref{spinotime}).
The case in which the Hubble constant is $h=0.1$ is shown explicitly.  Figure
15 shows $g\Sigma(\tau)$, which demonstrates the oscillatory behavior 
similar to what is seen in Minkowski space\cite{us1,mink}.  The correlation
length, $\xi(\tau)$, is shown in fig.16; its asymptotic value is again
approximately given by $1/h$.

\begin{figure}
\epsfig{file=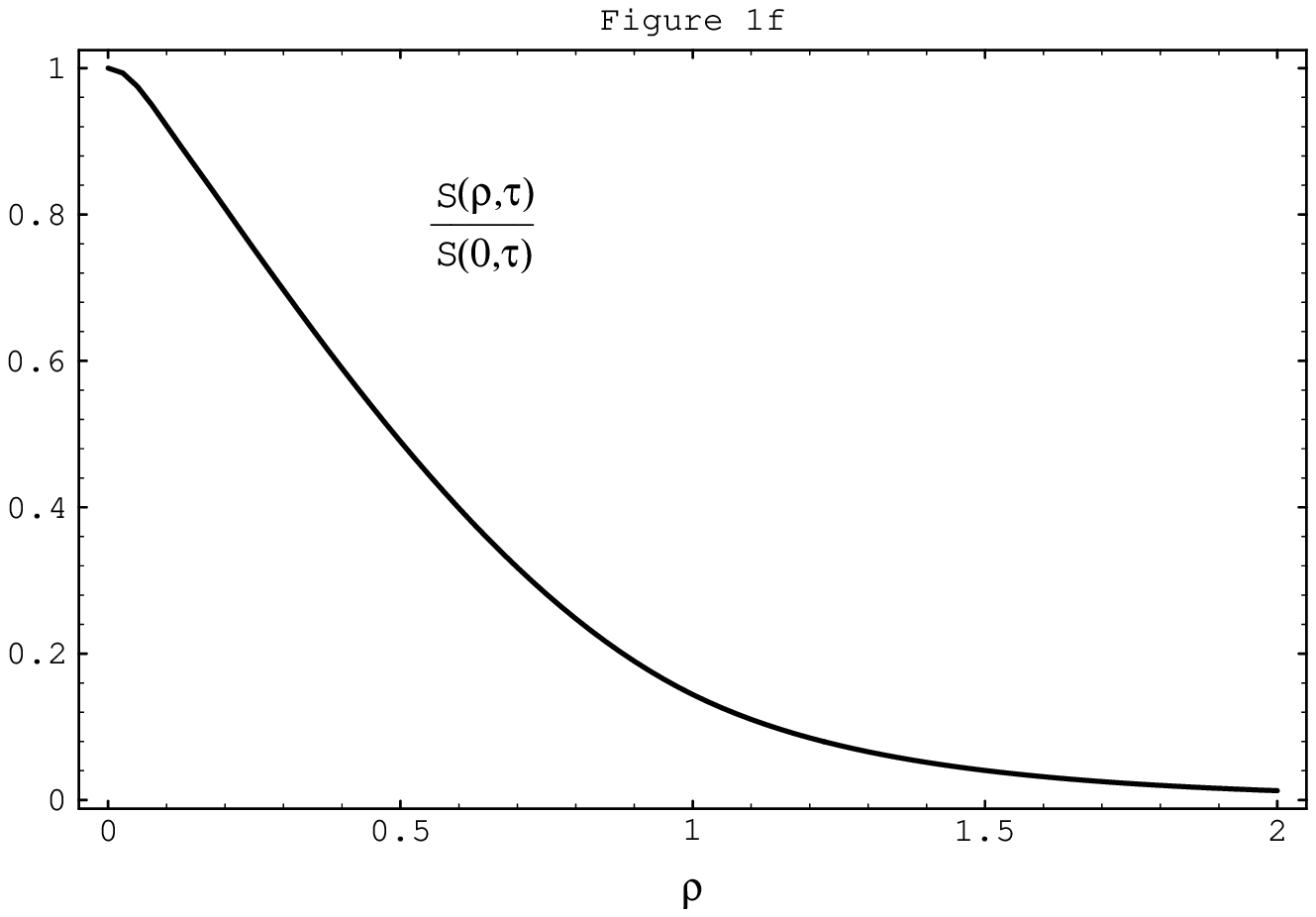,width=10.5cm,height=5.5cm}
\caption{$S(\rho,\tau) / S(0,\tau)$ vs. $\rho$ for
$\tau \ge 2$. Same parameters as in fig.10.
\label{fig1f}}
\epsfig{file=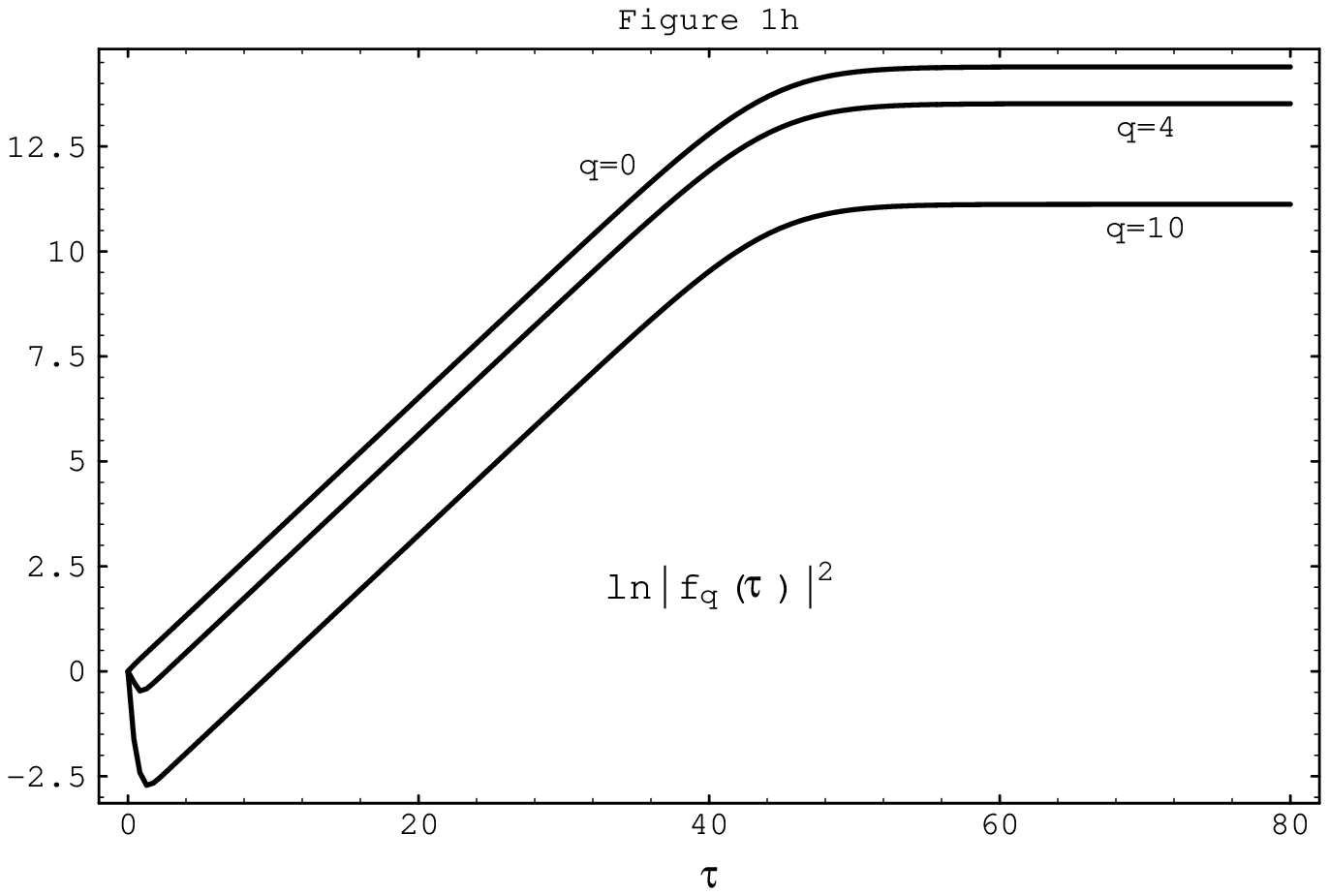,width=10.5cm,height=5.5cm}\
\caption{$\ln\left[|f_q(\tau)|^2\right]$ vs. $\tau$ for
$q=0,4,10$. Same parameters as in fig.10.
\label{fig1h}}
\epsfig{file=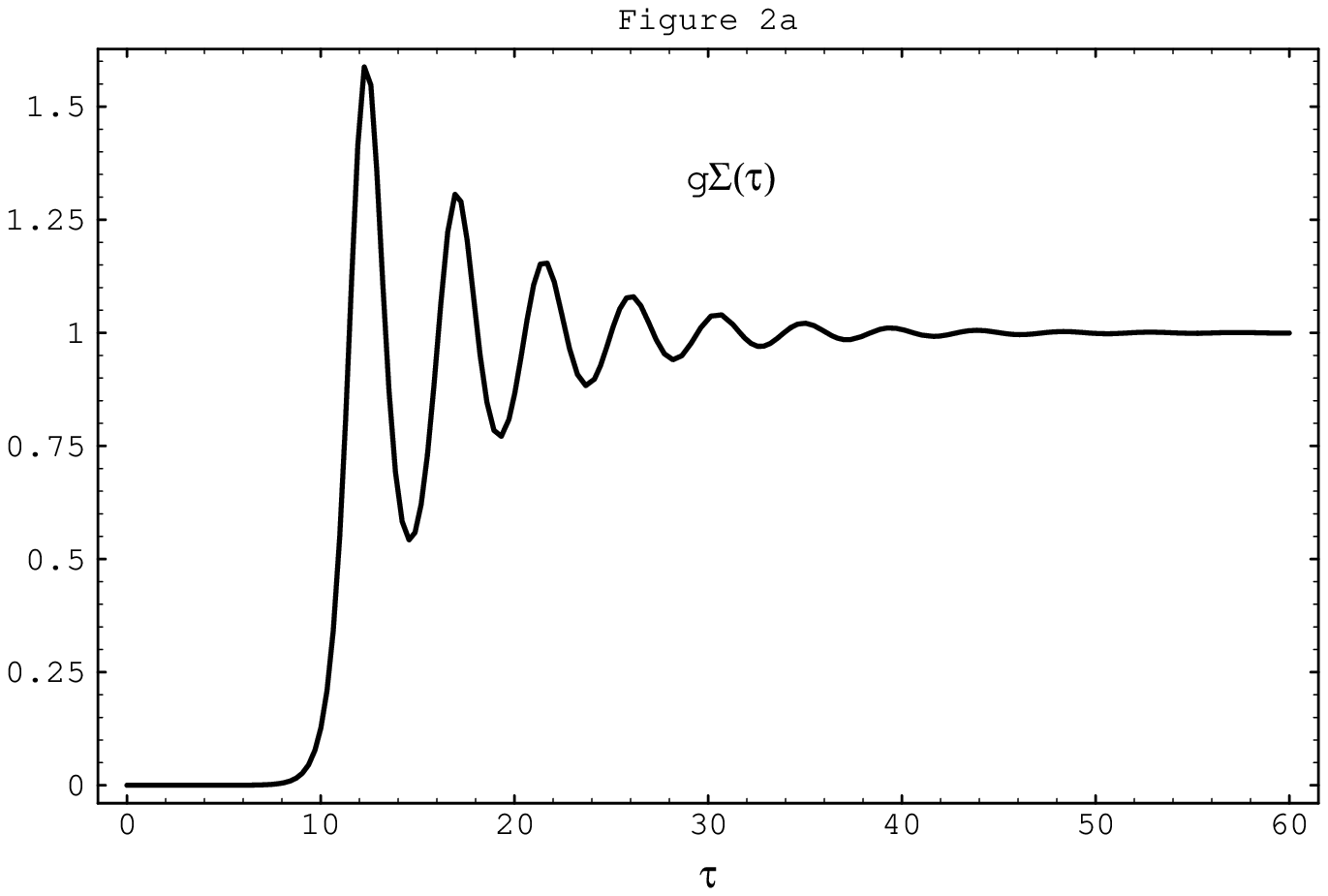,width=10.5cm,height=5.5cm}
\caption{$g\Sigma(\tau)$ vs. $\tau$ for
; $\eta(0)=\dot{\eta}(0)=0; \quad ; \lambda=10^{-12} \quad r=2; \quad h=0.1$.
\label{fig2a}}
\epsfig{file=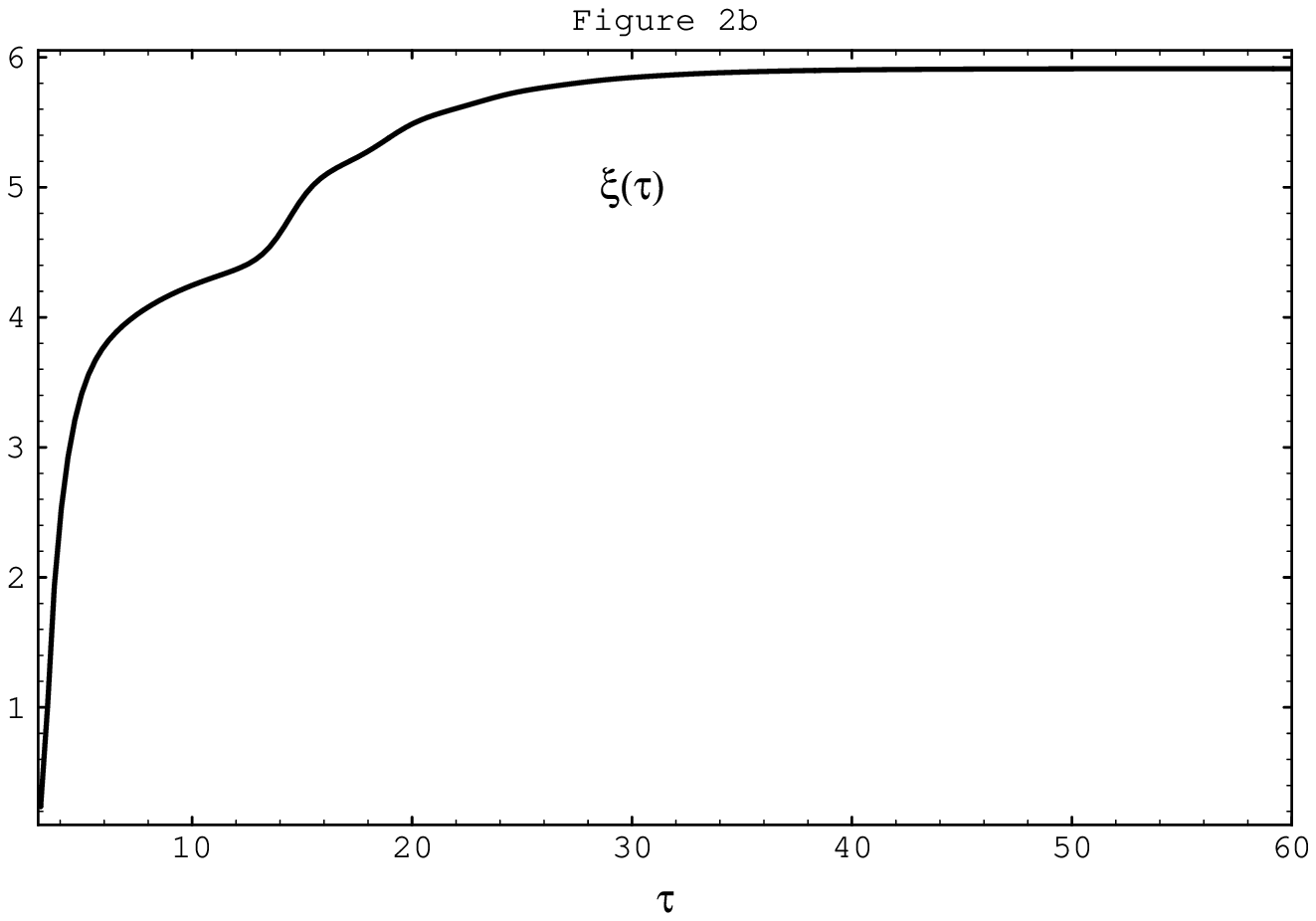,width=10.5cm,height=5.5cm}
\caption{$\xi(\tau)$ vs $\tau$. Same parameters as in fig.15.
\label{fig2b}}
\end{figure}
\subsection{\bf The late time limit}

For times $\tau > \tau_s \approx 40$ (for the values of the parameters used in
figs. (1)) we see from figs. 10-14 that the dynamics freezes out. The
fluctuation, $g\Sigma(\tau) =1$, and the mode functions effectively describe
free, minimally coupled, massless particles.  The sum rule,
\begin{equation}
-1+g\Sigma(\infty) =0, \label{sumrule1}
\end{equation}
is obeyed exactly in the large $N$ limit as in the Minkowski
case\cite{us1,mink,late}.

For the Hartree case $g \rightarrow 3g$, but the physical phenomena are the
same, with the only difference that the sum rule now becomes $g\Sigma(\infty) =
1/3$. We now show that this value is a self-consistent solution of the
equations of motion for the mode functions, and the {\em only} stationary
solution for asymptotically long times.

In the late time limit, the effective time dependent mass term,
$-1+\eta^2+g\Sigma$, in the equation for the mode functions,
(\ref{modcr}), vanishes (in this case with $\eta =0$).  Therefore,
these mode equations asymptotically become,
\begin{equation} 
\left[\frac{d^2}{d \tau^2}+3h \frac{d}{d \tau}+\frac{q^2}{a^2(\tau)}
\right]f_q(\tau)=0 \; . 
\end{equation}
The general solutions are given by,
\begin{equation}\label{asi}
f_q^{asy}(\tau)= \exp\left[-\frac{3}{2}h \tau \right] \left[c_+(q) \;
J_{3/2}\left( \frac{q}{h}e^{-h\tau}\right) -i c_-(q) \; N_{3/2}\left(
\frac{q}{h}e^{-h\tau}\right)\right] \; ,
\end{equation}
where $ J_{3/2}(z) $ and $ N_{3/2}(z) $ are the Bessel and Neumann functions,
respectively.  The coefficients, $ c_{\pm}(q) $  can be computed for
large $ q $ by matching $ f_q^{asy}(\tau) $ with the WKB approximation to the
exact mode functions $ f_q(\tau) $ that obey the initial conditions
(\ref{modkr}). The WKB approximation to $f_q(\tau) $ has been computed
in ref.\cite{frw}, and we find for large $q $,
\begin{equation}\label{abdek}
c_{\pm}(q) = \sqrt{{\pi\, q}\over {2\, h}}\; \left[ 1 - {i\over q}(h +
\Delta) + 
{\cal O}(q^{-2}) \right]\;e^{-iq/h} \pm \sqrt{{\pi\, h}\over {8\, q}}\; \left[ 1
+ {\cal O}\left({1\over q}\right) \right]\;e^{iq/h} \; ,
\end{equation}
where
\begin{equation}
\Delta \equiv \int_0^{\infty} d\tau \; e^{h\tau} \; M^2(\tau) \; .
\end{equation}

In the $ \tau \to \infty $ limit, we have for fixed $ q $,
\begin{equation}\label{uasi}
f_q^{asy}(\tau) \buildrel{ \tau \to \infty}\over=  i \sqrt{2 \over
{\pi}}\left({h \over q }\right)^{3/2}\; c_-(q) \; \; .
\end{equation}
which are independent of time asymptotically, and explains why the power
spectrum of quantum fluctuations freezes at times larger than the
spinodal. This behavior is confirmed numerically: fig.14 shows
$ \ln|f_q(\tau)|^2 $ vs. $\tau$ for $q=0,4,10$. Clearly at early
times the mode functions grow exponentially, and at times of the order of
$\tau_s$, when $g\Sigma(\tau) \approx 1$ the mode functions freeze-out and
become independent of time.  Notice that the largest $q$ modes have grown the
least, explaining why the
integral is dominated by $q \leq 10-20$.

For asymptotically large times, $g\Sigma(\tau)$ is given by,
\begin{equation}
g\Sigma(\infty) =g \; h^2 \; \int_0^{+\infty} {{dq}\over q } \;
\coth\left(\frac{ \omega_q}{2  T} \right) \; \left[ {{2h}\over {\pi }}
\mid c_-(q) \mid^2 - \, q \right] \;, \label{asinto}
\end{equation}
where only one term in the UV subtraction survived in the $\tau =\infty $
limit.  The factor $ \coth\left(\frac{ \omega_q}{2 T} \right) $ in 
eq.(\ref{asinto}) takes into account the nonzero initial temperature $ T $.

For consistency, this integral must converge and be equal to $1$ as
given by the sum rule.  For this to be the case and to avoid the potential
infrared divergence in (\ref{asinto}), the coefficients $ c_-(q) $ must vanish at
$ q = 0 $. The mode functions are finite in the $ q \to 0 $ limit provided,
\begin{equation}
c_-(q) \buildrel{ q \to 0 }\over= {\cal C} \; q^{3/2} \; ,
\end{equation}
where $ {\cal C} $ is a constant.

The numerical analysis and fig.12 clearly show that the mode functions
remain finite as $q \rightarrow 0$, and the coefficient ${\cal C}$ can be read
off from these figures.  This is a remarkable result. It is well known that for
{\em free} massless minimally coupled fields in de Sitter space-time with
Bunch-Davies boundary conditions, the fluctuation contribution $\langle
\psi^2(\vec x, t) \rangle$ grows linearly in time as a consequence of the
logarithmic divergence in the
integrals\cite{linde2}. However, in our case, although the
asymptotic mode functions are free, the coefficients that multiply the Bessel
functions of order $3/2$ have all the information of the interaction and
initial conditions and must lead to the consistency of the sum rule. Clearly
the sum rule and the initial conditions for the mode functions prevent the
coefficients $ c_{\pm}(q) $ from describing the Bunch-Davies vacuum. These
coefficients are completely determined by the initial conditions and the
dynamics.  This is the reason why the fluctuation freezes at long times unlike
in the free case in which they grow linearly\cite{linde2}.

It is easy to see from eqs.(\ref{hubble})-(\ref{ppluse}) and (\ref{uasi})
that the energy and pressure vanish for $\tau \to \infty$.

Analogously, the two point correlation function can be computed in the late
time regime using the asymptotic results obtained above.  Inserting
eq.(\ref{asi}) for the mode functions in eq.(\ref{rhocorr}) yields the
asymptotic behavior:
\begin{equation}\label{Slargo}
S(\rho, \tau) \buildrel{\tau \to \infty}\over = { 1 \over {4\pi^2 \rho}} \;
\int_0^{\infty} {q \, dq} \;\sin(q\rho)\; \coth\left(\frac{ \omega_q}{2 
T}\right) \; { {2 h^3} \over { \pi \, q^3}} \; \mid c_-(q) \mid^2 \; .
\end{equation}
The asymptotic behavior in time of the equal time correlation function is thus
solely a function of $ r $.  The large $ r $ behavior of $ S(\vec{r}, +\infty)
$ is determined by the singularities of $ \mid c_-(k) \mid^2 $ in the
complex $ k $ plane. We find an exponential decrease,
\begin{equation}
S(\rho, +\infty) \buildrel{ \rho \to \infty }\over \simeq C \; {{e^{- \rho/ \xi
  } }\over \rho}\; ,
\end{equation}
where $ \rho = i / \xi $ is the pole nearest to the real axis and $ C $ is some
constant. Thus we see that the freeze-out of the mode functions leads to the 
freeze-out of the correlation length $\xi$. The result of the numerical
analysis is shown in fig.16 which confirms this behavior and provides the
asymptotic value for $ \xi \approx 1/ h $. From these figures it is also clear
that the freeze-out time is given by the expansion time scale, $1/h$.  More
precisely, the numerical values for $ \xi $ can be accurately
reproduced by the following formula obtained by a numerical fit 
$$
h \xi \simeq 1.02 + 0.2 \ln h + 0.06 h + \ldots \; .
$$

This situation must be contrasted with that in Minkowski space-time
\cite{boylee} where the correlation length grows as $\xi(\tau)
\approx \sqrt{\tau}$ during the stage of spinodal growth. Eventually, this
correlation length saturates to a fairly large value that is typically several
times larger than the zero temperature correlation 
length\cite{boylee}.
We see that in the de Sitter case the domains are always horizon-sized.

\subsection{ Inflaton rolling down ($\eta(0) \neq 0$): 
classical or quantum behavior?}

Above we have analyzed the situation when $\eta(0) =0$ (or in dimensionful
variables $\phi(0)=0$). The typical analysis of inflaton dynamics in
the literature involves the {\em classical} evolution of $ \phi(t) $ 
with an initial condition in which $ \phi(0) $ is very close to zero (i.e.
the top of the potential hill) in the `slow-roll' regime, for which
$ \ddot{\phi} \ll 3H\dot{\phi}$. Thus, it is important
to quantify the initial conditions on $ \phi(t) $ for which the
dynamics will be 
determined by the classical evolution of $ \phi(t) $ and those for
which the quantum 
fluctuations dominate the dynamics. We can provide a criterion to
separate classical from quantum dynamics by analyzing the relevant time
scales, estimated by neglecting
non-linearities and backreaction effects. We consider the evolution 
of the zero mode in terms of dimensionless variables, and choose
$ \eta(0) \neq 0 $ and $\dot{\eta}(0) = 0 $.
($\dot{\eta}(0) \neq 0$ simply corresponds
to a shift in origin of time). We  assume  $ \eta(0)^2 << 1
$ which is the relevant case where  spinodal instabilities are important. 
We find
\begin{equation}
\eta(\tau) \approx \eta(0) \; e^{(\nu - \frac{3}{2})h_0\tau}\; .
\end{equation}

The non-linearities will become important and eventually terminate
inflation when $\eta(\tau) \approx 1$. This corresponds to a time scale
given by
\begin{equation}
\tau_c \approx \frac{\ln\left[1/ \eta(0)\right]}{(\nu - \frac{3}{2})\;
h_0}\; .\label{classtime}
\end{equation}

Comparing this time scale to the spinodal time scale given by
(\ref{spinotime}), for which quantum fluctuations grow to be of ${\cal O}(1)$,
we see that when,
\begin{equation}
\eta(0) << g^{1/4}, \label{smallzero}
\end{equation}
the quantum fluctuations will grow to be ${\cal O}(1)$ much {\em earlier} than
the zero mode for $T_i > T_c$ (for $T_i=0$ the bound becomes $\eta(0) <<
g^{1/2} $). In this case the dynamics will be driven completely
by the quantum fluctuations, as the zero mode will be rolling down the
potential hill very slowly and will not grow enough to compete with the quantum
fluctuations before the fluctuations grow to overcome the tree level terms in
the equations of motion.  In this case, as argued previously, the large $N$ and
Hartree approximations will be completely equivalent during the time scales of
interest.

On the other hand, if
\begin{equation}
\eta(0) >> g^{1/4}, \label{bigzero}
\end{equation}
then the zero mode will roll and become ${\cal O}(1)$ {\em before} the
fluctuations have enough time to grow to ${\cal O}(1)$ ($\eta(0) >>
g^{1/2} $ for $T_i=0$ ). In this case, the dynamics will be
dominated by the rolling of the zero mode and is mostly {\em classical}. The
quantum fluctuations remain perturbatively small throughout the inflationary
stage which will end when the velocity of the zero mode modifies the equation
of state to terminate de Sitter expansion.

For $\eta(0) \approx g^{1/4}$ (or $\eta(0) \approx
g^{1/2}$ for $T_i=0$), both the rolling of the zero mode {\em
and} the quantum fluctuations will give contributions of the same order to the
dynamics. In this case, the quantum fluctuations will be large for the
long-wavelength modes and the classical approximation to the inflationary
dynamics will not be accurate.

Since the scenario in which $\eta(0) >> g^{1/4}$, in which the
dynamics is basically driven by the classical evolution of the zero mode has
received a great deal of attention in the literature, we will {\em not} focus
on this case, but instead analyze numerically the cases in which $\eta(0)
\neq 0$ but such that $\eta(0) \leq g^{1/4}$.

\subsection{\bf Numerical Analysis:}

We have evolved the set of equations of motion given by
(\ref{modcr}) numerically with
initial conditions (\ref{modkr}) for the large $N$ case, and (\ref{hartphieq})
and (\ref{hartukeq}), with the corresponding
initial conditions (\ref{initcond}) on the mode functions for the Hartree
case.  The numerical code is the same as in the previous section with the same
relative errors.

{\bf Large N case:}

Figs. 17-18 show $\eta(\tau)$ and $g\Sigma(\tau)$ vs. $\tau$ for the values
$ g=10^{-12}; \quad T_i/T_c=2; \quad \eta(0)= 10^{-5} ; \quad
\dot{\eta}(0)=0$.  Clearly the dynamics is dominated by the fluctuations; the
zero mode grows but is always negligible small compared to $g\Sigma(\tau)$. The
time scale at which $g\Sigma(\tau)$ grows to be of order one is about the same
as in the case, $\eta(0)=0$, and all the behavior for the mode functions,
correlation length, energy density, pressure, etc. is similar to the case
analyzed in the previous section.

Asymptotically, we find that the sum rule (\ref{sumrule}) 
is satisfied to our numerical accuracy. This is the same as the situation in
Minkowski space-time\cite{us1,mink}, and when $\eta \neq 0$, this sum rule
is nothing but the Ward identity associated with Goldstone's theorem. The
fluctuations are Goldstone bosons, minimally coupled, and the symmetry is
spontaneously broken with a very small expectation value for the order
parameter as can be read off from fig.17.  For $\tau > \tau_s$, the
dynamics freezes completely and the zero mode and the fluctuations achieve
their asymptotic values much in the same way as in the case $\eta =0$ studied
in the previous section.  Again, the correlation length becomes independent of
time with $\xi(\infty) \approx 1/h$ in a time scale given by $1/h$.

Because there is a damping term in the zero mode equation, it is reasonable to
assume that asymptotically there will be a solution with a constant value of
$\eta$.  Then the Ward-identity, $\eta(\infty)\left[
-1+\eta^2(\infty)+g\Sigma(\infty)\right]=0$, must be fulfilled. In the large
$N$ case, the {\em only} stationary solutions are i) $\eta =0 ; \;
g\Sigma(\infty)=1$, or, ii) $\eta(\infty) \neq 0 ; \;
-1+\eta^2(\infty)+g\Sigma(\infty)=0$. To have a consistent solution of the mode
functions, it must be that the effective mass term $\left[
-1+\eta^2(\infty)+g\Sigma(\infty)\right]$ vanishes asymptotically, leading to
the mode equations for massless, minimally coupled modes which are
asymptotically independent of time as shown in the previous section (see
eq.(\ref{uasi})). Furthermore, from fig.18 it is clear that $g\Sigma(\tau)$
remains constant at long times, again unlike the case of free massless fields
with Bunch-Davies boundary conditions in which case the fluctuation grows
linearly in time\cite{linde2}.

{\bf Hartree case:} 

Figure 17 also shows the evolution of the zero mode in the large
$N$ and Hartree case. Although there is a quantitative difference in the
amplitude of the zero mode, in both cases it is extremely small and gives a
negligible contribution to the dynamics.  In the Hartree case, however, there
is no equivalent of the large $N$ sum rule; the {\em only} stationary solution
for $\eta \neq 0$ is, $\eta^2(\infty)=1 ; g\Sigma(\infty)=0$.  Such a solution
leads to mode equations with a {\em positive} mass term and mode functions that
vanish exponentially fast for $ \tau \to \infty $ for all momenta.  However,
whether the asymptotic behavior of the Hartree solution is achieved within the
interesting time scales is a matter of initial conditions. For example in
fig.17 the initial condition is such that the time scale for growth of the
quantum fluctuations is much shorter than the time scale for which the
amplitude of the zero mode grows large and the non-linearities become
important. 
In the large $N$ case the sum rule is satisfied with a large value of the
quantum fluctuations. In the Hartree case the equivalent sum rule
$-1+3\eta^2_H+3g\Sigma_H=0$ is satisfied for a very small $\eta_H$ and a
$g\Sigma_H \approx 1/3$.  The modes become effectively massless and they stop
growing.

\begin{figure}
\epsfig{file=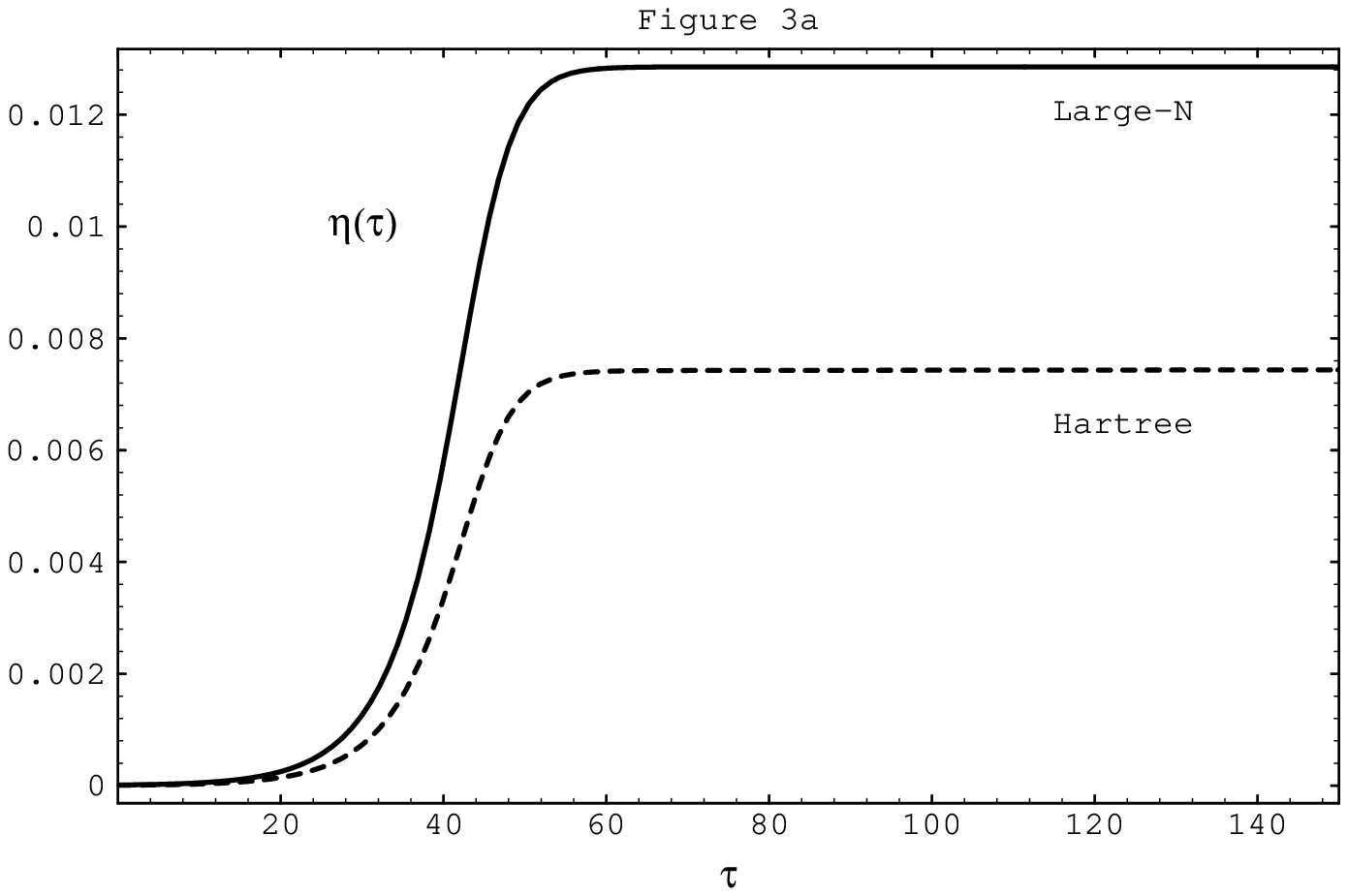,width=10.5cm,height=5.5cm}
\caption{ $\eta(\tau)$ vs. $\tau$ for $\lambda=10^{-12};
\quad r=2; \quad 
h=2; \quad \eta(0)=10^{-5};\quad \dot{\eta}(0)=0$ for large $N$ (solid curve)
and Hartree (dashed curve). 
\label{fig3a}}
\epsfig{file=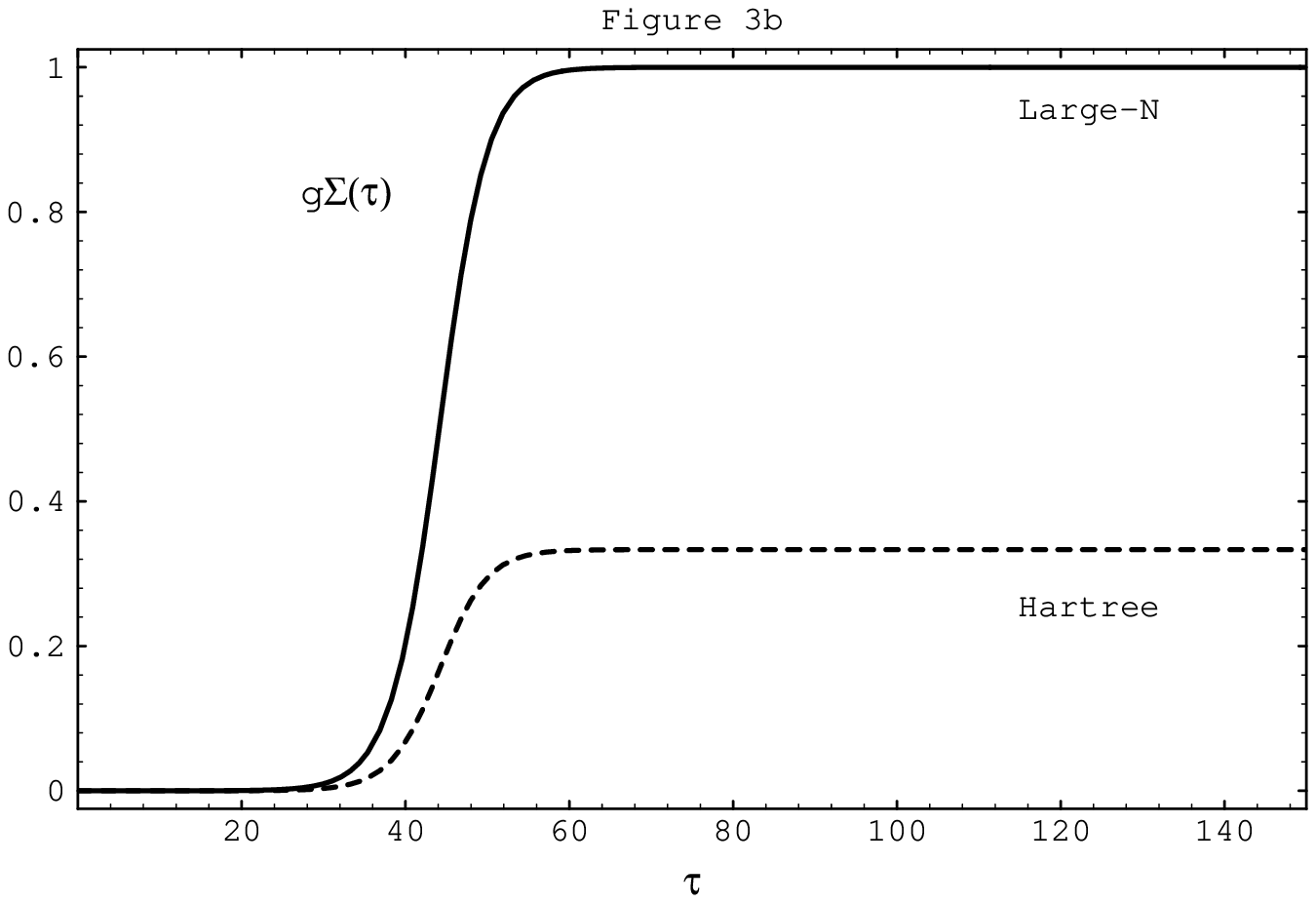,width=10.5cm,height=5.5cm}
\caption{$g\Sigma(\tau)$ vs. $\tau$ for the same values of the
parameters
as in fig.17, for large $N$ (solid curve) and Hartree (dashed curve).
\label{fig3b}}
\end{figure}

The equation for the zero mode (see eq.(\ref{hartphieq})) still has an
uncancelled piece of the non-linearity, $-2\eta^2$; however the derivatives and
the amplitude of $\eta$ are all extremely small and though the zero mode
still evolves in time, it does so extremely slowly.  In fact the Hartree curve
in fig.17 has an extremely small positive slope asymptotically, and while
$\eta_H$ grows very slowly, $g\Sigma_H$ diminishes at the same rate. In the
case shown in fig.17, we find numerically that $\dot{\eta}_H / \eta_H
\approx 10^{-7}$ at $\tau= 150$.  Before this time most of the interesting
dynamics that can be captured with a fixed de Sitter background had already
taken place, and the backreaction of the fluctuations on the metric becomes
substantial requiring an analysis that treats the scale factor dynamically. 

The conclusion of our analysis is that in the region of initial conditions for
which the quantum fluctuations dominate the dynamics, that is for $\eta(0) <<
g^{1/4}$, both large $N$ and Hartree give the same answer on
 the relevant time scales. 
The figures for ${\cal H}(\tau) / {\cal H}(0)$ are
numerically indistinguishable from the case of figs. 1.

We see that in the large $N$ case the zero mode rolls to a final amplitude
which is $ {\cal O}(1) $ and of the same order as $g\Sigma(\infty)$ and the sum
rule is satisfied.  However, the Hartree case clearly shows the asymptotics
analyzed above with $ \eta_H(\infty)=1 ; g\Sigma_H(\infty)=0 $.

This particular borderline case is certainly not generic and would imply some
fine tuning of initial conditions. Finally the case in which $ \eta(0) >>
g^{1/4} $ (or $ g^{1/2} $ for $ T_i=0 $) is basically classical in that
the dynamics is completely given by the classical rolling of the zero mode and the fluctuations are always perturbatively small.

\subsection{Discussion and Conclusions for the de Sitter background}

We have identified analytically and numerically two distinct regimes for the
dynamics determined by the initial condition on the expectation value of the
zero mode of the inflaton \cite{De Sitter}.

\begin{enumerate}
\item{When $\eta(0) << g^{1/4}$ (or $ g^{1/2}$ for $T_i=0$), the
dynamics is driven by quantum (and thermal) fluctuations. Spinodal
instabilities grow and eventually compete with tree level terms at a time
scale, $\tau_s \geq -3h\ln[g]/2$.  The growth of spinodal fluctuations
translates into the growth of spatially correlated domains which attain a
maximum correlation length (domain size) of the order of the horizon.  For very
weak coupling and $h \geq 1$ this time scale can easily accommodate enough
e-folds for inflation to solve the flatness and horizon problems. The quantum
fluctuations modify the equation of state dramatically providing a
means for a graceful exit to the inflationary stage without slow-roll.

This non-perturbative description of the non-equilibrium effects in this
regime in which quantum (and thermal) fluctuations are most important
is borne out by both the large $ N $ and Hartree approximations. Thus
our analysis provides a reliable understanding of the relevant 
non-perturbative, non-equilibrium effects of the fluctuations that have
not been revealed before in this setting\cite{us1} - \cite{din}.

These initial conditions are rather natural if the de Sitter era arises during
a phase transition from a radiation dominated high temperature phase in local
thermodynamic equilibrium, in which the order parameter and its time derivative
vanish.}

\item{When $\eta(0) >> g^{1/4}$ (or $ g^{1/2} $ for $T_i=0$), the
dynamics is driven solely by the classical evolution of the inflaton zero
mode. The quantum and thermal fluctuations are always perturbatively small
(after renormalization), and their contribution to the dynamics is negligible
for weak couplings. The de Sitter era will end when the kinetic contribution to
the energy becomes of the same order as the `vacuum' term. This is the realm of
the slow-roll analysis whose characteristics and consequences have been
analyzed in the literature at length. These initial conditions, however,
necessarily imply some initial state either with a biasing field that favors a
non-zero initial expectation value, or that in the radiation dominated stage,
prior to the phase transition, the state was strongly out of equilibrium with
an expectation value of the zero mode different from zero. Although such a
state cannot be ruled out and would naturally arise in chaotic scenarios, the
description of the phase transition in this case requires further input on the
nature of the state prior to the phase transition.}
\end{enumerate}

\section{Self-consistent  Evolution of Matter Fields with a dynamical
cosmological background}

We present in this section the full self-consistent matter-geometry 
dynamics\cite{din}. That is,
the scale factor $ a(t) $ is here a dynamical variable determined by the
Einstein-Friedman eq.(\ref{eif})-(\ref{hubble}) coupled with the matter 
evolution eqs.(\ref{modcr})-(\ref{modkr}).

\bigskip

In order to provide the full
solution we now must provide the values of $\eta(0)$, $\dot{\eta}(0)$,
and $h_0$. Assuming that the 
inflationary epoch is associated with a phase transition at the GUT scale,
this requires that $ N m^4_R/g \approx 
(10^{15}\mbox{ Gev })^4 $ and assuming the bound on the scalar
self-coupling $ g \approx 10^{-12}-10^{-14}$ (this will be seen
later 
to be a compatible requirement), we find that $h_0 \approx N^{1/4}$ which
we will take to be reasonably given by $h_0 \approx 1-10$ (for example
in popular GUT's $ N \approx 20 $ depending on particular representations). 

We will begin by studying the case of most interest from the point of view
of describing the phase transition: $\eta(0)=0$ and $\dot{\eta}(0)=0$,
which are the initial conditions that led to puzzling questions. With
these initial conditions, the evolution equation for the zero mode
eq.(\ref{modcr}) determines that $\eta(\tau) = 0$ by symmetry.

\subsection{Early time dynamics:}
Before engaging in the numerical study, it proves illuminating to
obtain an estimate of the relevant time scales and an intuitive idea of
the main features of the dynamics. Because the coupling is so weak 
($ g \sim 10^{-12} \ll 1 $) and after
renormalization the contribution from the quantum fluctuations to 
the equations of motion is finite, we can neglect  all the terms
proportional to $ g $ in eqs.(\ref{hubble}) and (\ref{modcr}). 

For the case where we choose $\eta(\tau) = 0$ and
the evolution equations for the 
mode functions are those for an inverted oscillator in De Sitter space-time,
which have been studied in sec. VIII \cite{guthpi}. One
obtains the approximate solutions (\ref{bessel})-(\ref{coefb}).

After the physical wavevectors cross the horizon, i.e. when $qe^{-h_0
\tau}/h_0 \ll 1$ we find that the mode functions factorize: 
\begin{equation}
f_q(\tau) \approx  {{B_q} \over {\Gamma(1-\nu)}} 
\; \left( {{2h_0\, }\over q}\right)^{\nu}e^{(\nu-3/2)h_0 \tau}. \label{factor}
\end{equation}
 This  result reveals a very
important feature: because of the negative mass squared
term in the matter Lagrangian leading to symmetry breaking (and $\nu > 3/2$), we see
that all of the mode functions {\em grow exponentially} after horizon
crossing (for positive mass squared $ \nu < 3/2 $, and  
they would {\em decrease
exponentially} after horizon crossing). This exponential growth is a 
consequence of the spinodal instabilities which 
is a hallmark of the process of phase separation that
occurs to complete the phase transition. 
 We note, in addition that the time 
dependence is exactly given by that of the $ q=0 $ mode, i.e. the zero 
mode, which is a consequence of the redshifting of the wavevectors and 
the fact that after horizon crossing the contribution of the term
$q^2/a^2(\tau)$ in the equations of motion become negligible.
 We clearly  see that the quantum fluctuations grow exponentially and
they will begin to be of the order of the tree level terms in the
equations of motion when $g\Sigma(\tau) \approx 1$. At large
times 
$$
\Sigma(\tau) \approx {\cal F}^2(h_0) h_0^2
e^{(2\nu-3)h_0 \tau} \; ,  
$$
with
${\cal F}(h_0)$ a finite constant that depends on the initial conditions and
is found numerically to be of ${\cal O}(1)$ [see fig.\ref{fofh}].  

In terms of the initial dimensionful variables, the
 condition  $ g\Sigma(\tau) \approx 1$ translates
to $ <\psi^2(\vec x,t)>_R \approx 2m^2_R/g $, i.e. the quantum
fluctuations sample the minima of the (renormalized) tree level potential.
We find that the
time at which the contribution of the 
quantum fluctuations becomes of the same order as the tree level terms is
estimated to be\cite{De Sitter}
\begin{equation}
\tau_s \approx \frac{1}{(2\nu-3)h_0}
\ln\left[\frac{1}{g\, h_0^2 {\cal F}^2(h_0)}\right] 
= \frac32 h_0 \ln\left[\frac{(2\pi)^2}{g\, h_0^2 {\cal F}^2(h_0)}\right]
+ {\cal O}(1/h_0).
\label{spinodaltime}
\end{equation}
At this time, the contribution of the quantum fluctuations makes the
back reaction very important and, as will be seen numerically, this
translates into the fact that $\tau_s$ also determines the end of the
De Sitter era and the end of inflation. The total number of e-folds during
the stage of exponential expansion of the scale factor (constant
$h_0$) is  given by   
\begin{equation}
N_e \approx \frac{1}{2\nu-3}\;
\ln\left[\frac{1}{g\; h_0^2\; {\cal F}^2(h_0)}\right] 
= \frac32\; h_0^2\; \ln\left[\frac{(2\pi)^2}{q\;  h_0^2 \;{\cal
F}^2(h_0)}\right]  
+ {\cal O}(1)\label{efolds}
\end{equation}
For large $h_0$ we see that the number of e-folds scales as $h^2_0$ as well
as with the logarithm of the inverse coupling. 
These results (\ref{factor}-\ref{efolds}) will be
confirmed numerically below and will be of paramount importance for the
interpretation of the main consequences of the dynamical evolution. 

As discussed in sec. VIII.D, the early time dynamics is dominated by classical
or quantum effects depending on the ratio between the time scales
 $ \tau_c $ and $ \tau_s $.

If $ \tau_c $ is much smaller than the spinodal time $ \tau_s $ given
by eq.(\ref{spinodaltime}) then the {\em classical} evolution of the
zero mode will dominate the dynamics and the quantum fluctuations will
not 
become very large, although they will still undergo spinodal growth. On 
 the other hand, if $\tau_c \gg \tau_s$ the quantum fluctuations will
grow to be very large well before the zero mode reaches the non-linear
regime. In this case the dynamics will be determined completely by
the quantum fluctuations. Then the criterion for the classical or quantum
dynamics is given by
\begin{eqnarray} 
\eta(0) & \gg & \sqrt{g}\;h_0 \Longrightarrow \mbox{ classical dynamics }
\nonumber \\
\eta(0) & \ll & \sqrt{g}\;h_0 \Longrightarrow \mbox{ quantum dynamics }
\label{classquandyn} 
\end{eqnarray}
or in terms of dimensionful variables $\phi(0) \gg H_0$ leads to 
{\em classical dynamics} and $\phi(0) \ll H_0$ leads to 
{\em quantum dynamics}. 

However, even when the classical evolution of the
zero mode dominates the dynamics, the quantum fluctuations grow
exponentially after horizon crossing unless the value of $\phi(t)$ is
very close to the minimum of the tree level potential. In the large $
N $ approximation the spinodal line, that is the values of $\phi(t)$ for  
which there are spinodal instabilities, reaches all the way to the minimum
of the tree level potential as can be seen from the equations of motion for
the mode functions. 
Therefore even in the
classical case one must understand how to deal with quantum fluctuations
that grow after horizon crossing.  

\subsection{Numerics}
The time evolution is carried out by means of a fourth order Runge-Kutta
routine with adaptive step-sizing while the momentum 
integrals are carried out using an 11-point Newton-Cotes integrator.  
The relative errors in both
the differential equation and the integration are of order $10^{-8}$.
We find that the energy is covariantly conserved throughout the evolution
to better than a part in a thousand. Figs. \ref{gsigma}--\ref{modu} 
show $g\Sigma(\tau)$ vs. $\tau$,
$h(\tau)$ vs. $\tau$ and $ \ln|f_q(\tau)|^2 $ vs. $\tau$ for several values
of $q$ with larger $q's$ corresponding to successively lower curves. 
Figs. \ref{povere},\ref{hinverse} show $p(\tau)/\varepsilon(\tau)$ 
and the horizon size $h^{-1}(\tau)$ for 
$g = 10^{-14} \; ; \; \eta(0)=0 \; ; \; \dot{\eta}(0)=0$
and we have chosen the representative value $h_0=2.0$.

\begin{figure}
\epsfig{file=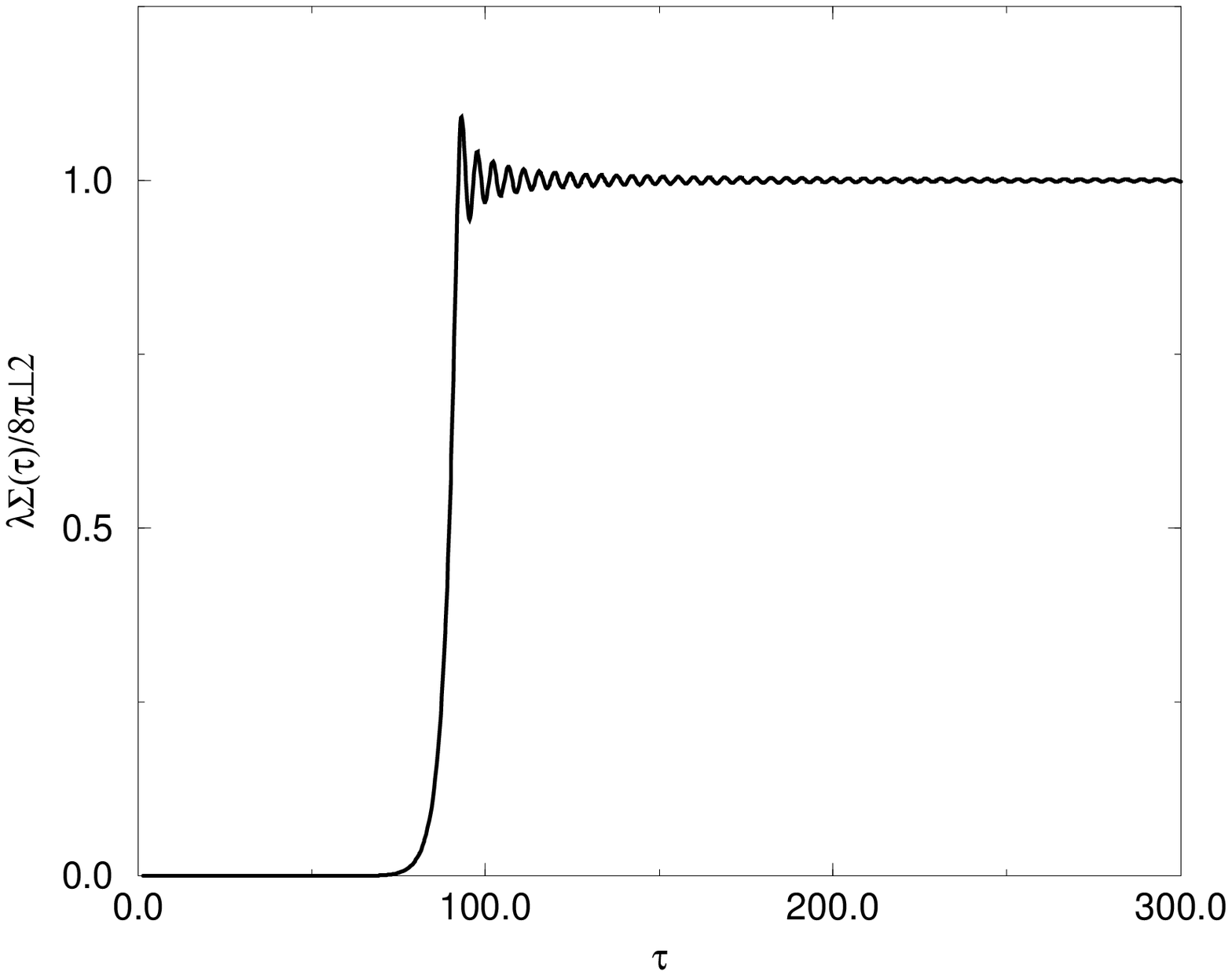,width=10.5cm,height=5.5cm}
\caption{ $ g\Sigma(\tau) $ vs. $\tau$, for $\eta(0)=0, \dot{\eta}(0)=0,
g = 10^{-14}, h_0 = 2.0$. }
\label{gsigma}
\epsfig{file=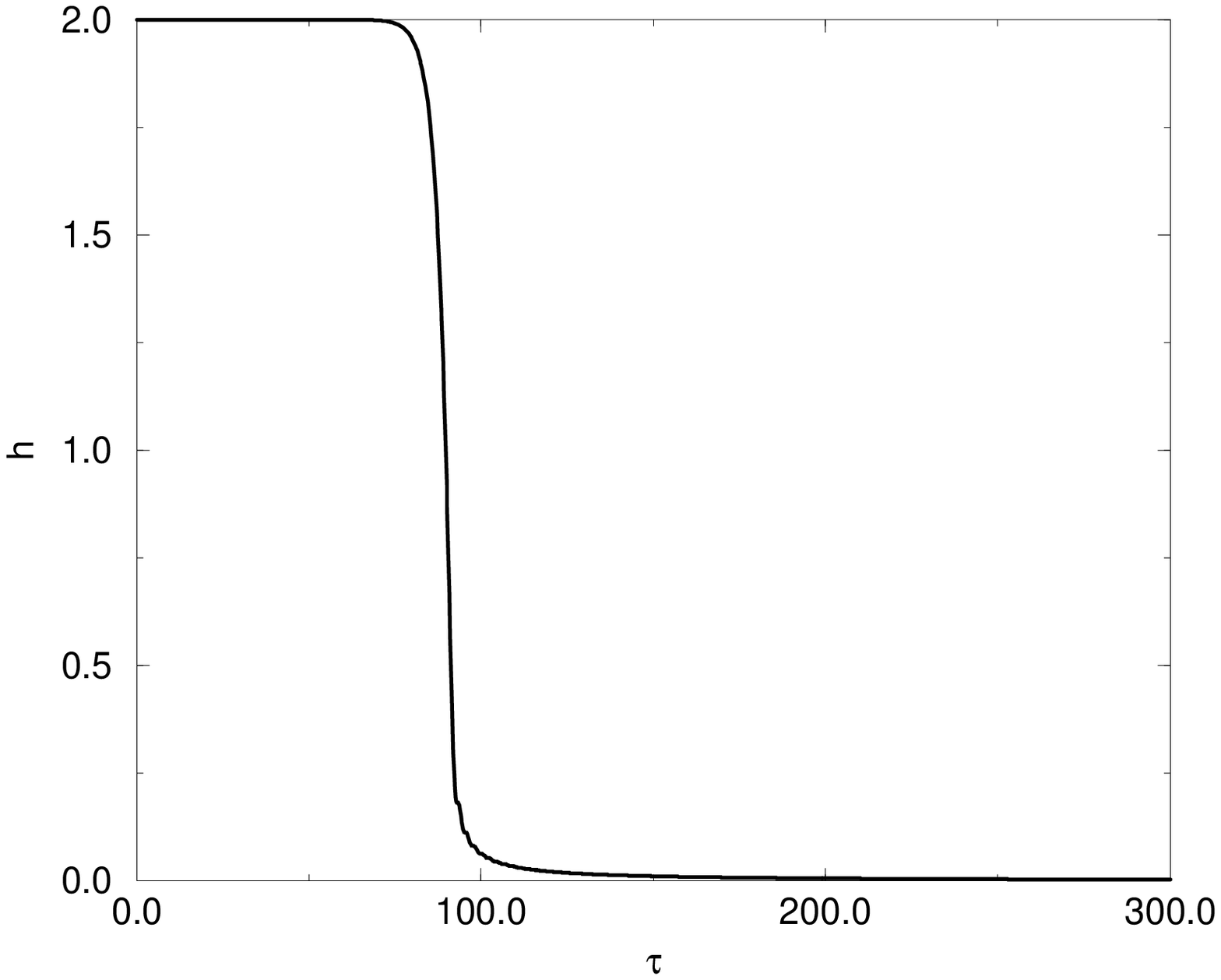,width=10.5cm,height=5.5cm}
\caption{$ H(\tau) $ vs. $ \tau $, for $ \eta(0)=0, \dot{\eta}(0)=0, 
g = 10^{-14}, h_0 = 2.0 $. }
\label{hubblefig}
\epsfig{file=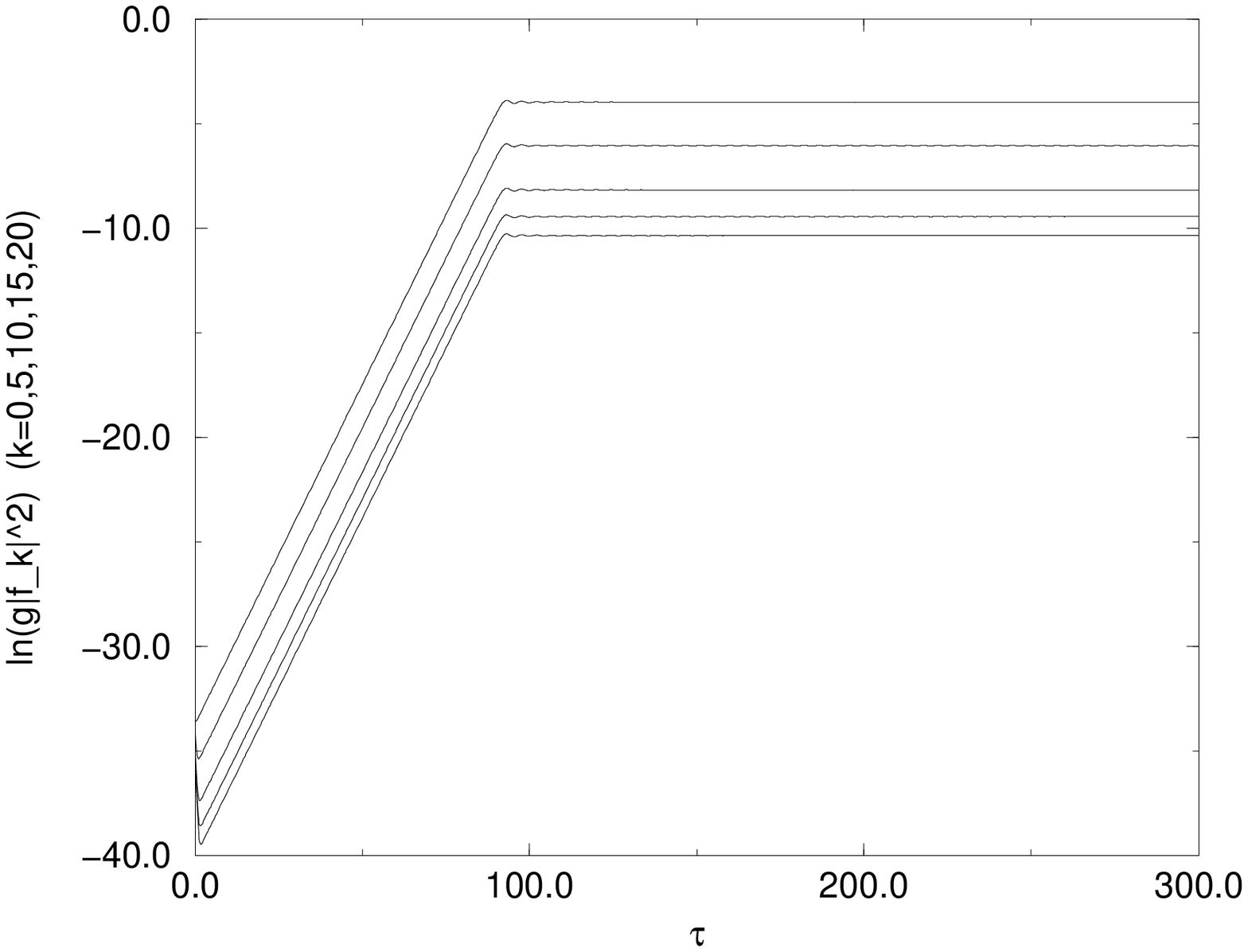,width=10.5cm,height=5.5cm}
\caption{$ \ln|f_q(\tau)|^2 $ vs. $\tau$, for $\eta(0)=0,
\dot{\eta}(0)=0,  g = 10^{-14}, h_0=2.0$ for
$q=0.0,5,10,15,20$ with smaller 
$q$ corresponding to larger values of $\ln|f_q(\tau)|^2 $.}
\label{modu}
\end{figure}

Figs. \ref{gsigma} and \ref{hubblefig} show clearly that 
when the contribution of the quantum
fluctuations $ g\Sigma(\tau) $ becomes of order 1 inflation ends,
and the time scale for $ g\Sigma(\tau) $ to reach ${\cal O}(1)$ is very well
described by  the estimate (\ref{spinodaltime}). From fig.19 we see
that this happens for $\tau =\tau_s\approx 90$, leading to a number of
e-folds  
$N_e \approx 180$ which is correctly estimated by  (\ref{spinodaltime},
\ref{efolds}). 

Fig. \ref{modu} shows clearly the factorization of the modes after they
cross the horizon as described by eq.(\ref{factor}).
 The slopes of all the curves after they become
straight lines in fig.\ref{modu} is given exactly by $(2\nu-3)$,
whereas the 
intercept depends on the initial condition on the mode function and
the larger the value of $ q $ the smaller the intercept because the
amplitude of the mode function is smaller initially. Although the
intercept depends on the initial conditions on the long-wavelength
modes, the slope is independent of the value of $q$ and is the same as
what would be obtained in the linear approximation for the {\em
square} of the zero mode at times 
long enough that the decaying solution can be neglected but short enough
that the effect of the non-linearities is very small.
 Notice from the figure that when inflation ends and
the non-linearities become important all of the modes effectively saturate.
This is also what one would expect from the solution of the zero mode:
exponential growth in early-intermediate times (neglecting the
decaying solution), with a growth exponent
given by $(\nu - 3/2)$ and an asymptotic behavior of small oscillations
around the equilibrium position, which for the zero mode is $\eta =1$, but
for the $q \neq 0$ modes depends on the initial conditions. 
All of the mode functions have this behavior once they cross the horizon.
We have also studied the phases of the mode functions and we found that 
they freeze after horizon crossing in the sense that they become independent
of time. This is natural since both the
real and imaginary parts of $ f_q(\tau) $ obey the same equation but
with different 
boundary conditions. After the physical wavelength crosses the horizon, the
dynamics is insensitive to the value of $q$ for real and imaginary parts and
the phases become independent of time. Again, this is a consequence of the
factorization of the modes.  

The growth of the quantum fluctuations is sufficient to end inflation
at a time given by $\tau_s$ in eq.(\ref{spinodaltime}). Furthermore fig.
\ref{povere} shows that during the inflationary epoch
$p(\tau)/\varepsilon(\tau)  
\approx -1$ and the end of inflation is rather sharp at $\tau_s$ with
$p(\tau)/\varepsilon(\tau)$ oscillating between $\pm 1$ with zero average
over the cycles, resulting in matter domination. Fig. \ref{hinverse}
shows this  
feature very clearly; $h(\tau)$ is constant during the de Sitter epoch and
becomes matter dominated after the end of inflation with $h^{-1}(\tau) 
\approx \frac32 (\tau -\tau_s)$. There are small oscillations around
this value because both $p(\tau)$ and $\varepsilon(\tau)$
oscillate. These oscillations 
are a result of small oscillations of the mode functions after they 
saturate, and are also a
feature of the solution for a zero mode. 

All of these features hold for a variety of initial conditions.  As an
example, we show in ref.\cite{din} the
case of an initial Hubble parameter of $h_0=10$.

\begin{figure}
\epsfig{file=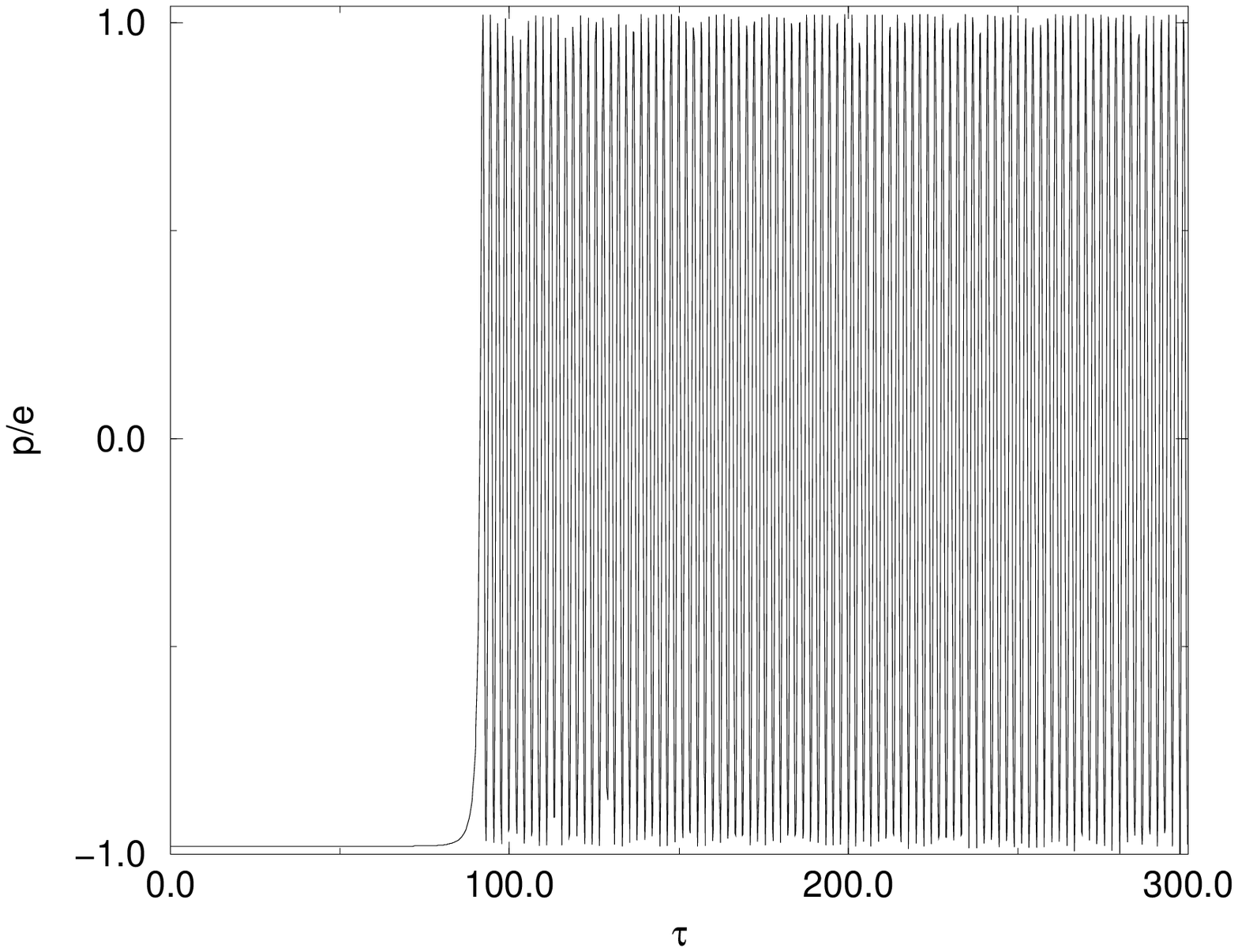,width=10.5cm,height=5.5cm}
\caption{$p/\varepsilon$ vs. $\tau$, for $\eta(0)=0, \dot{\eta}(0)=0,
g = 10^{-14}, h_0=2.0$.}
\label{povere}
\epsfig{file=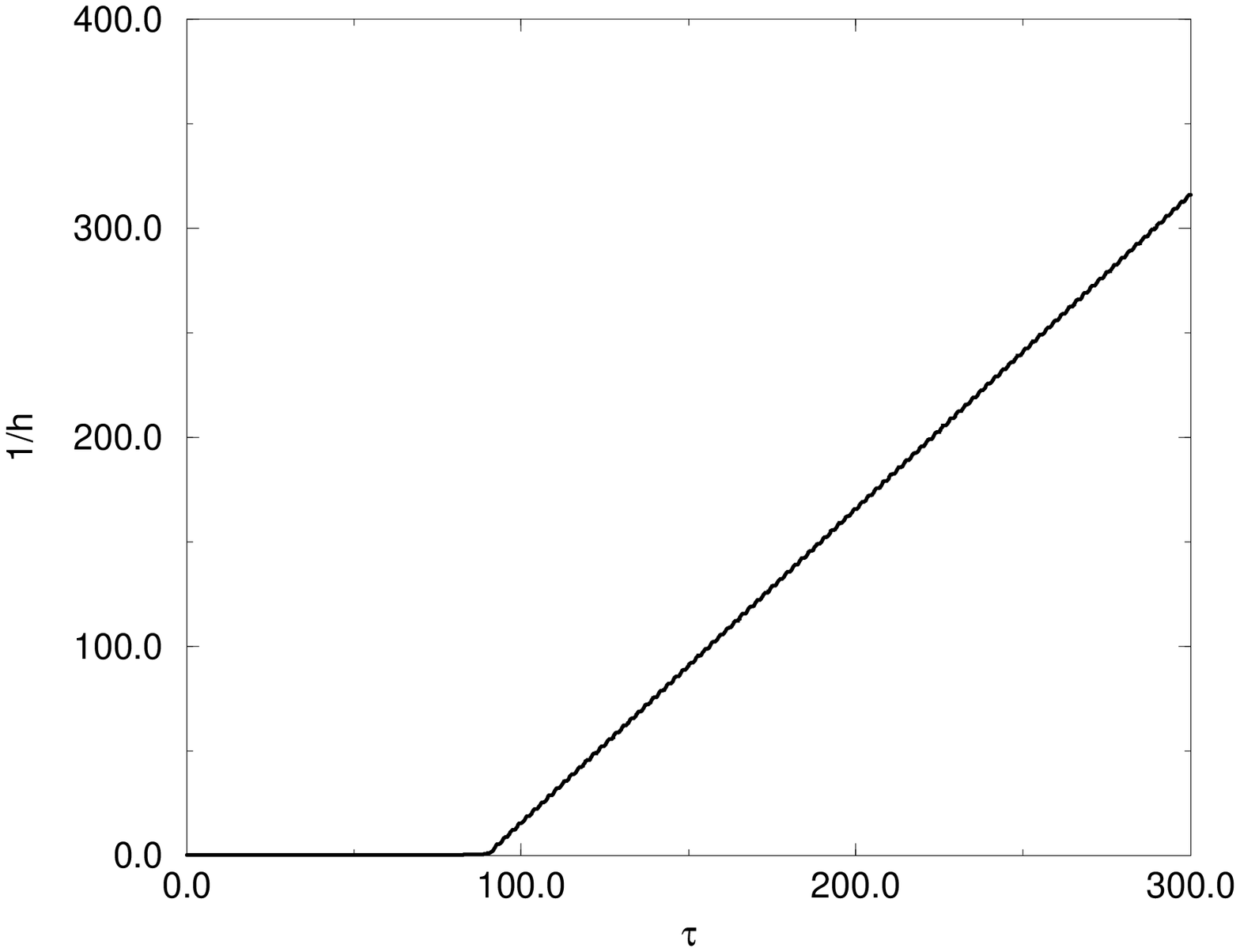,width=10.5cm,height=5.5cm}
\caption{$1/h(\tau)$ vs. $\tau$, for $\eta(0)=0, \dot{\eta}(0)=0,
g = 10^{-14}, h_0=2.0$. }
\label{hinverse}
\end{figure}

\subsection{Zero Mode Assembly:}
This remarkable feature of factorization of the mode functions after
horizon crossing can be elegantly summarized as
\begin{equation}
f_k(t)|_{k_{ph}(t) \ll H} = g(q,h)f_0(\tau),\label{factor2}
\end{equation}
with $k_{ph}(t) = k\,e^{-Ht}$ being the physical momentum, 
$ g(q,h)$ a complex constant, and $f_0(\tau)$ a {\em real} function
of time that satisfies the mode equation with $q=0$ and real initial
conditions which will be inferred later.
 Since the factor $g(q,h)$ depends solely on the initial
conditions on the mode functions, it turns out that for two mode
functions corresponding to momenta $k_1,k_2$ that have crossed the
horizon at times $t_1 > t_2$, the ratio of the two mode functions  at
time $t$, ($t_s>t >t_1 >t_2$) is 
$$
{{f_{k_1}(t)}\over {f_{k_2}(t)}} \propto
e^{(\nu-\frac{3}{2})h (\tau_1 - \tau_2)} > 1 \; . 
$$
Then if we consider the 
contribution of these modes to the  {\em renormalized} quantum
fluctuations a long time after the beginning of inflation (so as to
neglect the decaying solutions), we find that 
$$g\Sigma(\tau) \approx {\cal C}e^{(2\nu-3)h \tau} + \mbox{ small} \; , 
$$
where 
`small' stands for
the contribution of mode functions associated with momenta that have not
yet crossed the horizon at time $\tau$, which give a perturbatively
small (of order $ g $) contribution.  
 We find that several e-folds
after the beginning of inflation but well before inflation ends, this
factorization of superhorizon modes implies the following:

\begin{eqnarray}
g\int q^2 dq \; |f^2_q(\tau)| & \approx & 
 |C_0|^2 f^2_0(\tau), \label{int1} \\ 
g\int q^2 dq \; |\dot{f}^2_q(\tau)| & \approx & |C_0|^2 \dot{f}^2_0(\tau),
\label{int2} \\
g\int \frac{q^4}{a^2(\tau)} dq \; |f^2_q(\tau)| & \approx & 
 \frac{|C_1|^2 }{a^2(\tau)}f^2_0(\tau), \label{int3}
\end{eqnarray}
where we have neglected the weak time dependence arising from the perturbatively small
contributions of the short-wavelength modes that have not yet crossed the
horizon, and the integrals above are to be understood as the fully
renormalized (subtracted), finite integrals. For $\eta = 0$, we note that (\ref{int1}) and the fact that $f_0(\tau)$ obeys the equation of motion for the mode with $q=0$ leads at once to the conclusion that 
in this regime $\left[g\Sigma(\tau)\right]^{\frac{1}{2}} = |C_0|f_0(\tau)$ obeys the zero mode equation of motion
\begin{equation}
\left[\frac{d^2}{d \tau^2}+ 3h \frac{d}{d\tau}-1+
(|C_0|f_0(\tau))^2\right]|C_0|f_0(\tau) = 0 \; . 
\label{zeromodeeff}
\end{equation}

It is clear that 
several e-folds after the beginning of inflation, we can define an
  effective zero mode   as 
\begin{equation}
\eta^2_{eff}(\tau) \equiv g\Sigma(\tau), \mbox{ or in dimensionful
variables, } \phi_{eff}(t) \equiv \left[\langle \psi^2(\vec x, t)
\rangle_R \right]^{\frac{1}{2}} 
\label{effectivezeromode}
\end{equation}
Although this identification seems natural, we emphasize that it
is by no means a trivial or ad-hoc statement. There are several
important features that allow an {\em unambiguous} identification:
i) $\left[\langle \psi^2(\vec x, t) \rangle_R \right]$ is a fully 
renormalized operator product and hence finite, ii) because of  the
factorization 
 of the superhorizon modes that enter in the evaluation of 
$\left[\langle \psi^2(\vec x, t) \rangle_R \right]$,  
$\phi_{eff}(t)$ (\ref{effectivezeromode}) 
{\em obeys the equation of motion for the zero mode}, iii) this identification 
is valid several e-folds after the beginning of inflation,
after the transient decaying solutions have died away and the integral
in $\langle \psi^2(\vec x,t) \rangle$
is dominated by the modes with wavevector $k$ that have crossed the horizon at 
$t(k) \ll t$.
Numerically we see that this identification holds throughout the
dynamics except for a very few e-folds at the beginning of inflation. This
factorization determines at once the initial conditions of the effective
zero mode that can be extracted numerically: after the first few e-folds and
long before the end of inflation we find
\begin{equation}
\phi_{eff}(t) \equiv \phi_{eff}(0)\; e^{(\nu-
\frac{3}{2})Ht} \; \; , 
\label{effzeromodein} 
\end{equation}
where we parameterized 
$$
\phi_{eff}(0) \equiv \frac{H}{2\pi} \; {\cal F}(H/m)
$$
to make contact with the literature. 
As is shown in fig.~(\ref{fofh}), we find numerically that $ {\cal
F}(H/m)  \approx 
{\cal O}(1) $ for a large range of $ 0.1 \leq H/m \leq 50 $ and that
this quantity 
depends on the initial conditions of the long wavelength modes.  

\begin{figure}
\epsfig{file=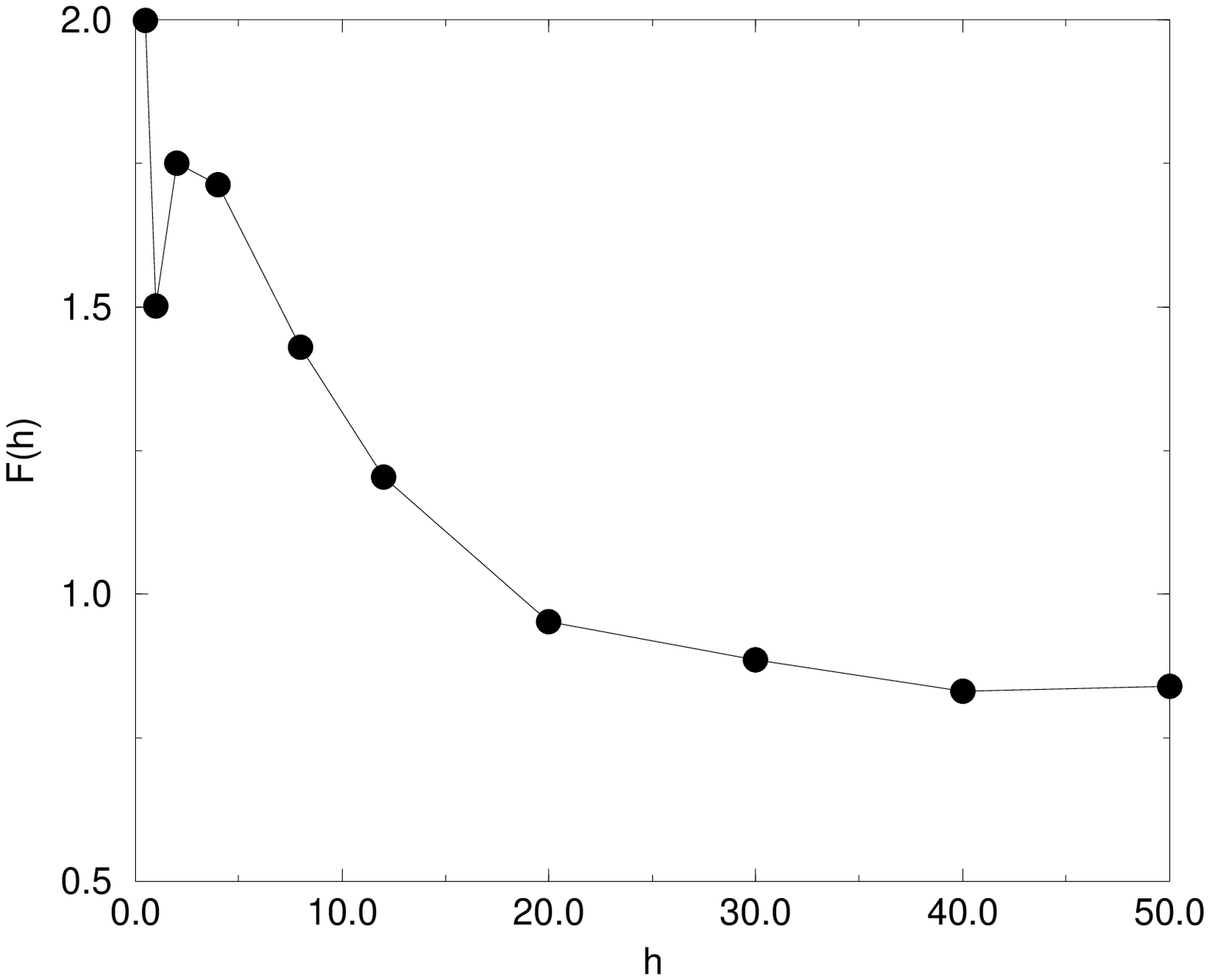,width=10.5cm,height=5.5cm}
\caption{${\cal F}(H/m)$ vs. $H$, where ${\cal F}(H/m)$
is defined by the relation
$\phi_{eff}(0) = (H/2\pi) {\cal F}(H/m)$ (see eqs.
(\ref{effectivezeromode}) and (\ref{effzeromodein})).}
\label{fofh}
\epsfig{file=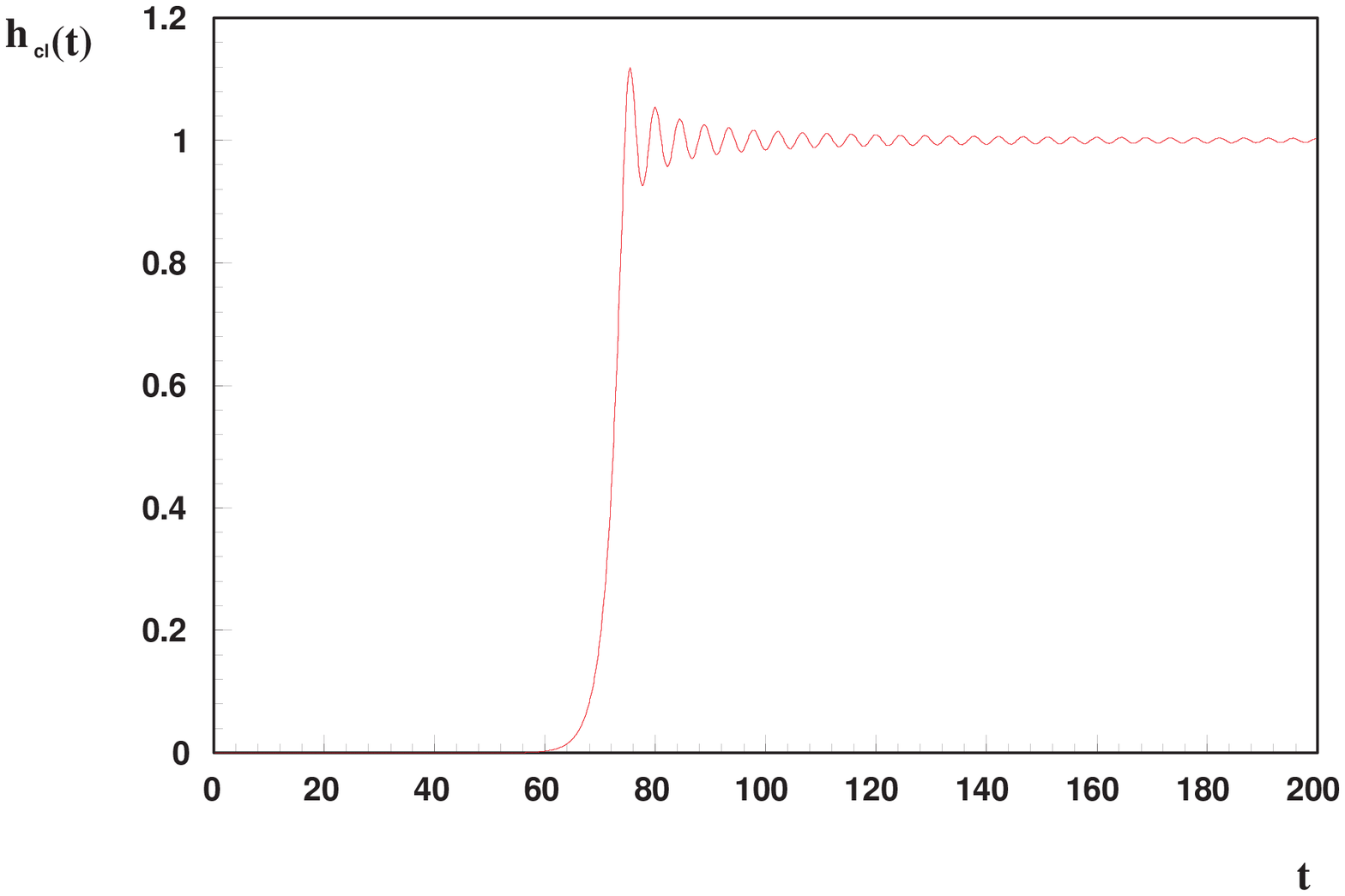,width=10.5cm,height=5.5cm}
\caption{ $\eta_{eff}^2(\tau)$ vs. $\tau$, for $\eta_{eff}(0)=3.94 \times
10^{-7}, \dot{\eta}_{eff}(0)=0.317\eta_{eff}(0),
g = 10^{-14}, h_0 = 2.0$. The initial conditions were obtained by
fitting the intermediate time regime of $g\Sigma(\tau)$ in
fig.\ref{gsigma}. $\eta_{eff}(\tau)$ is the solution of
eq.(\ref{effzeromode}) 
with these initial conditions.}
\label{etaclas}
\epsfig{file=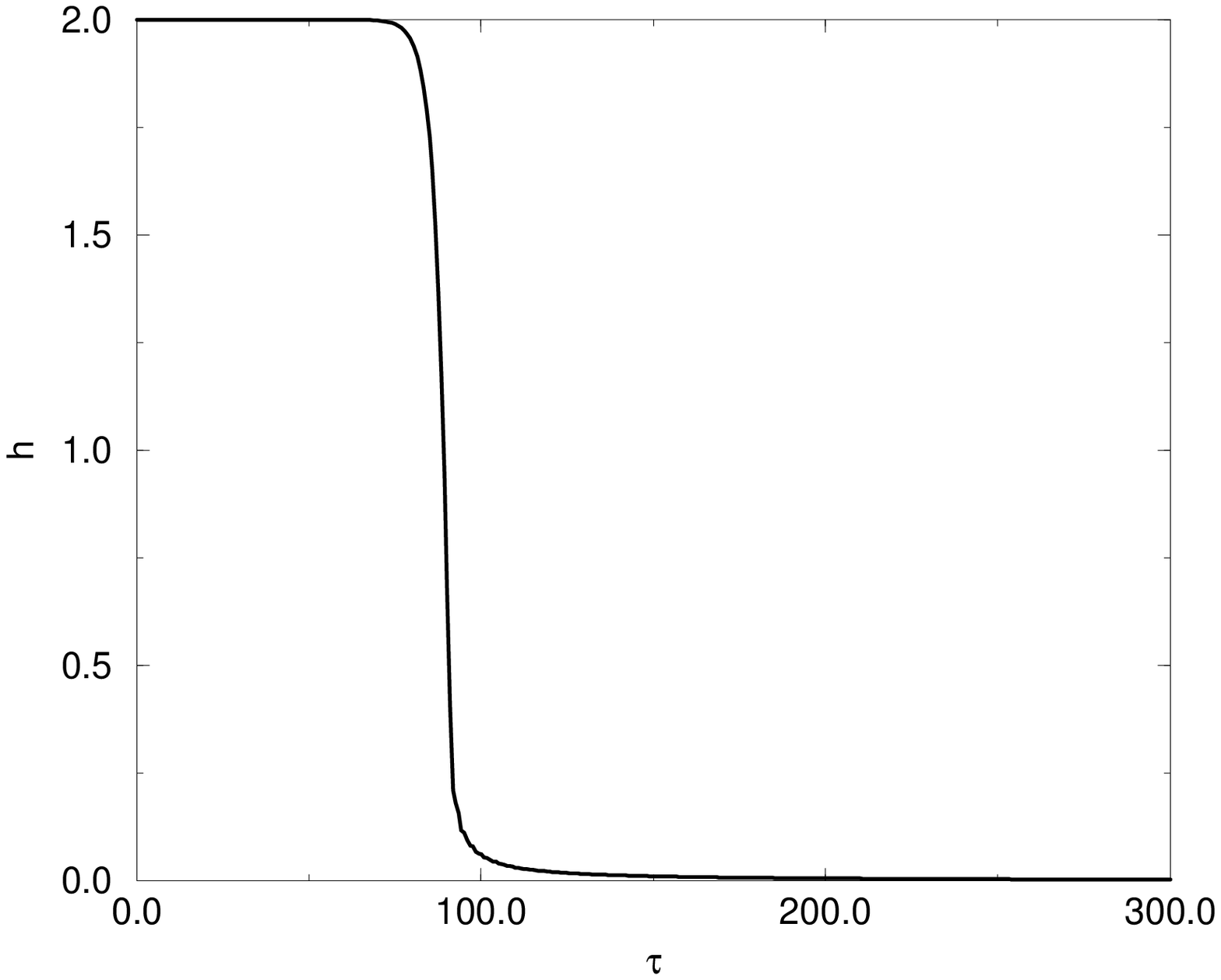,width=10.5cm,height=5.5cm}
\caption{$h(\tau)$ vs. $\tau$, obtained from the solution of
eqs. (\ref{effzeromode}) and (\ref{effscalefactor}) 
with the conditions of fig.\ref{etaclas}.}
\label{hubclas}
\end{figure}

Therefore, in summary, the effective composite zero mode obeys
\begin{equation}
\left[\frac{d^2}{d \tau^2}+ 3h \frac{d}{d\tau}-1+
\eta^2_{eff}(\tau)\right]\eta_{eff}(\tau) = 0 
\; ; \; \dot{\eta}_{eff}(\tau = 0) = (\nu -
\frac{3}{2}) \; \eta_{eff}(0) \; , \label{effzeromode}
\end{equation}
where $ \eta_{eff}(0)
 \equiv {{\sqrt{\lambda_R/2}}\over {m_R}} \; \phi_{eff}(0) $
 is obtained numerically for a given $ h_0 $ by fitting
the intermediate time behavior of  $ g\Sigma(\tau) $ with the growing zero
mode solution. Recall that $ \lambda_R = 8\pi^2 \, g $.

 Furthermore, this analysis
shows that in the case $\eta = 0$,  the renormalized energy and
pressure in this regime in which the renormalized integrals are
dominated by the superhorizon modes are given by  
\begin{eqnarray}
 \varepsilon_R(\tau) &  \approx   & \frac{2Nm^4_R}{\lambda_R} \left\{
\frac{1}{2}\dot{\eta}^2_{eff}+\frac{1}{4}\left(-1+\eta^2_{eff}\right)^2
\right\},
\label{effenergy} \\
 (p+\varepsilon)_R  & \approx & \frac{2Nm^4_R}{\lambda_R}\left\{
\dot{\eta}^2_{eff}\right\} \label{ppluseeff},
\end{eqnarray}
where we have neglected the contribution proportional to $1/a^2(\tau)$ 
because it is effectively red-shifted away after just a few e-folds.
We found numerically that this term is negligible after the interval
of time necessary for the superhorizon modes to dominate the contribution
to the integrals. 
Then the dynamics of the scale factor is given by 
\begin{equation}
h^2(\tau) = 4 h^2_0 \left\{
\frac{1}{2}\dot{\eta}^2_{eff}+\frac{1}{4}\left(-1+\eta^2_{eff}\right)^2
\right\}.
\label{effscalefactor}
\end{equation}

We have numerically evolved the set of effective equations (\ref{effzeromode}, \ref{effscalefactor}) by extracting the initial
condition for the effective zero mode from the intermediate time behavior
of $g\Sigma(\tau)$. We found a remarkable agreement  between the
evolution of $\eta^2_{eff}$ and $g\Sigma(\tau)$ and between 
the dynamics of the scale factor in terms of the evolution of
$\eta_{eff}(\tau)$, and the {\em full}
dynamics of the scale factor and quantum fluctuations within our numerical
accuracy. Figs. \ref{etaclas} and \ref{hubclas} show the evolution
of $\eta^2_{eff}(\tau)$ and $h(\tau)$ respectively from the {\em
classical} evolution 
eqs. (\ref{effzeromode}) and (\ref{effscalefactor}) using the initial
condition  $ \eta_{eff}(0) $ extracted from the exponential fit of
$ g\Sigma(\tau) $ in the intermediate regime. These figures should be 
compared to figs. \ref{gsigma} and \ref{hubblefig}. We have also
numerically compared  
$p/\varepsilon$ given solely by the dynamics of the effective zero mode
and it is again numerically indistinguishable from that obtained with the
full evolution of the mode functions. 

This is one of the main results of our work. In summary: the modes that
become superhorizon sized and grow through the spinodal instabilities assemble
themselves into an effective composite zero mode a few e-folds after
the beginning of inflation. This effective zero mode drives the dynamics
of the FRW scale factor, terminating inflation when the non-linearities
become important. In terms of the underlying fluctuations, the spinodal
growth of superhorizon modes gives a non-perturbatively large contribution
to the energy momentum tensor that drives the dynamics of the scale factor.
Inflation terminates when the mean square root fluctuation probes the
equilibrium minima of the tree level potential. 

This phenomenon of   zero mode assembly, i.e. the `classicalization'
of quantum mechanical fluctuations that grow after horizon crossing is
very similar to the interpretation of `decoherence without decoherence'
of Starobinsky and Polarski\cite{polarski}.

The extension of this analysis to the case for which $\eta(0) \neq 0$
is straightforward.  Since both $\eta(\tau)$ and 
$\sqrt{g\Sigma(\tau)} = |C_0|f_0(\tau)$ obey the equation for the
zero mode, eq.(\ref{modcr}), it is clear that we can generalize
our definition of the effective zero mode to be
\begin{equation}
\eta_{eff}(\tau) \equiv \sqrt{\eta^2(\tau)+g\Sigma(\tau)}\; .
\label{effeta}
\end{equation}
which obeys the equation of motion of a {\em classical} zero mode:
\begin{equation}
\left[\frac{d^2}{d\tau^2}+3h\frac{d}{d\tau}-1+\eta_{eff}(\tau)^2\right]
\eta_{eff}(\tau) = 0 \; . \label{effzeroeqn}
\end{equation}
If this effective zero mode is to drive the FRW expansion, then the
additional condition
\begin{equation}
\dot{\eta}^2 f_0^2 - 2\eta\dot{\eta}f_0\dot{f_0} + \eta^2 \dot{f_0}^2 = 0
\end{equation}
must also be satisfied.  One can easily show that this relation is indeed
satisfied if the mode functions factorize as in (\ref{factor2}) and if
the integrals (\ref{int1}) -- (\ref{int3}) are dominated by the 
contributions of the superhorizon mode functions.  This leads to the
conclusion that the gravitational dynamics is given by eqs. 
(\ref{effenergy}) -- (\ref{effscalefactor}) with $ \eta_{eff}(\tau) $ defined
by (\ref{effeta}).

We see that in {\em all} cases, the full large $N$ quantum dynamics in these 
models of inflationary phase transitions is well approximated by the
equivalent dynamics of a homogeneous, classical scalar field with initial
conditions on the effective field 
$\eta_{eff}(0) \geq \sqrt{g} h_0 {\cal F}(h_0)$.  
We have verified these
results numerically for the field and scale factor dynamics, finding that
the effective classical dynamics reproduces the results of the full
dynamics to within our numerical accuracy.  
We have also checked numerically
that the estimate for the classical to quantum crossover given by eq.(\ref{classquandyn}) is quantitatively correct. Thus in the classical case in
which $\eta(0) \gg \sqrt{g}\; h_0$ we find that 
$\eta_{eff}(\tau) = \eta(\tau)$,
 whereas in the opposite, quantum case $\eta_{eff}(\tau) =
\sqrt{g\Sigma(\tau)}$. 

This remarkable feature of zero mode assembly of long-wavelength,
spinodally unstable modes is a consequence of the presence of the horizon.
It also explains why, despite the fact that asymptotically the
fluctuations sample the broken symmetry state, the equation of state is
that of matter. 
Since the excitations in the broken symmetry state are massless Goldstone
bosons one would expect radiation domination. However, the assembly
phenomenon, i.e. the redshifting of the wave vectors, makes these modes behave
exactly like zero momentum modes that give an equation of state of matter
(upon averaging over the small oscillations around the minimum).  

Subhorizon modes at the end of inflation with $ q > h_0 \, e^{h_0
\tau_s} $ do not participate in the zero mode assembly. The behavior of such
modes do depend on $ q $ after the end of inflation. Notice that these
modes have extremely large comoving $ q $ since $  h_0 \, e^{h_0
\tau_s} \geq 10^{26} $. As discussed in sec. VII such modes decrease
with time after inflation as $ \sim 1/a(\tau) $\cite{frw2}. 

\subsection{Making sense of small fluctuations:}
Having recognized the effective classical variable that can be interpreted
as the component of the field that drives the FRW background and rolls
down the classical potential hill, we want to recognize unambiguously
the small fluctuations. We have argued above that after horizon crossing,
all of the mode functions evolve proportionally to the zero mode,
and the question arises: which modes are assembled into the effective
zero mode whose dynamics drives the evolution of the FRW scale factor
and which modes are treated as perturbations? In principle every
$k\neq 0$ mode provides some spatial inhomogeneity, and assembling these
into an effective homogeneous zero mode seems in principle to do away with
the very inhomogeneities that one wants to study. However, scales of
cosmological importance today first crossed the horizon during the 
last 60 or so e-folds of inflation. Recently Grishchuk\cite{grishchuk}
 has argued that the
sensitivity of the measurements of $ \Delta T/T $ probe inhomogeneities on
scales $\approx 500$ times the size of the present horizon. Therefore scales
that are larger than these and that have first crossed the horizon
much earlier than the last 60 e-folds of inflation are unobservable
today and 
can be treated as an effective homogeneous component, whereas the scales that
can be probed experimentally via the CMB inhomogeneities today must be treated 
separately as part of the inhomogeneous perturbations of the CMB. 

Thus a consistent description of the dynamics in terms of an effective
zero mode plus `small' quantum fluctuations can be given provided
the following requirements are met:
a) the total number of e-folds $N_e \gg 60$, b) all the modes that have
crossed the horizon {\em before} the last 60-65 e-folds are assembled into
an effective {\em classical} zero mode via $\phi_{eff}(t) = 
\left[\phi^2_0(t)+ \langle \psi^2(\vec x,t) \rangle_R
\right]^{\frac{1}{2}}$, c) the modes that cross the horizon during the
last 60--65 e-folds are accounted as `small' perturbations. The reason
for the requirement a) is that in the separation 
$\phi(\vec x, t) = \phi_{eff}(t)+\delta \phi(\vec x,t)$ one requires that
$\delta \phi(\vec x,t)/\phi_{eff}(t) \ll 1$. As argued above, after the 
modes cross the horizon, the ratio of amplitudes of the mode functions remains
constant and given by $e^{(\nu - \frac{3}{2})\Delta N}$ with $\Delta N$ 
being the number of e-folds between the crossing of the smaller $ k $ and the
crossing of the larger $ k $. Then for $\delta \phi(\vec x, t)$ to be much
smaller than the effective zero mode, it must be that the Fourier components
of $\delta \phi$ correspond to very large $k$'s at the beginning of inflation,
so that the effective zero mode can grow for a long time before the components
of $\delta \phi$ begin to grow under the spinodal instabilities. 
In fact requirement a) is not very severe; in the figs.(19-23) we have taken
$h_0 = 2.0$ which is a very moderate value and yet for $ g = 10^{-12}$
the inflationary stage lasts for well over 100 e-folds, 
and as argued above, the
larger $h_0$ for fixed $ g $, the longer is the inflationary stage. 
Therefore under this set of conditions, the classical dynamics of the effective zero mode $\phi_{eff}(t)$  drives the FRW background, whereas
the inhomogeneous fluctuations $\delta \phi(\vec x,t)$, which are made up
of Fourier components with wavelengths that are much smaller than the
horizon at the beginning of inflation and that cross the horizon during
the last 60 e-folds, provide the inhomogeneities that seed density
perturbations.

\subsection{Scalar  Metric Perturbations:}
Having identified the effective zero mode and the `small perturbations',
we are now in position to provide an estimate for the amplitude and spectrum
of scalar metric perturbations. We use the clear formulation in
ref.\cite{mukhanov} in terms of gauge invariant
variables. In particular we focus on the dynamics of the Bardeen
potential\cite{bardeen}, which in longitudinal gauge is identified
with the 
Newtonian potential. The equation of motion for the Fourier components (in
terms of comoving wavevectors) for this variable in terms of the effective
zero mode is\cite{mukhanov}
\begin{equation}
\ddot{\Phi}_k +
\left[H(t)-2\frac{\ddot{\phi}_{eff}(t)}{\dot{\phi}_{eff}(t)}\right] 
\dot{\Phi}_k+\left[\frac{k^2}{a^2(t)}+ 2\left(\dot{H}(t)-H(t)
\frac{\ddot{\phi}_{eff}(t)}{\dot{\phi}_{eff}(t)}\right)\right]\Phi_k =
0 .
\label{bardeen}
\end{equation}

We are interested in determining the dynamics of $\Phi_k$ for those
wavevectors that cross the horizon during the last 60 e-folds before the
end of inflation. During the inflationary stage the numerical analysis
yields  to a very good approximation
\begin{equation}
H(t) \approx H_0  \; ; \; \phi_{eff}(t) = \phi_{eff}(0)\; e^{(\nu-
\frac{3}{2})H_0t}, \label{infla}
\end{equation}
where $H_0$ is the value of the Hubble constant during inflation, leading to 
\begin{equation}
\Phi_k(t) = e^{(\nu -2)H_0t}\left[a_k\;
H^{(1)}_{\beta}\left(\frac{ke^{-H_0t}}{H_0}\right) 
+b_k \; H^{(2)}_{\beta}\left(\frac{ke^{-H_0t}}{H_0}\right)\right]
\; ; \; \beta= \nu-1 \; .
\label{solbardeen}
\end{equation}
The coefficients $a_k,b_k$ are determined by the initial conditions.

Since we are interested in the wavevectors that cross the horizon during
the last 60 e-folds, the consistency for the zero mode assembly and
the interpretation of `small perturbations' requires that there must be
many e-folds before the {\em last} 60. We are then considering wavevectors
that were deep inside the horizon at the onset of
inflation. $\Phi_k(t)$ is related to the canonical `velocity field'
that determines scalar  perturbations 
of the metric and which is quantized with Bunch-Davies initial
conditions for the large $k$-mode functions. The relation between
$\Phi_k$ and $v$ and the  
initial conditions on $v$ lead at once to a determination of the
coefficients $a_k$ and $b_k$ for $k >> H_0$\cite{mukhanov} 
\begin{equation}
a_k = -\frac{3}{2} \left[\frac{8\pi}{3M^2_{Pl}}\right] \dot{\phi}_{eff}(0)
\sqrt{\frac{\pi}{2H_0}} \frac{1}{k} \quad ; \quad b_k = 0\; .
\label{coeffs}
\end{equation} 

Thus we find that the amplitude of scalar metric perturbations after 
horizon crossing is given by
\begin{equation}
|\delta_k(t)| = k^{\frac{3}{2}}|\Phi_k(t)| \approx
\frac{3}{2} \left[\frac{8\sqrt{\pi}}{3M^2_{Pl}}\right] \dot{\phi}_{eff}(0)
\left(\frac{2H_0}{k}\right)^{\nu -\frac{3}{2}}
e^{(2\nu-3)H_0t}\; .
\label{perts}
\end{equation}
The power spectrum per logarithmic $k$  interval is given by
$ |\delta_k(t)|^2 $. The time dependence of $ |\delta_k(t)| $ displays the
unstable growth associated with the spinodal instabilities of
super-horizon modes and 
is a hallmark of the phase transition.
This time dependence can be also understood from the constraint equation
that relates the Bardeen potential to the gauge invariant field fluctuations\cite{mukhanov}, which in longitudinal gauge are identified
with $\delta \phi(\vec x,t)$. 
The constraint equation and the evolution equations for the
gauge invariant scalar field fluctuations are\cite{mukhanov}
\begin{equation}
\frac{d}{dt}(a \Phi_k) = \frac{4\pi}{M^2_{Pl}}\; a\; \delta \phi^{gi}_k
\; \dot{\phi}_0\; ,
\label{constraint}
\end{equation}

\begin{equation}
\left[\frac{d^2}{dt^2}+3H \frac{d}{dt}+\frac{k^2}{a^2}+{\cal M}^2 \right]
\delta \phi^{gi}_k-4 \;
\dot{\phi}_{eff}\;\dot{\Phi}_k+2V'(\phi_{eff})\;\Phi_k=0\; . 
\label{gauginv}
\end{equation}

Since the right hand side of (\ref{constraint}) is proportional to
$\dot{\phi}_{eff}/M^2_{Pl} \ll 1 $ during the inflationary epoch in this
model,  we can neglect the terms proportional
to $\dot{\Phi}_k $ and $\Phi_k$ on the left hand side of (\ref{gauginv}),
in which case the equation for the gauge invariant scalar field
fluctuation is the same as for the mode functions. In fact, since $ 
\delta \phi^{gi}_k$ is gauge invariant we can evaluate it in the longitudinal gauge wherein it is identified with the mode functions
$f_k(t)$. Then absorbing a constant of integration in the initial
conditions for the Bardeen variable, we find
\begin{equation}
\Phi_k(t) \approx \frac{4\pi}{M^2_{Pl}\; a(t)}\int_{t_o}^t
a(t')\;\phi_{eff}(t')\; f_k(t')\; dt' + {\cal O}\left(\frac{1}{M^4_{Pl}}\right) ,
\label{bard}
\end{equation}
and using that $\phi(t) \propto e^{(\nu-3/2)H_0t} $ and that
 after horizon crossing $f_k(t) \propto e^{(\nu-3/2)H_0t}$, one obtains
at once the time dependence of the Bardeen variable after horizon
crossing. In particular the time dependence is found to be $\propto 
e^{(2\nu-3)H_0t}$. It is then clear that the time dependence is a reflection of
the spinodal (unstable) growth of the superhorizon field fluctuations. 

 To obtain the amplitude and spectrum
of density perturbations at {\em second} horizon crossing we use the
conservation law associated with the gauge invariant variable\cite{mukhanov}
\begin{equation}
\xi_k = \frac{2}{3}
\frac{\frac{\dot{\Phi}_k}{H}+\Phi_k}{1+p/\varepsilon} + \Phi_k
\; \; ; \; \; \dot{\xi}_k =0\; , \label{xivar}
\end{equation}
which is valid after horizon crossing of the mode with wavevector $ k $.
Although this conservation law is an exact statement of superhorizon mode
solutions of eq.(\ref{bardeen}), 
we have obtained solutions assuming that during
the inflationary stage $H$ is constant and have neglected the $\dot{H}$ term in
Eq. (\ref{bardeen}). Since during the inflationary stage,
\begin{equation}
\dot{H}(t) = -\frac{4\pi}{M^2_{Pl}}\, \dot{\phi}^2_{eff}(t) \propto
H^2_0 \; \left(\frac{d\eta_{eff}(\tau)}{d\tau}\right)^2\ll H^2_0 \label{Hdot}
\end{equation}
and $\ddot{\phi}/\dot{\phi} \approx H_0$, the above approximation is
justified. We then see that $\phi^2_{eff}(t) \propto
e^{(2\nu-3)H_0t}$ which is the same time dependence as that of
$\Phi_k(t)$. Thus the
term proportional to $1/(1+p/\varepsilon)$ in Eq. 
(\ref{xivar}) is indeed constant in time after horizon crossing. On the other
hand, the term that
does not have this denominator evolves in time but is of order 
$(1+p/\varepsilon) =
-2\dot{H}/3H^2 \ll 1$ with respect to the constant term and therefore can be
neglected. Thus, we confirm that the variable $\xi$ is conserved
up to the small term proportional to $(1+p/\varepsilon)\Phi_k$ which
is negligible during the inflationary stage. 
This small time dependence is consistent with the fact
that we neglected the $\dot{H}$ term in the equation of motion for $\Phi_k(t)$.
The validity of the conservation law has been recently studied and confirmed in
different contexts\cite{caldwell}. Notice that we do not have to assume
that $\dot{\Phi}_k$ vanishes, which in fact does not occur.

 However, upon second horizon crossing
it is straightforward to see that $\dot{\Phi}_k(t_f) \approx 0$. The 
reason for this assertion can be seen as follows: eq.(\ref{gauginv}) shows that
at long times, when the effective zero mode is oscillating around the minimum
of the potential with a very small amplitude and when the time dependence of
the fluctuations has saturated (see fig.3), $\Phi_k$ will redshift
as $\approx 1/a(t)$\cite{frw2} and its derivative becomes extremely small. 

 Using
this conservation law,  assuming matter domination at second horizon 
crossing,  and $\dot{\Phi}_k(t_f)\approx 0$\cite{mukhanov}, we find
\begin{equation}
|\delta_k(t_f)| = \frac{12 \, \Gamma(\nu)\,\sqrt{\pi}}{5\, (\nu-\frac{3}{2})\, 
{\cal F}(H_0/m)} \left(\frac{2H_0}{k}\right)^{\nu-\frac{3}{2}},
\label{amplitude}
\end{equation}
where ${\cal F}(H_0/m)$ determines the initial amplitude of the effective
zero mode (\ref{effzeromodein}). 
We can now read the power spectrum per logarithmic $k$ interval
\begin{equation}
{\cal P}_s(k) = |\delta_k|^2 \propto k^{-2(\nu-\frac{3}{2})}.
\end{equation}
leading to the index for scalar density perturbations
\begin{equation}
n_s = 1-2\left(\nu-\frac{3}{2}\right) \; . \label{index}
\end{equation}

For $H_0/m \gg 1$, we can expand $\nu-3/2$ as a series in $m^2/H_0^2$ in
eq.(\ref{amplitude}).  Given that the comoving wavenumber of the mode which 
crosses the horizon $n$ e-folds before the 
end of inflation is $k=H_0 e^{(N_e-n)}$ 
where $N_e$ is given by (\ref{efolds}), we arrive at the following expression
for the amplitude of fluctuations on the scale corresponding to $n$ 
in terms of the De Sitter Hubble constant and the coupling 
$ \lambda = 8\pi^2 \, g $:
\begin{equation}
|\delta_n(t_f)| \simeq 
\frac{9 H^3}{5\sqrt{2} m^3} \left(2e^n\right)^{m^2/3H_0^2}
\sqrt{\lambda} \left[1+\frac{2m^2}{3H_0^2} \left(\frac76 - 
\ln 2 - \frac{\gamma}{2}
\right) + {\cal O}\left( \frac{m^4}{H_0^4} \right) \right] \; .
\label{amplitude_n}
\end{equation}
Here, $\gamma$ is Euler's constant.  Note the explicit dependence of the 
amplitude of density perturbations on $\sqrt{g}$.  For $n \approx 60$,
the factor $\exp(nm^2/3H_0^2)$ is ${\cal O}(100)$ for $H_0/m = 2$, while
it is ${\cal O}(1)$ for $H_0/m \geq 4$.  Notice that for $H_0/m$ large,
the amplitude increases approximately as $(H_0/m)^3$, which will place 
strong restrictions on $ g $ in such models.

We remark that we have not included the small corrections to the dynamics
of the effective zero mode and the scale factor arising from the
non-linearities. We have found numerically that these nonlinearities 
are only significant for the
modes that cross about 60 e-folds before the end of inflation for
values of the Hubble parameter $H_0/m_R > 5$.  The effect of these
non-linearities in the large $N$ limit is to slow somewhat the exponential
growth of these modes, with the result of shifting the power spectrum
closer to an exact Harrison-Zeldovich spectrum with $n_s =1$.  Since
for $H_0/m_R > 5$ the power spectrum given by (\ref{index}) differs from
one by at most a few percent, the effects of the non-linearities are
expected to be observationally unimportant.
The spectrum given by (\ref{amplitude}) is
similar to that obtained in references\cite{turner,guthpi} although
the amplitude differs from that obtained there. In addition, we do not
assume slow roll for which $(\nu - \frac{3}{2})\ll 1$, although 
this would be the case if $N_e \gg 60$.

We emphasize an important
feature of the spectrum: it has more power at {\em long
wavelengths} because $\nu-3/2 > 0$. This is recognized to be a
consequence 
of the spinodal instabilities that result in the growth of long wavelength
modes and therefore in more power for these modes.  
This seems to be a robust prediction of new inflationary scenarios in
which the potential has negative second derivative in the region of field
space that produces inflation.  

 It is at this
stage that we recognize the consistency of our approach for separating
the composite effective zero mode from the small fluctuations. We have
argued above that many more than 60 e-folds are required for consistency,
and that the small fluctuations correspond to those modes that cross
the horizon during the last 60 e-folds of the inflationary stage. For these
modes $H_0/k = e^{-H_0 t^*(k)}$ where $t^*(k)$ is the time since the beginning 
of inflation of horizon crossing of the mode with wavevector $k$. 
The scale that  corresponds to the Hubble radius today $\lambda_0
=2\pi/k_0$ is the first to cross during the last 60 or so e-folds
before the end of 
inflation. Smaller scales today will correspond to $k > k_0$ at the
onset of inflation since they will cross the first horizon later and
therefore will reenter earlier. The bound on $|\delta_{k_0}| \propto  
\Delta T/ T \leq  10^{-5}$ on
these scales provides a lower bound on the number of e-folds required for
these type of models to be consistent:
\begin{equation}
N_e >
60+\frac{12}{\nu-\frac{3}{2}}-\frac{\ln(\nu-\frac{3}{2})}{\nu-\frac{3}{2}}\; ,
\label{numbofefolds}
\end{equation}
where we have written the total number of e-folds as $N_e=H_0\; t^*(k_0)+60$.
This in turn can be translated into a bound on the coupling constant using
the estimate given by eq.(\ref{efolds}).

The four year COBE  DMR Sky Map\cite{gorski} gives $n \approx 1.2 \pm 0.3$
thus providing an upper bound on $\nu$
\begin{equation}
0 \leq \nu-\frac{3}{2} \leq 0.05 \label{cobebound}
\end{equation}
corresponding to $h_0 \geq 2.6$. We then find that these values of $h_0$ and
$\lambda \approx 10^{-12}-10^{-14}$ provide sufficient e-folds to satisfy
the constraint for scalar density perturbations. 

\subsection{Tensor Metric Perturbations:} 
The scalar field does not couple to the tensor (gravitational wave)
modes directly, and the tensor perturbations are gauge invariant from
the beginning. Their dynamical evolution is completely determined by
the dynamics of the scale factor\cite{mukhanov,grishchuk2}. 
Having established numerically that the inflationary epoch is
characterized by $\dot{H}/H^2_0 \ll 1$ and that scales of cosmological
interest cross the 
horizon during the stage in which this approximation is excellent, we can
just borrow the known result for the power spectrum of gravitational waves
produced during inflation extrapolated to the matter
era\cite{mukhanov,grishchuk2} 
\begin{equation}
{\cal P}_T(k) \approx \frac{H^2_0}{M^2_{Pl}}k^0\; .
\end{equation}
Thus the spectrum to this order is scale invariant (Harrison-Zeldovich)
with an amplitude of the order $m^4/\lambda M^4_{Pl}$. Then, for values
of $m \approx 10^{12}-10^{14} \mbox{ Gev }$ and 
$\lambda \approx 10^{-12}-10^{-14}$
one finds that the amplitude is $\leq 10^{-10}$ which is much smaller than the
amplitude of scalar density perturbations. 
As usual the amplification of scalar perturbations
is a consequence of the equation of state during the inflationary epoch.

\subsection{Contact with the Reconstruction Program:}
The program of reconstruction of the inflationary potential seeks to
establish a relationship between features of the inflationary scalar
potential and the spectrum of scalar and tensor perturbations.
This program, in combination with measurements of scalar and tensor
components either from refined measurements of temperature inhomogeneities
of the CMB or through galaxy correlation functions will then offer a 
glimpse of the possible realization of the 
inflation\cite{reconstruction,lyth}. 
Such a reconstruction program is based on the slow roll approximation
and the spectral index of scalar and tensor perturbations are obtained
in a perturbative expansion in the slow roll
parameters\cite{reconstruction,lyth} 
\begin{eqnarray}
\epsilon(\phi) & = &
\frac{\frac{3}{2}\dot{\phi}^2}{\frac{\dot{\phi}^2}{2}+V(\phi)}\; ,
\label{epsifi}  \\
\eta(\phi) & = & -\frac{\ddot{\phi}}{H \dot{\phi}}\; . \label{etafi}
\end{eqnarray}
We can make contact with the reconstruction program by identifying $\phi$
above with our $\phi_{eff}$ after the first few e-folds of inflation needed
to assemble the effective zero mode from the quantum
fluctuations. We have numerically established that for the weak scalar
coupling required 
for the consistency of these models, the cosmologically interesting scales
cross the horizon during the epoch in which $H \approx H_0 \; ; \;
\dot{\phi}_{eff} \approx (\nu - 3/2)\; H_0 \; \phi_{eff} \; ; \; V \approx
m_R^4/\lambda \gg \dot{\phi}^2_{eff}$. In this case we find 
\begin{equation}
\eta(\phi_{eff}) = -(\nu - \frac{3}{2}) \; ; \;  \epsilon(\phi_{eff}) \approx
{\cal O}(\lambda) \ll  \eta(\phi_{eff}).
\end{equation}

With these identifications, and in the notation of\cite{reconstruction,lyth}
the reconstruction program predicts  the index for scalar density
perturbations $n_s$ given by
\begin{equation}
 n_s-1 = -2\left(\nu - \frac{3}{2}\right)+ {\cal O}(\lambda),
\end{equation}
which coincides with the index for  the power
spectrum per logarithmic interval $|\delta_k|^2$ with $|\delta_k|$
given by eq.(\ref{amplitude}).  
We must note however that our treatment did not
assume slow roll for which $(\nu - \frac{3}{2})\ll 1$. Our
self-consistent, non-perturbative study of the dynamics plus the
underlying 
requirements for the identification of a composite operator acting as an
effective zero mode, validates the reconstruction program in weakly
coupled new inflationary models.

\section{DECOHERENCE: QUANTUM TO CLASSICAL TRANSITION DURING INFLATION} 

An important aspect of cosmological perturbations is that 
they are of quantum origin but eventually they become classical as
they are responsible for the small classical metric perturbations. This
quantum to classical crossover is associated with a decoherence process and
has received much attention\cite{polarski,salman}. 

In this section we study
the quantum to classical transition of superhorizon modes for the Bardeen 
variable by relating these to the field mode functions through
eq.(\ref{bard}) and analyzing the full time evolution of the density
matrix of the matter field. Eq.(\ref{bard}) establishes that  in the
models under consideration the classicality of the Bardeen variable is
determined by the classicality of the scalar field modes. 

In the situation under consideration, long-wavelength field modes become
spinodally unstable and grow exponentially after horizon crossing. The 
factorization (\ref{factor}) results in the phases of these modes `freezing
out'. This feature and the growth in amplitude entail that these modes become
classical. Eq.(\ref{bard}) in turn implies that these features 
also apply to the superhorizon modes of the  Bardeen potential. 

 Therefore we can address the quantum
to classical transition of the Bardeen variable (gravitational potential) by
analyzing the evolution of the density matrix for the matter field. 

To make contact with previous work\cite{polarski,salman} we choose to study
the evolution of the field density matrix in conformal time, although the
same features will be displayed in comoving time. 

In the large $N$ or Hartree (also in the self-consistent one-loop)
approximation, the density matrix 
is Gaussian, and defined by a normalization factor, a complex covariance that
determines the diagonal matrix elements, and a real covariance that determines
the mixing in the Schr\"odinger representation as discussed in
ref.\cite{frw} (and references therein).

That is, the density matrix takes the form
\begin{eqnarray}\label{matden}
\rho[\Phi,\tilde{\Phi},{\cal T}] & = & \prod_{\vec{k}} {\cal{N}}_k({\cal T})
\exp\left\{ 
- \frac12 A_k({\cal T}) \; \vec{\eta}_{\vec{k}}({\cal T})\cdot
\vec{\eta}_{-{\vec{k}}}({\cal T})- 
\frac12 A^*_k({\cal T}) \;
\tilde{\vec{\eta}}_{\vec{k}}({\cal T})\cdot 
\tilde{\vec{\eta}}_{-{\vec{k}}}({\cal T}) 
\right. \nonumber \\  &   & \left.
- \;  B_k({\cal T}) \; \vec{\eta}_{\vec{k}}({\cal
T})\cdot\tilde{\vec{\eta}}_{-{\vec{k}}}({\cal T}) 
+i\, \vec{\pi}_{\vec{k}}({\cal T})\cdot\left(\vec{\eta}_{-{\vec{k}}}({\cal T})-
\tilde{\vec{\eta}}_{-{\vec{k}}}({\cal T})\right) \right\} ;\ , \\
\vec{\eta}_{\vec{k}}({\cal T})          & = &
\vec{\chi}_{\vec{k}}({\cal T})-\chi_0({\cal T})\,\delta_{i,1} \;\delta(\vec{k}) 
\; \; ; \;  \; 
\tilde{\vec{\eta}}_k({\cal T})        = 
{\tilde{\vec{\chi}}}_k({\cal T})-\chi_0({\cal T})\, 
\delta_{i,1}\; \delta(\vec{k}) \; . \nonumber
\end{eqnarray}
$\vec{\pi}_{\vec{k}}({\cal T}) $ is the Fourier 
transform of $ \Pi_{\chi}({\cal T},\vec{x}) $. This form of the density matrix
is dictated by the hermiticity condition 
$$
\rho[\Phi,\tilde{\Phi},{\cal T}]
=\rho^*[\tilde{\Phi},\Phi,{\cal T}]\; ;
$$
as a result of this, $ B_k({\cal T}) $ is real.
The kernel $ B_k({\cal T}) $ determines the amount of `mixing' in the
density matrix since if $ B_k=0 $, the density matrix corresponds to a pure
state because it factorizes into  a wave functional depending only on
$ \Phi(\cdot) $ times its complex conjugate taken at $ \tilde{\Phi}(\cdot) $. This is
the case under consideration, since the initial conditions correspond to a
pure state and under time evolution the density matrix remains that of a pure
state\cite{frw}. 

In conformal time quantization the evolution of the density matrix
is via the conformal time Hamiltonian (\ref{confham}). The evolution equations
for the covariances  (given by equation (2.20) in \cite{frw}; see 
also equation (2.44) of \cite{frw}) are obtained from those given in
ref.\cite{frw} by setting $a(t) = 1$ and using the frequencies
$ \omega^2_k({\cal T}) = k^2+{\cal{M}}^2({\cal T}) $. In particular,
by setting 
\begin{equation}\label{FpF}
A_k({\cal T}) = -i \, \frac{F'^*_k({\cal T})}{F^*_k({\cal T})}.
\end{equation}

More explicitly \cite{frw},
\begin{eqnarray}\label{coero}
{\cal{N}}_k({\cal T}) &=&{\cal{N}}_k({\cal T}_0)\;
\exp\left[\int_{{\cal T}_0}^{{\cal T}} A_{Ik}({\cal T}')\;d{\cal T}' \right] = 
{{{\cal N }_k({\cal T}_0)} \over { \sqrt{\omega_k({\cal T}_o)} \;
|F_k({\cal T})|}}  
\; , \cr \cr
A_{Ik}({\cal T}) &=& - {d \over {d{\cal T}}}\log|F_k({\cal T})| = -{\dot a}(t)-
a(t)\;{d \over {dt}}\log|f_k(t)| \; , \cr \cr
A_{Rk}({\cal T}) &=& {1 \over { |F_k({\cal T})|^2}} \, = { 1\over 
{ a(t)^2 \; |f_k(t)|^2 }} \; ,  \;
B_k({\cal T}) \equiv  0 \; , \nonumber
\end{eqnarray}
where $A_{Rk}$ and $A_{Ik}$ are respectively the real and imaginary parts
of $A_k$ and we have used the value of the Wronskian (\ref{wro}) in
evaluating (\ref{coero}).

The coefficients $ A_k({\cal T}) $ and $ {\cal{N}}_k({\cal T}) $
in the gaussian density matrix (\ref{matden}) are
completely determined by  the conformal mode functions $ F_k({\cal T}) $
(or alternatively the comoving time mode functions $f_k(t)$).

Let us study the time behavior  of these coefficients. 
During inflation, $ a(t) \approx  e^{h_0t}$,
and the mode functions  factorize after horizon crossing, and superhorizon
modes  grow in cosmic time  as in 
Eq.~(\ref{factor}):
$$
a^2(t) |f_k(t)|^2 \approx { 1 \over  {\cal D}_k} \; e^{(2\nu -1)h_0t}
$$ 
where the coefficient $ {\cal D}_k $ can be read from eq.~(\ref{factor}).

We emphasize that this is a {\em result of the full evolution} as
displayed from the numerical solution in fig.\ref{modu}. These mode
functions 
encode all of the self-consistent and non-perturbative features of the
dynamics. This should be contrasted with other studies in which typically
free field modes in a background metric are used.  

 Inserting this expression in eqs.(\ref{coero}) yields, 
\begin{eqnarray}\label{roasi}
A_{Ik}({\cal T})&\buildrel{t \to \infty}\over =&-h_0 \;  e^{h_0t} \left(\nu -
\frac12 \right)+ {\cal O}(e^{-h_0t})    \; , \cr \cr
A_{Rk}({\cal T}) &\buildrel{t \to \infty}\over =& {\cal D}_k\; \; 
 e^{-(2\nu -1)h_0t}  . \nonumber
\end{eqnarray}

Since $ \nu -\frac12 > 1 $,
we see that the imaginary part of the covariance $  A_{Ik}({\cal T}) $
{\it grows} very fast. Hence, the off-diagonal elements of $
\rho[\Phi,\tilde{\Phi},{\cal T}] $ oscillate wildly after a few e-folds of
inflation. In particular their contribution to expectation values of operators
will be washed out. That is, we quickly reach a {\it classical} regime where
only 
the diagonal part of the density matrix is relevant:
\begin{equation}\label{claden}
\rho[\Phi,\Phi,{\cal T}] = \prod_{\vec{k}} {\cal{N}}_k({\cal T})
\exp\left\{ -  A_{Rk}({\cal T}) 
\; \eta_{\vec{k}}({\cal T})\;{\eta}_{-{\vec{k}}}({\cal T}) \right\}.
\end{equation}

The real part of the covariance $ A_{Rk}({\cal T}) $ (as well as any non-zero
mixing kernel $ B_k({\cal T}) $\cite{frw}) {\it decreases} as $ e^{-(2\nu
-1)h_0t} $. Therefore, characteristic field configurations $ \eta_{\vec k} $
are very large (of order $ e^{(\nu -\frac12)h_0t} $). Therefore configurations
with field amplitudes up to ${\cal O}(e^{(\nu -\frac12)h_0t})$ will have a
substantial probability of occurring and being represented in the density
matrix.

Notice that $ \chi \sim  e^{(\nu -\frac12)h_0t} $  corresponds to
field configurations $ \Phi $ with amplitudes of order  $  e^{(\nu
-\frac32)h_0t} $ [see eq.(\ref{camco})]. It is the fact that $ \nu
-\frac32 > 0 $ which in this 
situation is responsible for the `classicalization', which is  seen to
be a consequence of the spinodal growth of long-wavelength fluctuations.

The equal-time field correlator is given by 
\begin{eqnarray}
\langle \bar{\chi}(\vec x,{\cal T})\; \bar{\chi}({\vec x}',{\cal T}) \rangle 
&=& \int \frac{d^3k}{2(2\pi)^3}\; |F_k({\cal T})|^2\; e^{i{\vec k}.({\vec x}-
{\vec x}')} \quad   , \cr \cr
& = & a(t)^2 \; \int \frac{d^3k}{2(2\pi)^3}\; 
|f_k(t)|^2\; e^{i{\vec k}.({\vec x}-{\vec x}')} \quad .
\end{eqnarray}
and is seen to be dominated by the superhorizon mode functions and to grow as 
$  e^{(2\nu -1)h_0 t} $, whereas the
field commutators remain fixed showing the emergence of a classical behavior.
As a result we obtain
\begin{equation}
\langle \bar{\chi}(\vec x,{\cal T})\; \bar{\chi}({\vec x}',{\cal T}) \rangle 
\propto a^2(t) \; \phi_{eff}(t) \;\phi_{eff}(t') \; G(|\vec x - \vec{x}'|)+
\mbox{ small }
\end{equation}
where $G(|\vec x - \vec{x}'|)$ falls off exponentially for distances larger than
the horizon\cite{De Sitter} and `small' refers to terms that are 
smaller in magnitude. This factorization of the correlation functions
is another indication of classicality.

Therefore, it is possible to  describe the physics by using
classical field theory. 
More precisely, one can use a classical statistical (or stochastic) field theory
described by the functional probability distribution (\ref{claden}).

These results generalize 
the decoherence treatment given in  ref.\cite{ps}
for a free massless field in pure quantum states to the case of interacting
fields with broken symmetry.
Note that the formal decoherence or classicalization in the density matrix
appears after the modes with wave vector $k$ become superhorizon sized i.e. 
when the factorization of the mode functions becomes effective.  

\section{Conclusions} 

Since there are a number of articles in the literature treating the problem of
preheating, it is useful to review the unique features of the present work.
First, we have treated the problem {\em dynamically}, without using the
effective potential (an equilibrium construct) to determine the evolution.
Second, we have provided consistent non-perturbative calculations of the
evolution to bring out some of the most relevant aspects of the late time
behavior.  In particular, we found that the quantum backreaction naturally
inhibits catastrophic growth of fluctuations and provides a smooth transition
to the late time regime in which the quantum fluctuations decay as the zero
mode approaches its asymptotic state.  Third, the dynamics studied obeys the
constraint of covariant conservation of the energy momentum tensor.

The Hartree approximation and the large $ N $ limit are physically 
inequivalent. The Hartree approximation considers a single field with 
self-coupling $ g $ while in the large $ N $ limit one considers 
$ N \to \infty $ fields with an $ O(N) $ invariant self-coupling $ g/N $. 
Since grand unified theories contain a large number of scalar fields,
the large $ N $ limit seems a more realistic effective model than the Hartree
approximation for the inflationary universe. 
Particle production during preheating turns to be much
more effective in  the large $ N $ limit than within the Hartree approximation
(see sec. VII and \cite{frw2}). In addition, the field evolution is definitely
model dependent. For example, in the fine-tuned two-fields model (such that 
the Mathieu equation applies) studied in ref.\cite{stei}, parametric
resonance in Hartree approximation is significative only for a limited range 
of parameters.

We have considered in these Lectures single-mass inflaton models.
The Hartree approximation for a two-field model (with unequal masses)
is presented in the second reference under \cite{us1}. In such cases,
one must  carefully apply the Hartree approximation to {\bf all}
fields in the model self-consistently. Applying Hartree to one field
and just doing perturbation theory in the other field leads to inconsistencies
as noticed in ref.\cite{kls}.

It can be argued that the inflationary paradigm as currently understood is one
of the greatest applications of quantum field theory. The imprint of quantum
mechanics is everywhere, from the dynamics of the inflaton, to the
generation of 
metric perturbations, through to the reheating of the universe. It is clear
then that we need to understand the quantum mechanics of inflation in as deep a
manner as possible so as to be able to understand what we are actually testing
via the CMBR temperature anisotropies, say.

What we have found in our work is that the quantum mechanics of inflation is
extremely subtle. We now understand that it involves both non-equilibrium as
well as non-perturbative dynamics and that what you start from may {\it not} be
what you wind up with at the end!

In particular, we see now that the correct interpretation of the
non-perturbative growth of quantum fluctuations via spinodal decomposition is
that the background zero mode must be redefined through the process of zero
mode reassembly that we have discovered. When this is done (and {\it only}
when!) we can interpret inflation in terms of the usual slow-roll approach with
the now small quantum fluctuations around the redefined zero mode driving the
generation of metric perturbations. 

We have studied the non-equilibrium dynamics of a `new inflation' scenario in a
self-consistent, non-perturbative framework based on a large $ N $ expansion,
including the dynamics of the scale factor and backreaction of quantum
fluctuations. Quantum fluctuations associated with superhorizon modes grow
exponentially as a result of the spinodal instabilities and contribute to the
energy momentum tensor in such a way as to end inflation consistently.

Analytical and numerical estimates have been provided that establish the regime
of validity of the classical approach.  We find that these superhorizon modes
re-assemble into an effective zero mode and unambiguously identify the
composite field that can be used as an effective expectation value of the
inflaton field whose {\em classical} dynamics drives the evolution of the scale
factor. This identification also provides the initial condition for this
effective zero mode.

A consistent criterion is provided to extract small fluctuations that will
contribute to cosmological perturbations from large non-perturbative
spinodal fluctuations. This is an important ingredient for a consistent
calculation and interpretation of cosmological perturbations.  This criterion
requires that the model must provide many more than 60 e-folds to identify the
`small perturbations' that give rise to scalar metric (curvature)
perturbations. We then use this criterion combined with the gauge invariant
approach to obtain the dynamics of the Bardeen variable and the spectrum for
scalar perturbations.

We find that during the inflationary epoch, superhorizon modes of the Bardeen
potential grow exponentially in time reflecting the spinodal
instabilities. These long-wavelength instabilities are manifest in the spectrum
of scalar density perturbations and result in an index that is less than
one, i.e. a `red' power spectrum, providing more power at long wavelength.  We
argue that this red spectrum is a robust feature of potentials that lead to
spinodal instabilities in the region in field space associated with inflation
and can be interpreted as an imprint of the phase transition on the
cosmological background. Tensor perturbations on the other hand, are not
modified by these features, they have much smaller amplitude and a
Harrison-Zeldovich spectrum.

We made contact with the reconstruction program and validated the results for
these type of models based on the slow-roll assumption, despite the fact that
our study does not involve such an approximation and is non-perturbative.

Finally we have studied the quantum to classical crossover and decoherence of
quantum fluctuations by studying the full evolution of the density matrix, thus
making contact with the concept of `decoherence without
decoherence'\cite{polarski} which is generalized to the interacting case. In
the case under consideration decoherence and classicalization is a consequence
of spinodal growth of superhorizon modes and the presence of a horizon. The
phases of the mode functions freeze out and the amplitudes of the
superhorizon modes grow exponentially during the inflationary stage, again as a
result of long-wavelength instabilities.  As a result field configurations with
large amplitudes have non-vanishing probabilities to be represented in the
dynamical density matrix.  In the situation considered, the quantum to
classical crossover of cosmological perturbations is directly related to the
classicalization of superhorizon matter field modes that grow exponentially
upon horizon crossing during inflation. The diagonal elements of the density
matrix in the Schroedinger representation can be interpreted as a classical
distribution function, whereas the off-diagonal elements are strongly
suppressed during inflation.

\section{Acknowledgements:}

The authors thank J. Baacke, K. Heitman, L. Grishchuk,  A. Singh and
M. Srednicki, E. Weinberg, E. Kolb,
A. Dolgov and D. Polarski, for conversations and discussions and
especially to C. Destri in collaboration with whom sec. II was
written. D. B. thanks the
N.S.F for partial support through the grant awards: PHY-9605186 and
INT-9216755, the Pittsburgh Supercomputer Center for grant award No: PHY950011P
and LPTHE for warm hospitality.  R. H., D. C. and S. P. K. were supported by
DOE grant DE-FG02-91-ER40682. H. J. de V.  acknowledges partial support by
NATO.


\begin{thebibliography}{99}


\bibitem{guth} A. H. Guth, Phys. Rev. {\bf D23}, 347 (1981).

\bibitem{kolb}For   thorough reviews of standard and inflationary 
cosmology see: E. W. Kolb and M. S. Turner, {\em The Early Universe}
(Addison Wesley, Redwood City, C.A. 1990). A. Linde, {\em Particle
Physics and Inflationary Cosmology}, (Harwood 1990) and ref.\cite{rev}.

\bibitem{turner} For more recent reviews see: M. S. Turner, astro-ph-9703197;
astro-ph-9703196; astro-ph-9703174; astro-ph-9703161; astro-ph-9704062;
astro-ph-9704024.  A. Linde, in Current Topics in Astrofundamental Physics,
Proceedings of the Chalonge Erice School, N.  S\'anchez and A. Zichichi
Editors, Nato ASI series C, vol. 467, 1995, Kluwer Acad. Publ.

A. R. Liddle, astro-ph-9612093, Lectures at the
Casablanca  School Morocco, 1996. 

\bibitem{lyth1}G. Smoot,   in the Proceedings of the  Vth. Erice
Chalonge School on  Astrofundamental Physics, p. 407-484, N. S\'anchez
and A. Zichichi eds., World Scientific, 1997.

A. R. Liddle and D. H. Lyth, Phys. Rep. 231, 1 (1993). 

\bibitem{rev} for reviews of inflation, see R. Brandenberger, Rev. of
Mod. Phys. {\bf 57}, 1 (1985); Int. J. Mod. Phys. {\bf A2}, 77 (1987) and
ref.\cite{kolb}.


\bibitem{origreheat} A. D. Dolgov and A. D. Linde, Phys. Lett. {\bf B116}, 329
(1982); 

L. F. Abbott, E. Farhi and M. Wise, Phys. Lett. {\bf B117}, 29 (1982).

\bibitem{branden}J. Traschen and R. Brandenberger, Phys Rev {\bf D42}, 
2491 (1990); 

Y. Shtanov, J. Traschen and R. Brandenberger, Phys. Rev. {\bf D51},
5438 (1995). 

\bibitem{kls} L. Kofman, A. Linde and A. Starobinsky,
Phys. Rev. Lett. {\bf 73}, 3195 (1994) and {\bf 76}, 1011 (1996);
Phys. Rev. D56, 3258 (1997); gr-qc/9508019 (1995).  
L. Kofman,  astro-ph/9605155 (1996).

\bibitem{barrabajo} D. Boyanovsky and H. J. de Vega,
Phys. Rev.  {\bf D47}, 2343 (1993).

\bibitem{us1} D. Boyanovsky, H. J. de Vega, R. Holman, D.-S. Lee and A. Singh,

Phys. Rev. {\bf D51}, 4419 (1995); 

D. Boyanovsky, M. D'Attanasio, H. J. de
Vega, R. Holman and D. S. Lee, 

Phys. Rev. {\bf D52}, 6805 (1995);

For reviews see, D. Boyanovsky, H. J. de Vega and R. Holman,
in the Proceedings of the Second Paris Cosmology Colloquium, Observatoire de
Paris, June 1994, p. 127-215, H. J. de Vega and N. S\'anchez Editors, World
Scientific, 1995; D. Boyanovsky, M. D'Attanasio, H. J. de Vega, R. Holman and
D.-S. Lee, `New aspects of reheating', in the Proceedings of the
Erice Chalonge School, `String Gravity and Physics at the Planck
Energy Scale', NATO ASI, N. S\'anchez and A. Zichichi Editors, Kluwer
1996, p. 451-492. 

\bibitem{mink} D. Boyanovsky, H.J. de Vega, R. Holman, J.F.J. Salgado,
Phys. Rev. {\bf D54}, 7570 (1996). 

D. Boyanovsky, H. J. de
Vega and R. Holman in the Proceedings of the  Vth. Erice Chalonge School on
Astrofundamental Physics, p. 183-270, N. S\'anchez and A. Zichichi eds., World
Scientific, 1997.

\bibitem{late} D. Boyanovsky, C. Destri, H.J. de Vega, R. Holman and
J.F.J. Salgado, `Asymptotic Dynamics in Scalar Field Theory: Anomalous
       Relaxation', hep-ph/9711384.

\bibitem{frw} D. Boyanovsky, H. J. de Vega, and R. Holman, Phys. Rev.  {\bf
D49}, 2769 (1994).

\bibitem{frw2} D. Boyanovsky, D. Cormier, H. J. de Vega, R. Holman,
A. Singh, M. Srednicki,  

Phys. Rev. {\bf D56}, 1939 (1997).

\bibitem{De Sitter} D. Boyanovsky, D. Cormier, H. J. de Vega and R. Holman, 

Phys. Rev. {\bf D55}, 3373 (1997).

\bibitem{din} `Non-Perturbative Quantum Dynamics of a New Inflation
Model', D. Boyanovsky, D. Cormier, H. J. de Vega,
R. Holman and S. P. Kumar, hep-ph/9709232 to appear in Phys. Rev. {\bf
D}, 15 february 1998.

\bibitem{par} D. Boyanovsky, H.J. de Vega, R. Holman and J. F. J. Salgado,
astro-ph/9609007, to appear in the Proceedings of the Paris Euronetwork Meeting
`String Gravity'.

\bibitem{boylee}  D. Boyanovsky, D-S. Lee, and A. Singh, Phys. Rev.
{\bf D48}, 800 (1993).

\bibitem{datan} D. Boyanovsky, M. D'Attanasio, H. J. de Vega and R. Holman, 

Phys. Rev. {\bf D54}, 1748 (1996), and references therein.

\bibitem{grure} H.J. de Vega and J. F. J. Salgado, Phys. Rev. {\bf D 56},
    6524 (1997). 

\bibitem{tkachev} S. Yu. Khlebnikov and I.I. Tkachev, Phys. Rev. Lett. {\bf
77}, 219 (1996); S. Yu. Khlebnikov and I.I. Tkachev, hep-ph/9608458 (1996).

\bibitem{son} D.T. Son, Phys. Rev. {\bf D54}, 3745 (1996);
hep-ph/9601377.  

\bibitem{symrest} L. Kofman, A. Linde and A. Starobinsky, Phys. Rev. Lett. {\bf
76}, 1011, (1996); I.I Tkachev, Phys. Lett. {\bf B376}, 35 (1996); A. Riotto
and I.I. Tkachev, Phys. Lett. {\bf B385}, 57 (1996); E.W. Kolb and A. Riotto,
astro-ph/9602095 (1996).

\bibitem{kaiser} D.I. Kaiser, Phys. Rev {\bf D53}, 1776 (1996) and
{\bf D56}, 706 (1997), hep-ph/9707516.

\bibitem{yoshimura} H. Fujisaki, K. Kumekawa, M. Yamaguchi and M.Yoshimura,
Phys. Rev. {\bf D53}, 6805 (1996); M. Yoshimura, Progr. Theor. Phys. 
{\bf 94}, 873 (1995); hep-ph/9605246.

\bibitem{ctp}J. Schwinger, J. Math. Phys. {\bf 2}, 407 (1961); P. M. Bakshi and
K. T. Mahanthappa, J. Math. Phys. {\bf 4}, 1 (1963); {\it ibid}, 12;
L. V. Keldysh, Sov. Phys. JETP {\bf 20}, 1018 (1965); A. Niemi and G. Semenoff,
Ann. of Phys. (N.Y.)  {\bf 152}, 105 (1984); Nucl. Phys. B [FS10], 181 (1984);
E. Calzetta, Ann. of Phys. (N.Y.) {\bf 190}, 32 (1989); R. D. Jordan,
Phys. Rev. {\bf D33}, 444 (1986); N. P. Landsman and C. G. van Weert,
Phys. Rep. {\bf 145}, 141 (1987); R. L. Kobes and K. L. Kowalski,
Phys. Rev. {\bf D34}, 513 (1986); R. L. Kobes, G. W. Semenoff and N. Weiss,
Z. Phys. {\bf C29}, 371 (1985).

\bibitem{hu} For a thorough exposition of non-equilibrium methods in cosmology
see, for example: E. Calzetta and B.-L. Hu, Phys. Rev. {\bf D35}, 495 (1988);
{\it ibid } {\bf D37}, 2838 (1988); J. P. Paz, Phys. Rev. {\bf D41},
1054 (1990); 
{\it ibid } {\bf D42}, 529 (1990); B.-L. Hu in {\it Bannf/Cap Workshop on 
Thermal Field Theories: proceedings}, edited by F.C. Khanna, R. Kobes, 
G. Kunstatter, H. Umezawa (World Scientific, Singapore, 1994), p.309 and in the
{\it Proceedings of the Second Paris Cosmology Colloquium,
Observatoire de Paris}, 
edited by H. J. de Vega and N. S\'anchez (World Scientific, Singapore, 1995), 
p.111 and references therein.


\bibitem{leutwyler} H. Leutwyler and S. Mallik, Ann. of Phys. (N.Y.) {\bf 205},
1 (1991).


\bibitem{jackiwetal} O. Eboli, R. Jackiw and S.-Y. Pi, Phys. Rev. {\bf D37},
3557 (1988); 

M.Samiullah, O. Eboli and S-Y. Pi, Phys.  Rev. {\bf D44}, 2335 (1991).

 J. Guven, B. Liebermann and C. Hill, Phys. Rev. {\bf D39}, 438
(1989).

\bibitem{birrell} N. D. Birrell and P.C.W. Davies, {\it Quantum fields
in curved space} (Cambridge Univ. Press, Cambridge, 1986).


\bibitem{baacke} J. Baacke, K. Heitmann and C. P\"atzold, Phys. Rev. {\bf D55},
2320 (1997), hep-ph/9706274, hep-th/9711144 and hep-ph/9712506.

\bibitem{aands} M. Abramowitz and I.E. Stegun (eds.), Handbook of Mathematical
Functions (National Bureau of Standards, Washington, D.C., 1972), chapter 13.


\bibitem{defe} `The Scalar, Vector and Tensor Contributions to CMB
anisotropies from Topological Defects', by N. Turok, Ue-Li Pen, U. Seljak
astro-ph/9706250 (1997).
`The case against scaling defect models of cosmic
structure formation', A. Albrecht, R. A. Battye, J. Robinson, astro-ph/9707129
(1997). 
`CMB Anisotropy Induced by Cosmic Strings on Angular Scales
$>~ 15'$', by B. Allen, R. R. Caldwell, S. Dodelson, L. Knox,
E. P. S. Shellard, A. Stebbins, astro-ph/9704160 (1997).

\bibitem{lyth2} D. H. Lyth, hep-ph-9609431 (1996). 
S. Dodelson, W. H. Kinney and E. W. Kolb, astro-ph-9702166.

\bibitem{linde2} A.D. Linde, Phys. Lett. B116, 335 (1982).
 A. Vilenkin and L. H. Ford, Phys. Rev. D26, 1231 (1982).
A. Vilenkin, Nucl. Phys. B226, 504 (1983);
Nucl. Phys. B226, 527 (1986).

\bibitem{vilenkin} A. Vilenkin, Phys. Lett. B115, 91 (1982).

\bibitem{stein} P. J.  Steinhardt and M. S. Turner, Phys. Rev. D29,
2162, (1984). 

\bibitem{guthpi}A. Guth and S-Y. Pi, Phys. Rev. {\bf D32}, 1899 (1985).

\bibitem{motola} For non-equilibrium methods in different contexts see for
example: F. Cooper, J. M . Eisenberg, Y. Kluger, E. Mottola, B. Svetitsky,
Phys. Rev. Lett. 67, 2427 (1991); F. Cooper, J. M. Eisenberg, Y, Kluger,
E. Mottola, B. Svetitsky, Phys. Rev. D48, 190 (1993).

\bibitem{largen} F. Cooper and E. Mottola, Mod. Phys. Lett. A 2, 635 (1987);
F. Cooper, S. Habib, Y. Kluger, E. Mottola, J. P. Paz, P. R. Anderson,
Phys. Rev. D50, 2848 (1994). F. Cooper, S.-Y. Pi and P. N. Stancioff,
Phys. Rev. D34, 3831 (1986). F. Cooper and E. Mottola, Phys. Rev. D36, 3114
(1987). 

F. Cooper, Y. Kluger, E. Mottola, J. P. Paz,
Phys. Rev. D51, 2377 (1995). 

\bibitem{ramsey}S. A. Ramsey, B. L. Hu, 
Phys. Rev. D56, 678 (1997).

\bibitem{dcc} D. Boyanovsky, H. J. de Vega and R. Holman,
Phys. Rev. {\bf D 51}, 734 (1995).

D. Boyanovsky, H. J. de Vega, R. Holman and S. Prem Kumar,

Phys. Rev. {\bf D56},  3929 and 5233 (1997).

The last reference under \cite{largen}.

\bibitem{polarski} D. Polarski and A. A. Starobinsky, Class. Quant. Grav. {\bf
13}, 377 (1996); 

J. Lesgourgues, D. Polarski and A. A. Starobinsky,
Nucl. Phys. B497, 479 (1997).

\bibitem{grishchuk} L. P. Grishchuk, Phys. Rev. D 45, 4717 (1992).

\bibitem{mukhanov} V. F. Mukhanov, H. A. Feldman and R. H. Brandenberger,
Phys. Rep. 215, 293 (1992). 

\bibitem{bardeen} J. Bardeen, Phys. Rev. D22, 1882 (1980).

\bibitem{caldwell} R. R. Caldwell, Class. Quant. Grav. 13, 2437 (1995).

J. Martin and D. J. Schwarz, gr-qc-9704049 (1997).

\bibitem{gorski} K. M. Gorski, A. J. Banday, C. L. Bennett, G. Hinshaw,
A. Kogut and G. F. Smoot, (astro-ph-9601063) (submitted to
ApJ. Lett.).


\bibitem{grishchuk2} L. P. Grishchuk, Phys. Rev. D. 52, 5549, (1995);
Proceedings of the Erice Chalonge School NATO ASI on `String Gravity and
Physics at the Planck Scale', Ed. N.  S\'anchez and A. Zichichi,
(Kluwer, 1996), p. 369; Phys. Rev. D53 (1996) 6784; Proceedings of
the NATO ASI on `Current Topics in Astrofundamental Physics', 
Ed. N.  Sanchez and A. Zichichi (Kluwer, 1995), p. 205.
 
\bibitem{reconstruction} J. E. Lidsey, A. R. Liddle, E. W. Kolb,
E. J. Copeland, T. Barreiro, M. Abney, 

Rev. Mod. Phys. 69, 373, (1997).  

\bibitem{lyth} E. D. Stewart and D. H. Lyth, Phys. Lett. B302, 171
(1993).

\bibitem{salman}  S. Habib, Phys. Rev. D 46, 2408 (1992); Phys. Rev. D
42, 2566, (1990); 

S. Habib and R. Laflamme, Phys. Rev. D42, 4056,
(1990), and references therein. 

\bibitem{ps} See eq.(53) in the first reference under \cite{polarski}.

\bibitem{stei} I. Zlatev, G. Huey and P. J. Steinhardt, astro-ph/9709006.
\end{thebibliography}
\end{document}